\documentclass[11pt]{article}
\pdfoutput=1
\usepackage{soul}
\usepackage{jheppub}
\usepackage{amsfonts}
\usepackage{amsmath}
\allowdisplaybreaks[4]         
\usepackage{amssymb}   
\usepackage{euscript}       
\usepackage{color}          
\usepackage{tensor}   
\usepackage{float}  
\usepackage{caption}
\usepackage{subcaption}   
\usepackage{tikz,tikz-cd}
\usepackage[T1]{fontenc} 

\newcommand{\req}[1]{(\ref{#1})} 

\newcommand{\bea}{\begin{eqnarray}}
\newcommand{\eea}{\end{eqnarray}}
\newcommand{\ba}{\begin{eqnarray}}
\usepackage{braket}
\newcommand{\ea}{\end{eqnarray}}
\newcommand{\nn}{\nonumber \\}

\newcommand{\beq}{\begin{equation}}
\newcommand{\eeq}{\end{equation} }
\newcommand{\beqa}{\begin{eqnarray}}

\newcommand{\eeqa}{\end{eqnarray}}
\newcommand{\beqar}{\begin{eqnarray*}}
\newcommand{\eeqar}{\end{eqnarray*}}

\newcommand{\cO}{\mathcal{O}}
\newcommand{\cT}{\mathcal{T}}
\def\th{\hbox{\char'336}}
\def\edth{\hbox{\char'360}}

\newcommand{\be}{\begin{equation}}
\newcommand{\ee}{\end{equation}}

\newcommand{\dvtag}{\\ &}
\newcommand{\dvvtag}{\right.\\ &\left.}

\renewcommand{\req}[1]{(\ref{#1})}

\newcommand{\Dbar}{\text{\DH}}

\title{ \boldmath  The universal Teukolsky equations and black hole perturbations in higher-derivative gravity}

\author[a]{Pablo A. Cano,}
\author[b]{Kwinten Fransen,}
\author[a]{Thomas Hertog}
\author[a]{and Simon Maenaut}

\affiliation[a]{Institute for Theoretical Physics, KULeuven.
Celestijnenlaan 200D, B-3001 Leuven, Belgium \vspace{0.1cm}}

\affiliation[b]{Department of Physics, University of California, Santa Barbara, CA93106, USA \vspace{0.1cm}}

\emailAdd{pabloantonio.cano@kuleuven.be}
\emailAdd{kfransen@ucsb.edu}
\emailAdd{thomas.hertog@kuleuven.be}
\emailAdd{simon.maenaut@kuleuven.be}

\date{\today}
\abstract{
We reduce the study of perturbations of rotating black holes in higher-derivative extensions of general relativity to a system of decoupled radial equations that stem from a set of universal Teukolsky equations. We detail a complete computational strategy to obtain these decoupled equations in general higher-derivative theories. We apply this to six-derivative gravity to compute the shifts in the quasinormal mode frequencies with respect to those of Kerr black holes in general relativity. At linear order in the angular momentum we reproduce earlier results obtained with a metric perturbation approach. In contrast with this earlier work, however, the method given here applies also to post-merger black holes with significant spin, which are of particular observational interest.
 }

\begin{document} 
\maketitle

\newpage

\section{Introduction}

Gravitational wave (GW) observations probe the characteristic spectrum of quasinormal modes (QNMs) of black holes \cite{LIGOScientific:2018mvr,LIGOScientific:2020iuh,LIGOScientific:2020ibl,LIGOScientific:2021usb, Capano:2020dix, Capano:2021etf, Capano:2022zqm,Ma:2023cwe,Ma:2023vvr}. Moreover, they will do so with high precision in the not-too-distant future \cite{Hild:2010id,Punturo:2010zz,Somiya:2011np, aso2013interferometer,LIGOScientific:2014pky,LIGOScientific:2016emj,lisa2018lisa,Shoemaker:2019bqt,Cabero:2019zyt,Reitze:2019iox,abbott2020prospects,KAGRA:2020cvd,DiPace:2021hxc,Cahillane:2022pqm}. 

QNMs of black holes are only significantly excited in highly dynamical processes such as binary coalescences. However the spectrum of QNMs is independent of the specific excitation mechanism, making it a key strong-field fingerprint of the stationary state to which systems like binary collisions relax \cite{LIGOScientific:2021sio}. This is especially so since the QNM spectrum of black holes in general relativity is fully determined in terms of just two parameters: the black hole mass and angular momentum. 
Advanced GW observations therefore offer a promising route to constrain compact objects that are alternatives to black holes and even modifications to general relativity  \cite{Cardoso:2016oxy,Glampedakis:2017dvb,Chung:2018dxe,Berti:2018vdi,Carballo-Rubio:2018jzw,Chung:2021roh,Chakraborty:2022zlq}. 

Working in Einstein's theory, the precise predictions of QNMs in terms of the black hole mass and angular momentum were calculated long ago. This was first done for static black holes, based on metric perturbations exploiting spherical symmetry \cite{regge1957stability, vishveshwara1970stability, zerilli1970effective, press1971long, moncrief1974gravitational}. Later this was done for Kerr black holes using curvature perturbations, the algebraically special (Petrov type D) nature of the Kerr metric and its hidden symmetries \cite{Newman:1961qr, price1972nonspherical,bardeen1973radiation,teukolsky1972rotating, teukolsky1973, teukolsky1974perturbations,Frolov:2017kze}. A key property of the Kerr QNM spectrum in general relativity is that it is fully governed by the Teukolsky equation, a single (decoupled) separable second-order differential equation.\footnote{Notwithstanding these impressive results, high-precision numerical calculations, alternative approaches to various limits, and analytical properties of the spectrum of rotating black holes in vacuum four-dimensional general relativity continue to be active areas of investigation. See e.g. \cite{Cook:2014cta,Stein:2019mop,Aminov:2020yma,Tanay:2022era,Gregori:2022xks,Fransen:2023eqj}.}

In contrast with these results, we lack similarly detailed predictions of the gravitational spectrum of compact objects that differ from rotating black holes in general relativity, either by their very nature or on account of modifications to general relativity. Here, despite a wide array of results for non-rotating  \cite{Cardoso:2009pk,Molina:2010fb,Blazquez-Salcedo:2016enn,Blazquez-Salcedo:2017txk,Tattersall:2018nve,Konoplya:2020bxa,Moura:2021eln,Moura:2021nuh,deRham:2020ejn,Cardoso:2018ptl,McManus:2019ulj} and slowly-rotating compact objects \cite{Pani:2012bp,Pani:2013pma,Pierini:2021jxd,Wagle:2021tam,Srivastava:2021imr,Cano:2021myl,Pierini:2022eim}, it has remained a challenging open problem to obtain the spectrum of highly spinning black holes. This state-of-affairs is particularly problematic since it is precisely this regime that is of key observational interest \cite{LIGOScientific:2021djp,LIGOScientific:2021psn}.

The reason it is difficult to find the QNM spectrum in the presence of significant rotation is simple: the fortunate extended symmetry of Kerr black holes is generally lost. This being said, recently significant progress was made to extend the use of the Teukolsky equation to a more general setting \cite{Li:2022pcy,Hussain:2022ins}. Yet, the key challenge in the light of forthcoming GW observations remains: we need a sufficiently general theoretical framework that can be employed in explicit theories to extract specific QNM predictions that are precise enough to compare theory with (future) observations. The goal of this work is to provide a complete solution to this problem. 

We first construct a set of universal Teukolsky equations, along the same lines as \cite{Li:2022pcy,Hussain:2022ins}. That is, we formulate in full generality the specific combination of Bianchi identities that reduce to the usual Teukolsky equations when the background is a Ricci-flat Petrov type D spacetime. The universal Teukolsky equations apply to any background geometry and to any theory, since the theory-dependence only enters through the presence of an effective stress-energy tensor on the right-hand side of the Einstein equation. Next we describe in detail how to evaluate these equations for fluctuations around rotating black hole solutions that deviate perturbatively from Kerr. Finally, by following the approach of \cite{Cano:2020cao}, we show how to effectively separate these equations into a set of radial equations.

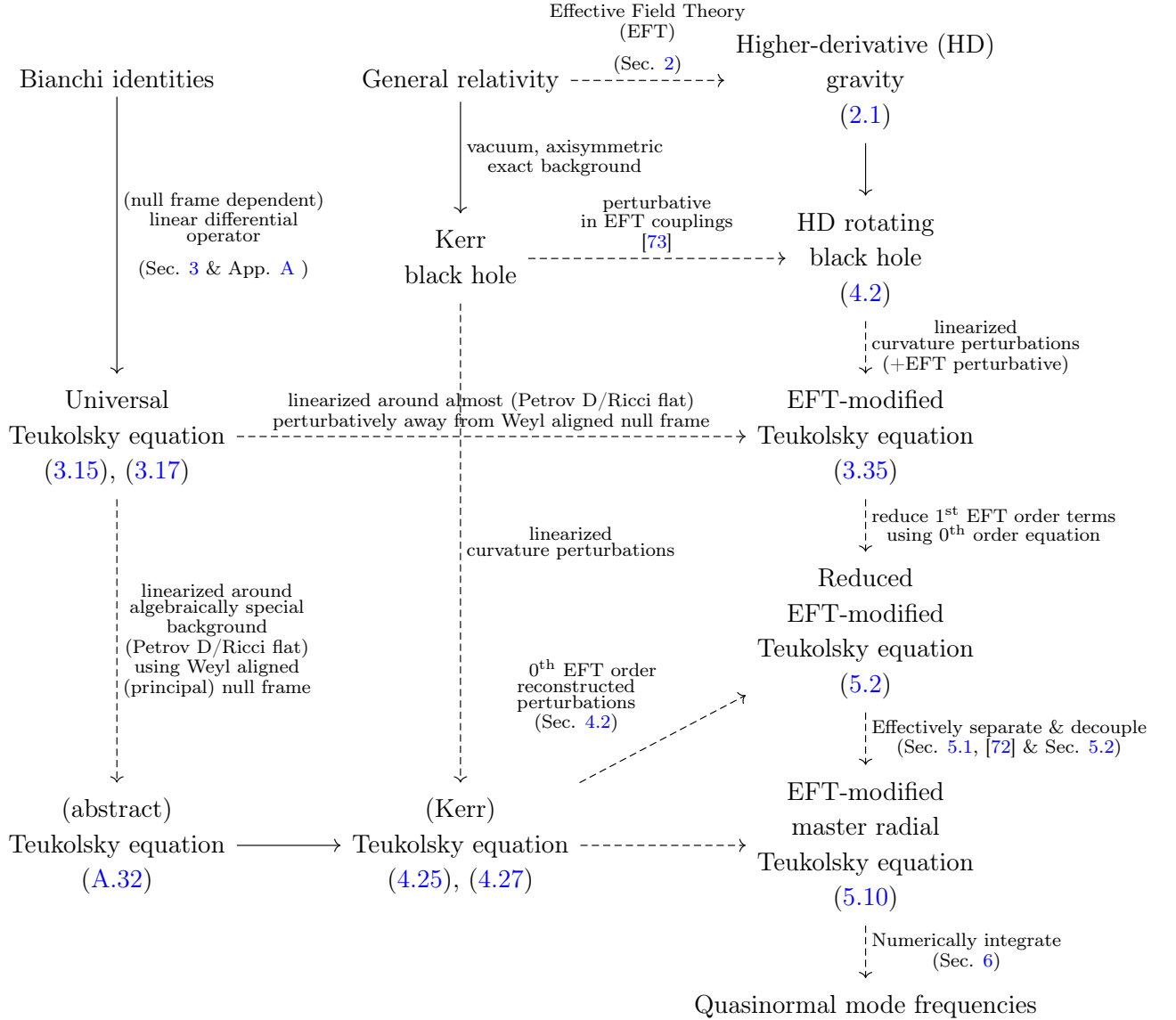
\begin{figure}[H]
	\centering
	\begin{tikzcd}[column sep=1.4cm, row sep=0.8cm]
		\begin{matrix}
			\text{pseudo-Riemannian} \\	
			\text{kinematics}
		\end{matrix} 
		\arrow[rr,dash,    start anchor={[xshift=-16ex, yshift=-8ex]},
		end anchor={[xshift=-0ex, yshift=-8ex]}] 
		& 		\begin{matrix}
			\phantom{testtest}	\text{Gravitational} \\	
			\phantom{testtest}	\text{dynamics}
		\end{matrix} 
		& \phantom{temp} \\
		\phantom{test} & & \\
		\text{Bianchi identities} \arrow[dd,"\substack{\text{(null frame dependent)} \\ \text{linear differential}\\ \text{operator} \\ \\ \text{(Sec. \ref{sec:Teukolsky} \& App. \ref{app:NPGHP} )}}"] 
		& \text{General relativity} \arrow[r,"\substack{\text{Effective Field Theory}\\ \text{(EFT)} \\ \\ \text{(Sec. \ref{sec:EFTgravity})}}",dashed] \arrow[d,"\substack{\text{vacuum, axisymmetric} \\ \text{exact background}}"] 
		& 
		\begin{matrix}
			\text{Higher-derivative (HD)} \\
			\text{gravity} \\ \text{\eqref{eq:EFTofGR}}
		\end{matrix} \arrow[d," "] \\ 
		& 	
		\begin{matrix}
			\text{Kerr} \\ \text{black hole}
		\end{matrix} 
		\arrow[ddd,"\substack{\text{linearized} \\ \text{curvature perturbations}}",dashed]  \arrow[r,"\substack{\text{perturbative} \\ \text{in EFT couplings} \\ \text{\cite{Cano:2019ore}}}",dashed] & 
		\begin{matrix}
			\text{HD rotating} \\ \text{black hole} \\ \text{\eqref{rotatingmetric}}
		\end{matrix}
		\arrow[d,"\substack{\text{linearized} \\ \text{curvature perturbations} \\ \text{(+EFT perturbative)}}",dashed]  \\ 
		\begin{matrix}
			\text{Universal} \\ \text{Teukolsky equation} \\ \text{\eqref{eqn:universalteukolsky0}, \eqref{eqn:universalteukolsky4}}
		\end{matrix} 
		\arrow[dd,"\substack{\text{linearized around} \\ \text{algebraically special} \\ \text{background} \\ \text{(Petrov D/Ricci flat)} \\ \text{using Weyl aligned} \\ \text{(principal) null frame}}",dashed] \arrow[rr,"\substack{\text{linearized around almost (Petrov D/Ricci flat)} \\ \text{perturbatively away from Weyl aligned null frame}}",dashed]
		& &  \begin{matrix}
			\text{EFT-modified} \\ \text{Teukolsky equation} \\ \text{\eqref{goalequations}}
		\end{matrix} \arrow[d,"\substack{\text{reduce 1\textsuperscript{st} EFT order terms} \\ \text{using 0\textsuperscript{th} order equation}}",dashed]
		\\
		\phantom{test} & & 	\begin{matrix}
			\text{Reduced} \\ \text{EFT-modified} \\ \text{Teukolsky equation} \\ \text{\eqref{nonradial}}
		\end{matrix} \arrow[d,"\substack{\text{Effectively separate \& decouple} \\ \text{(Sec. \ref{subsec:evalandsep}, \cite{Cano:2020cao} \& Sec. \ref{subsec:decouple})}}",dashed] \\
		\begin{matrix} 
			\text{(abstract)} \\ \text{Teukolsky equation} \\
			\text{\eqref{eqn:teukolsky}} 
		\end{matrix} \arrow[r,"\substack{\text{} \\ \text{}}"]
		& 	\begin{matrix} 
			\text{(Kerr)} \\ \text{Teukolsky equation} \\
			\text{\eqref{angularequation}, \eqref{Teukoperator}} 
		\end{matrix} 
		\arrow[ru,"\substack{\phantom{test} \text{0\textsuperscript{th} EFT order} \\ \text{reconstructed} \\ \text{perturbations} \\ (\text{Sec. \ref{subsec:reconstruction})}}",dashed] \arrow[r,"\substack{\text{} \\ \text{}}",dashed]
		& 	\begin{matrix}
			\text{EFT-modified} \\ \text{master radial} \\ \text{Teukolsky equation} \\ \text{\eqref{radial2}}  \arrow[d,"\substack{\text{Numerically integrate} \\ \text{(Sec. \ref{sec:QNM6})}}",dashed]
		\end{matrix}  \\
		\phantom{test} & & 	\begin{matrix}
			\text{Quasinormal mode frequencies} 
		\end{matrix}  \\
	\end{tikzcd}
	\caption{Schematic overview of the approach we develop in this paper to compute the spectrum of quasinormal modes of black holes with significant spin beyond Kerr black holes in general relativity. Dashed lines indicate an approximate or perturbative relation.}
	\label{fig:overview}
\end{figure}

This three-step strategy can be applied in full generality in any theory.\footnote{In practice, some steps of the computation rely on having an analytic expression for the corrected Kerr background expressed as a power series in the spin (cf. Section~\ref{subsec:background}), which is needed to obtain the radial equations analytically. Most extensions of general relativity allow for such solutions, including general higher-derivative gravities and scalar-tensor theories like Einstein-scalar-Gauss-Bonnet gravity and dynamical Chern-Simons theory \cite{Cano:2019ore}. Our method can also be applied even if an analytic solution is not known, but it becomes more involved in that case.} Here we apply it explicitly to the case of a general effective field theory (EFT) extension of general relativity with six-derivative corrections.
We obtain the modified radial Teukolsky equations analytically in a slow-spin expansion and we integrate these  to find the corrections to the Kerr QNM frequencies.  In addition to several consistency checks, we perform a highly nontrivial test of the validity of our approach by matching our results against earlier studies based on metric perturbations \cite{Cano:2021myl}. Contrary to the latter, the approach implemented here lends itself more readily to a high-order expansion in spin, hence allowing us to study perturbations of black holes with higher angular momentum. The entire approach is schematically summarized in Figure \ref{fig:overview}. 

The structure of the paper is as follows. First, in Section \ref{sec:EFTgravity}, we review the broad class of EFT extensions of general relativity for which we wish to compute QNMs. In the next three sections we devise a strategy to accomplish this goal. In Section \ref{sec:Teukolsky} we derive the universal Teukolsky equations and outline how to evaluate these perturbatively around a Kerr black hole, in the spirit of \cite{Li:2022pcy,Hussain:2022ins}. In Section \ref{sec:computation} we provide the required ingredients to evaluate these equations explicitly. In Section \ref{sec:masterradial} we explain how the equations can effectively be separated, and thereby reduced to a set of radial equations. 
In Section \ref{sec:QNM6} we obtain these equations for the six-derivative EFT extension of Einstein gravity introduced in Section \ref{sec:EFTgravity}. We solve these numerically in two different ways and obtain explicit results for the shifts in the QNM frequencies that pass several consistency tests. We provide some closing remarks in Section \ref{sec:conclusions}.

\newpage 

\section{Effective field theory of gravity}\label{sec:EFTgravity}

To compute quasinormal modes and make actual model-based predictions one needs to fix a theory to work with. This is an important aspect of the problem of moving beyond Einstein gravity and has the advantage over more phenomenological approaches that it can be folded into different aspects of the binary two-body problem, rather then merely parameterizing the ringdown. Although we will set-up our computational framework as generally as possible, part of our goal is to get explicit results. Therefore, we must fix a theory. To do so, we will work with the leading order corrections to general relativity within an EFT perspective on gravity. As an additional advantage, this approach aligns naturally with certain technical assumptions we will make for practical reasons. In particular, we will work perturbatively away from general relativity. 

The EFTs we consider consist of all possible covariant actions that can be constructed from the curvature alone. Assuming a characteristic length scale $\ell$ which is small compared to the characteristic scales of the binary problem, such a theory naturally organizes itself as a perturbative series in higher-derivative terms
\begin{equation}\label{eq:EFTofGR}
S=\frac{1}{16\pi G}\int d^4x\sqrt{-g}\left\{R+\ell^4\mathcal{L}_{(6)}+\ell^6\mathcal{L}_{(8)}+\ldots \right\}\, .
\end{equation}
In four dimensions, the four-derivative terms do not modify the vacuum Einstein gravity solutions, which is the reason why these do not appear in the EFT above.
In this paper we will focus on the leading corrections to Einstein gravity, corresponding to the general six-derivative Lagrangian \cite{Cano:2019ore}
\begin{align}\label{cubiclagrangian}
\mathcal{L}_{(6)} &= \lambda_{\rm ev}\tensor{R}{_{\mu\nu }^{\rho\sigma}}\tensor{R}{_{\rho\sigma }^{\delta\gamma}}\tensor{R}{_{\delta\gamma }^{\mu\nu }}+\lambda_{\rm odd}\tensor{R}{_{\mu\nu}^{\rho\sigma}}\tensor{R}{_{\rho\sigma }^{\delta\gamma }} \tensor{\tilde R}{_{\delta\gamma }^{\mu\nu }} \, ,
\end{align}
where 
\begin{equation}
{\tilde R}^{\mu\nu\rho\sigma}=\frac{1}{2}\epsilon^{\mu\nu\alpha\beta}\tensor{R}{_{\alpha\beta}^{\rho\sigma}}\, 
\end{equation}
is the dual Riemann tensor. Note that, precisely because of the appearance of ${\tilde R}^{\mu\nu\rho\sigma}$, the second cubic term violates parity. It is convenient for the discussion in the next section to write the equations of motion of this theory as
\begin{equation}\label{EinsteinEq}
G_{\mu\nu}=\ell^4 T_{\mu\nu}^{(6)}\, ,
\end{equation}
where $G_{\mu\nu}$ is the Einstein tensor and $T_{\mu\nu}^{(6)}$, playing the role of an effective stress-energy tensor, reads
\begin{equation}
T_{\mu\nu}^{(6)}=-\tensor{R}{_{\mu}^{\sigma\alpha\beta}}P^{(6)}_{\nu\sigma\alpha\beta}+\frac{1}{2}\mathcal{L}_{(6)}g_{\mu\nu}-2\nabla^{\alpha}\nabla^{\beta}P^{(6)}_{\mu\alpha\nu\beta}\, ,
\end{equation}
where 
\begin{equation}
P^{(6)}_{\mu\nu\rho \sigma}=3\lambda_{\rm ev} \tensor{R}{_{\mu\nu}^{\alpha \beta}}\tensor{R}{_{\alpha\beta\rho\sigma}}+\frac{3\lambda_{\rm odd}}{2}\left(\tensor{R}{_{\mu\nu}^{\alpha \beta}}\tensor{\tilde R}{_{\alpha\beta\rho\sigma}}+\tensor{R}{_{\rho\sigma}^{\alpha \beta}}\tensor{\tilde R}{_{\alpha\beta\mu\nu}}\right)\, .
\end{equation}
One may also include eight-derivative corrections, with a general Lagrangian given by \cite{Endlich:2017tqa}
\begin{align}
\mathcal{L}_{(8)}&=\epsilon_1\mathcal{C}^2+\epsilon_2\tilde{\mathcal{C}}^2+\epsilon_3\mathcal{C}\tilde{\mathcal{C}}\, ,
\end{align}
with
\begin{equation}
\mathcal{C}=R_{\mu\nu\rho\sigma} R^{\mu\nu\rho\sigma}\, ,\quad \tilde{\mathcal{C}}=R_{\mu\nu\rho\sigma} \tilde{R}^{\mu\nu\rho\sigma}\, .
\end{equation}
Since the main goal of this paper is to illustrate the validity of our approach to compute black hole QNMs, we will only consider the six-derivative theories. The study of perturbations in the eight-derivative theory will be carried out in a coming publication \cite{QNMbeyondKerr}.  

Although interesting, we will not include additional fields as in say \cite{Cardoso:2009pk,Molina:2010fb,Blazquez-Salcedo:2016enn,Blazquez-Salcedo:2017txk,Tattersall:2017erk,Franciolini:2018uyq,Tattersall:2018nve,Charmousis:2019fre,Okounkova:2019dfo,Pierini:2021jxd,Li:2022pcy,Hussain:2022ins,Garcia-Saenz:2022wsl,Pierini:2022eim}, but most of our results could straightforwardly be extended to those theories as well.  

\section{The universal Teukolsky equation}\label{sec:Teukolsky}

In this section, we derive what we call the ``universal Teukolsky equations''. This should simply be taken to mean that we write down in full generality a particular combinations of Bianchi identities that reduce to the Teukolsky equations for perturbations around a Petrov type D spacetime. This approach has previously been taken to higher-order perturbations of rotating black holes in general relativity \cite{Campanelli:1998jv}. As it is a particularly technical exercise, we shall only present the essential pieces here and refer the reader to Appendix \ref{app:NPGHP} for more details. The final expressions can be found in \eqref{eqn:universalteukolsky0} and \eqref{eqn:universalteukolsky4}\footnote{The expressions found here are, in part, described in \cite{Fransen:2022jhq}.}.

\subsection{Derivation}

The schematic form of the Teukolsky equation is as follows
\begin{equation}\label{eqn:teukolskyschematic}
	\hat{D}[e_a{}^{\mu},\gamma_{abc}] \left(\nabla_{[\gamma}R_{\alpha \beta] \sigma \rho}\right) = 0 \, ,
\end{equation}
with $\hat{D}[e_a{}^{\mu},\gamma_{abc}]$ a linear, first order differential operator depending on the choice of a null frame $e_a{}^{\mu}$ and the associated spin connection $\gamma_{abc}$. That is to say, the Teukolsky equation is in essence a linear differential operator acting on the differential Bianchi identities. The frame is often explicitly written as\footnote{Note that we use $\bar{m}^{\mu}$ to denote the conjugate of $m^{\mu}$ but in the rest of the quantities conjugation will be denoted by $^*$.}
\begin{equation}\label{eqn:frame}
	\tensor{e}{_{1}^{\mu}}= l^{\mu}\, ,\quad \tensor{e}{_{2}^{\mu}}=n^{\mu}\, ,\quad \tensor{e}{_{3}^{\mu}}=m^{\mu}\, ,\quad \tensor{e}{_{4}^{\mu}}=\bar{m}^{\mu}\,  ,
\end{equation}
and is chosen to satisfy
\begin{equation}
	g_{\mu\nu}=-2l_{(\mu}n_{\nu)}+2m_{(\mu}\bar{m}_{\nu)}\, .
\end{equation}
In terms of the frame, the spin connection is defined as
\begin{equation}
\gamma_{abc} = e_a{}^{\mu}e_c{}^{\nu}\nabla_{\nu}e_{b}{}_{\mu} \, .
\end{equation}
We shall follow the Newman-Penrose (NP) and Geroch-Held-Penrose (GHP) approaches to \eqref{eqn:teukolskyschematic} but without making any assumptions on the spacetime or NP-frame along the way. It should be noted that although these methods are well-suited to the task, they are notation-heavy. We mostly follow the convention of \cite{Pound:2021qin}. \\  

For the frame \eqref{eqn:frame}, the different components of the spin connection are denoted by
\begin{equation}\label{eqn:spincoefficients}
	\begin{aligned}
\kappa &= -m^{\mu}l^{\nu}\nabla_{\nu}l_{\mu} \, , \qquad 
\sigma = -m^{\mu}m^{\nu}\nabla_{\nu}l_{\mu} \, , \qquad 
 \sigma'  =n^{\mu}\bar{m}^{\nu}\nabla_{\nu}\bar{m}_{\mu} \, , \qquad
\kappa' = n^{\mu}n^{\nu}\nabla_{\nu}\bar{m}_{\mu}  \, , \\ \\ 
\rho	 &= -m^{\mu}\bar{m}^{\nu}\nabla_{\nu}l_{\mu} \, ,  \qquad
\rho' = n^{\mu}m^{\nu}\nabla_{\nu}\bar{m}_{\mu} \, ,  \qquad
\tau	 = -m^{\mu}n^{\nu}\nabla_{\nu}l_{\mu} \, , \qquad
\tau' = n^{\mu}l^{\nu}\nabla_{\nu}\bar{m}_{\mu} \, ,  \\ \\
\epsilon &= -\frac{1}{2}(n^{\mu}l^{\nu}\nabla_{\nu}l_{\mu} + m^{\mu}l^{\nu}\nabla_{\nu}\bar{m}_{\mu}  ) \, , \qquad 
\epsilon' = \frac{1}{2}(n^{\mu}n^{\nu}\nabla_{\nu}l_{\mu} + m^{\mu}n^{\nu}\nabla_{\nu}\bar{m}_{\mu}) \, , \\ \\
\beta' &= \frac{1}{2}(n^{\mu}\bar{m}^{\nu}\nabla_{\nu}l_{\mu}+ m^{\mu}\bar{m}^{\nu}\nabla_{\nu}\bar{m}_{\mu}) \, ,\qquad  \beta = -\frac{1}{2}(n^{\mu}m^{\nu}\nabla_{\nu}l_{\mu} + m^{\mu}m^{\nu}\nabla_{\nu}\bar{m}_{\mu}) \, . 
	\end{aligned}
\end{equation}
while writing the different curvature components as
\begin{equation} \label{eqn:appintro:curv} 
	\begin{aligned}
\phi_{00} &= \frac{1}{2} R_{\mu \nu} l^{\mu}l^{\nu}  \, , \quad \phi_{22}  =\frac{1}{2} R_{\mu \nu} n^{\mu}n^{\nu} \, , \quad \phi_{01} = \frac{1}{2} R_{\mu \nu} l^{\mu}m^{\nu} \, ,
\quad \phi_{21} = \frac{1}{2} R_{\mu \nu}  n^{\mu}\bar{m}^{\nu}\, , \\ \\ \phi_{02} &= \frac{1}{2} R_{\mu \nu} m^{\mu}m^{\nu}\, ,  \qquad
\phi_{11} = \frac{1}{4} (R_{\mu \nu} l^{\mu}n^{\nu} + R_{\mu \nu} m^{\mu}\bar{m}^{\nu})\, , \quad R = -2R_{\mu \nu} l^{\mu}n^{\nu} + 2R_{\mu \nu} m^{\mu}\bar{m}^{\nu} \, , 
	\end{aligned}
\end{equation}
and 
\begin{equation}\label{eqn:appintro:curv2} 
	\begin{aligned}
\Psi_0 &= C_{\alpha \beta \mu \nu} l^{\alpha}m^{\beta}l^{\mu}m^{\nu} \, , \qquad \Psi_1 = C_{\alpha \beta \mu \nu} l^{\alpha}n^{\beta}l^{\mu}m^{\nu} \, ,  \qquad
\Psi_2 = \Psi_2' = C_{\alpha \beta \mu \nu} l^{\alpha}m^{\beta} \bar m^{\mu}n^{\nu}\, , \\ \\ \qquad \Psi_3 &= \Psi_1' = C_{\alpha \beta \mu \nu} l^{\alpha}n^{\beta}\bar m^{\mu} n^{\nu} \, ,  \qquad \Psi_4 = \Psi_0' = C_{\alpha \beta \mu \nu} n^{\alpha}\bar{m}^{\beta}n^{\mu}\bar{m}^{\nu} \, .
	\end{aligned}
\end{equation}
We will, for practical computational purposes, re-express the final answers simply with this NP-notation, but for compactness, it is useful to work covariantly with respect to the local rescalings
	\begin{equation}\label{eqn:intro:typeIII}
	l^\mu \mapsto e^{\lambda}l^\mu,\qquad n^\mu \mapsto  e^{-\lambda} n^\mu,\qquad  m^\mu \mapsto e^{i \theta} m^\mu\qquad \bar m^\mu \mapsto e^{-i \theta} \bar m^\mu \, .
\end{equation}
where $\lambda,\theta \in \mathbb R$. More generally, one then says an object $X$ transforms with weight $w_{\rm GHP}(X) = \{p,q\}$ if it transforms under the change of frame \eqref{eqn:intro:typeIII} as
\begin{equation}
 X = e^{\frac{p}{2}\left(\lambda+i \theta\right)+\frac{q}{2}\left(\lambda-i \theta\right)}X \, ,
\end{equation}
such that the weights of the frame fields are
	\begin{equation}\label{eqn:intro:typeIIIweights}
	w_{\rm GHP}(l^\mu)=  \{1,1\} \, ,\quad w_{\rm GHP}(n^\mu)=  \{-1,-1\} \, ,\quad  w_{\rm GHP}(m^\mu)=  \{1,-1\} \, ,\quad  w_{\rm GHP}(\bar{m}^\mu)=  \{-1,1\} \, .
\end{equation}
On the other hand, the spin coefficients with definite weight are
\begin{equation}
		w_{\rm GHP}(\kappa)=  \{3,1\} \, ,\quad w_{\rm GHP}(\sigma)=  \{3,-1\} \, ,\quad  w_{\rm GHP}(\rho)=  \{1,1\} \, ,\quad  w_{\rm GHP}(\tau)=  \{1,-1\} \, ,
\end{equation}
while $\epsilon$, $\epsilon'$, $\beta$, $\beta'$ are used to construct the GHP derivatives of definite weight
\begin{equation}\label{eqn:GHPderivatives}
	\begin{aligned}
		\th   = (l^\alpha \nabla_\alpha - p \epsilon - q \epsilon^*),  \quad
		\th'  = (n^\alpha \nabla_\alpha + p  \epsilon' + q \epsilon'{}^*), \\ \\
		\edth  = (m^\alpha \nabla_\alpha - p \beta + q \beta'{}^*), \quad
		\edth' = (\bar{m}^\alpha \nabla_\alpha + p \beta' - q \beta^*) \, ,
	\end{aligned}
\end{equation} 
as acting on an object of weight $\{p,q\}$. In this notation, the Bianchi identities that we will use are \
\begin{subequations}\label{eqn:GHPbianchi39}
\begin{equation} \label{eqn:intro:bianchi39p}
\begin{aligned}
	\th'\Psi_0-\edth\Psi_1-\edth\phi_{01}+\th\phi_{02} &=  \rho'\Psi_0 - 4\tau \Psi_1+3\sigma \Psi_2+\sigma'{}^* \phi_{00} -2\tau'{}^*\phi_{01}-2\kappa \phi_{21}^*\\
	&+2\sigma \phi_{11}+\rho^* \phi_{02} \, , 
\end{aligned}
\end{equation}
\begin{equation} \label{eqn:intro:bianchi39}
	\begin{aligned}
	\th\Psi_4-\edth'\Psi_3-\edth'\phi_{21}+\th'\phi_{02}{}^* &= \rho\Psi_4 - 4\tau' \Psi_3+3\sigma' \Psi_2+\sigma{}^* \phi_{22} -2\tau{}^*\phi_{21}-2\kappa' \phi_{01}^*\\
	&+2\sigma' \phi_{11}+\rho{}'^* \phi_{02}^* \, , 
	\end{aligned}
\end{equation}
\end{subequations}
and
\begin{subequations}\label{eqn:GHPbianchi36}
	\begin{equation} \label{eqn:intro:bianchi36p}
		\begin{aligned}
			\th'\Psi_3-\edth\Psi_4-\th'\phi_{21}+\edth'\phi_{22} &= - \tau\Psi_4 + 4\rho' \Psi_3-3\kappa' \Psi_2+\tau{}^* \phi_{22} -2\rho'{}^*\phi_{21}-2\sigma' \phi_{21}^*\\
			&+2\kappa' \phi_{11}+\kappa'{}^* \phi_{02}{}^* \,  , \\
		\end{aligned}
	\end{equation}
	\begin{equation} \label{eqn:intro:bianchi36}
		\begin{aligned}
			\th\Psi_1-\edth'\Psi_0-\th\phi_{01}+\edth\phi_{00} &=- \tau'\Psi_0 + 4\rho \Psi_1-3\kappa \Psi_2+\tau'{}^* \phi_{00} -2\rho^*\phi_{01}-2\sigma \phi_{01}^*\\
			&+2\kappa \phi_{11}+\kappa^* \phi_{02} \, .
		\end{aligned}
	\end{equation}
\end{subequations}
These expressions (and the additional identities) can be found, for instance, as (4.12.36)-(4.12.41) in  \cite{penrose1984spinors}. We remark that this reference uses a ``mostly minus'' convention. In addition, the curvature tensor $R^{\mu}{}_{\nu \alpha \beta}$ is defined with opposite sign. Consequently, our definitions for the spin coefficients are made with opposite signs compared to (4.5.22) of \cite{penrose1984spinors} as are the curvature scalars, compared to (4.11.10). On the other hand, \eqref{eqn:appintro:curv2} are the same as (4.11.9) of \cite{penrose1984spinors}. The result of these choices is that the Bianchi identities in our GHP notation agree with \cite{penrose1984spinors}. Nevertheless, as a check, we have independently derived them. \\

Now, applying $\th$ and $\edth$ to respectively the Bianchi equations \eqref{eqn:intro:bianchi39p} and \eqref{eqn:intro:bianchi36}, one finds, upon summing, the result
\begin{equation}\label{eqn:universalteukolsky0}
\cO^{(0)}_{2}\left(\Psi_0\right) +\cO^{(1)}_{2}\left(\Psi_1\right)+ \cO^{(2)}_{2}\left(\Psi_0\right)   = 8 \pi \left(\cT^{(0)}_{2} +\cT^{(1)}_{2} +\cT^{(2)}_{2}\right) \, ,
\end{equation}
with
\begin{subequations}
	\bea
	\cO^{(0)}_{2} &=& 2 \left[(\th - 4 \rho - \rho^*)(\th'-\rho') - (\edth-4\tau-\tau'{}^*)(\edth'-\tau') -3\Psi_2\right] \, , \\
	\cO^{(1)}_{2} &=& 4 \left[2\kappa\left(\th'-\rho'^*\right)-2\sigma\left(\edth'-\tau^*\right)+2\left(\th'\kappa\right)-2\left(\edth'\sigma\right)+5\Psi_1 \right] \, , \\
	\cO^{(2)}_{2} &=& 6 \left[\kappa \kappa'-\sigma \sigma'\right] \, , \\  \nn
	\cT^{(0)}_{2}  &=& (\edth -\tau'{}^*-4\tau)[(\th-2\rho^*)T_{lm}-(\edth -\tau'{}^*)T_{ll}] \nn &+& (\th-4\rho-\rho^*)[(\eth - 2\tau'{}^*)T_{lm}-(\th-\rho^*)T_{mm}] \, , \\
	\cT^{(1)}_{2}  &=& \frac{1}{2}\left[\sigma \th -\kappa \edth \right]T - \left[3 \sigma \left(\th' -\rho'{}^*\right) -\sigma'{}^*\left(\th -4\rho-\rho^*\right)- \th\left(\sigma'{}^*\right)\right]T_{ll} \nn 
	&-& 2\left[\sigma \left(\edth -\tau-\tau'{}^*\right) +\edth\left(\sigma\right) \right] T_{l\bar{m}}+\left[3\sigma \left(\edth' -2\tau{}^*\right) +3\kappa\left(\th' -2\rho'{}^*\right)\right] T_{lm} \nn 
	&-& \left[3\kappa \left(\edth'-\tau^*\right) - \kappa^* \left(\edth-4\tau-\tau'{}^*\right) - \edth\left(\kappa^*\right) \right] T_{mm} \nn 
	&+& \left[\kappa \edth + \sigma \left(\th-2\rho-2\rho^*\right) +2 \th\left(\sigma\right)-\Psi_0 \right] \left(T_{ln}+T_{m\bar{m}}\right) \nn 
	&-& 2\left[\kappa \left(\th -\rho-\rho^*\right) + \th\left(\kappa\right)\right] T_{nm} \, , \\
	\cT^{(2)}_{2}  &=& 3\left[\kappa \kappa'{}^*T_{ll}+\sigma \sigma{}^* T_{mm}\right] \, , 
	\eea
and\footnote{Naturally, $T_{ll}$, $T_{lm}$, etc., are the components of the stress-energy tensor in the basis \req{eqn:frame}. This enters in Einstein equations as	$G_{\mu\nu}=8\pi T_{\mu\nu}$. In the case of the higher-derivative corrections we have the equations \req{EinsteinEq} and hence we must use $8\pi T_{\mu\nu}=\ell^4 T_{\mu\nu}^{(6)}$.}

\bea
8 \pi T_{ll} &=& 2 \phi_{00} \, , \quad 8 \pi T_{lm} = 2 \phi_{01} \, , \quad 8 \pi T_{mm} = 2 \phi_{02}  \, , \quad  8 \pi T_{n \bar{m}} = 2 \phi_{21}\, , \\ \nn 
8 \pi T &=& - R  \, ,  \quad 8 \pi \left(T_{ln}+T_{m\bar{m}}\right) = 4 \phi_{11} \, .
\eea
\end{subequations}

Here, \eqref{eqn:intro:bianchi39p} and \eqref{eqn:intro:bianchi36} were additionally used to replace $\edth \Psi_2$ and $\th \Psi_1$. Moreover,  several other relations between the spin coefficients and curvature scalars were used along the way. Relevant additional identities and details can be found in Appendix \ref{app:NPGHP}. Similarly, applying $\th'$ and $\edth'$ to respectively the Bianchi equations \eqref{eqn:intro:bianchi39} and \eqref{eqn:intro:bianchi36p}, one finds, upon summing, the result
\begin{equation}\label{eqn:universalteukolsky4}
	\cO^{(0)}_{-2}\left(\Psi_4\right) +\cO^{(1)}_{-2}\left(\Psi_3\right)+ \cO^{(2)}_{-2}\left(\Psi_4\right)   = 8 \pi \left(\cT^{(0)}_{-2} +\cT^{(1)}_{-2} +\cT^{(2)}_{-2}\right) \, ,
\end{equation} 
with
\begin{subequations}
	\bea
	\cO^{(0)}_{-2} &=& 2 \left[(\th' - 4 \rho' - \rho'{}^*)(\th-\rho) - (\edth'-4\tau'-\tau{}^*)(\edth-\tau) -3\Psi_2\right] \, , \\
	\cO^{(1)}_{-2} &=& 4 \left[2\kappa'\left(\th-\rho^*\right)-2\sigma'\left(\edth-\tau'{}^*\right)+2\left(\th\kappa'\right)-2\left(\edth\sigma'\right)+5\Psi_3 \right] \, , \\
	\cO^{(2)}_{-2} &=& 6 \left[\kappa \kappa'-\sigma \sigma'\right] \, , \\  \nn
	\cT^{(0)}_{-2}  &=& (\edth' -\tau{}^*-4\tau')[(\th'-2\rho'{}^*)T_{n \bar{m}}-(\edth' -\tau{}^*)T_{nn}] \nn &+& (\th'-4\rho'-\rho'{}^*)[(\eth' - 2\tau{}^*)T_{n \bar{m}}-(\th'-\rho'{}^*)T_{\bar{m} \bar{m}}] \, , \\
	\cT^{(1)}_{-2}  &=& \frac{1}{2}\left[\sigma' \th' -\kappa' \edth' \right]T - \left[3 \sigma' \left(\th -\rho{}^*\right) -\sigma{}^*\left(\th' -4\rho'-\rho'{}^*\right)- \th'\left(\sigma{}^*\right)\right]T_{nn} \nn 
	&-& 2\left[\sigma' \left(\edth' -\tau'-\tau{}^*\right) +\edth'\left(\sigma'\right) \right] T_{n m}+\left[3\sigma' \left(\edth -2\tau'{}^*\right) +3\kappa'\left(\th -2\rho{}^*\right)\right] T_{n \bar{m}} \nn 
	&-& \left[3\kappa' \left(\edth-\tau'{}^*\right) - \kappa'{}^* \left(\edth'-4\tau'-\tau{}^*\right) - \edth'\left(\kappa'{}^*\right) \right] T_{\bar{m} \bar{m}} \nn 
	&+& \left[\kappa' \edth' + \sigma' \left(\th'-2\rho'-2\rho'{}^*\right) +2 \th'\left(\sigma'\right)-\Psi_4 \right] \left(T_{ln}+T_{m\bar{m}}\right) \nn 
	&-& 2\left[\kappa' \left(\th' -\rho'-\rho'{}^*\right) + \th'\left(\kappa'\right)\right] T_{l \bar{m}} \, , \\
	\cT^{(2)}_{-2}  &=& 3\left[\kappa' \kappa{}^*T_{nn}+\sigma' \sigma'{}^* T_{\bar{m} \bar{m}}\right] \, , 
	\eea
	and
	\bea
	8 \pi T_{nn} &=& 2 \phi_{22} \, , \quad 8 \pi T_{lm} = 2 \phi_{01} \, , \quad 8 \pi T_{\bar{m} \bar{m}} = 2 \phi_{02}^*  \, , \quad  8 \pi T_{n \bar{m}} = 2 \phi_{21}\, , \\ \nn 
	8 \pi T &=& - R  \, ,  \quad 8 \pi \left(T_{ln}+T_{m\bar{m}}\right) = 4 \phi_{11} \, .
	\eea
\end{subequations}
More details can again be found in Appendix \ref{app:NPGHP} but suffice it to say that, aside from the key observations that were made long ago for the case of rotating black holes \cite{Newman:1961qr, price1972nonspherical,bardeen1973radiation,teukolsky1972rotating, teukolsky1973, teukolsky1974perturbations}, the derivation of these expressions is an exercise in translation and algebra. Yet, it is important to emphasize no assumptions are made on the spacetime or NP-frame to derive \eqref{eqn:universalteukolsky0} and \eqref{eqn:universalteukolsky4}. Nevertheless, we have suggestively grouped together terms based on quantities that would vanish in a Petrov D spacetime in an aligned frame.

In order to be entirely explicit, as is useful to do actual calculations, the NP versions of \eqref{eqn:universalteukolsky0} and \eqref{eqn:universalteukolsky4} are given in Appendix \ref{app:NPGHP} in \eqref{eqn:universalteukolskyNP0} and \eqref{eqn:universalteukolskyNP4}.

\subsection{Linearization}

Contrary to the Teukolsky equations, the equations derived in the previous section apply for any four-dimensional spacetime and any choice of NP-frame. Contrary to the Teukolsky equations, they are thus totally unpractical. Our goal is now to approach the middle ground, where we are close enough to the Teukolsky equations to use their nice properties while still being very general in terms of applicability. The key assumption going into this is the perturbative nature of the corrections. All observations so far indicate that this is likely a good assumption \cite{LIGOScientific:2021sio} but it should be stressed that non-perturbative corrections to the spectrum are also of interest \cite{Bueno:2017hyj,Oshita:2018fqu,Fransen:2020prl}. These will not be captured here. Moreover, in light of the pseudospectral instability of these modes such corrections could imply an entirely different spectrum \cite{Nollert:1996rf,Gasperin:2021kfv,Jaramillo:2021tmt}. However, the time-domain response relevant for observations is more robust \cite{Berti:2022xfj}. Therefore, we would still expect the perturbative corrections we compute to be relevant for the observed ringdown.

In general, although there are of course other relations between these variables, the equations \eqref{eqn:universalteukolsky0} and \eqref{eqn:universalteukolsky4} involve the frame $e_a$, spin connection $\gamma_{abc}$, the (Ricci) curvatures\footnote{Note however that Ricci curvature will be replaced by the effective stress-energy tensor by imposing the Einstein's equations. In the case of higher-derivative theories like in \req{EinsteinEq}, the effective stress-energy tensor is a function of the frame, spin connection and Weyl curvature, since at leading order in a perturbative expansion we only need to evaluate it on a solution of the   (uncorrected) vacuum Einstein's equations --- this is, on a Ricci-flat spacetime.} $\phi_{ab}$, and Weyl scalars $\Psi_{i}$, so they take the form
\begin{equation}\label{kwinteneq}
\mathcal{E}_{\pm 2}(e_{a},\gamma_{abc},\phi_{ab},\Psi_{i})=\mathcal{E}_{\pm 2}(\Phi,\Psi_{i})=0\, ,
\end{equation}
where we will denote the variables excluding $\Psi_i$ collectively as $\Phi$. Consider a perturbation over a solution, so we make $e_a\rightarrow \bar e_a+\delta e_a$, $\gamma_{abc}\rightarrow \bar\gamma_{abc}+\delta \gamma_{abc}$, $\phi_{ab}\rightarrow \bar\phi_{ab}+\delta\phi_{ab}$  (so in general $\Phi \to \bar\Phi + \delta \Phi$), and $\Psi_{i}\rightarrow \bar\Psi_{i}+\delta\Psi_{i}$, where the bar on top of a given quantity denotes that it is evaluated on the background. The equations above then yield two linear equations for the perturbations 
\begin{equation}\label{linearE1}
\mathcal{E}_{\delta\Psi,\pm 2}\left(\delta \Psi_{i}\right)+\mathcal{E}_{\delta\Phi,\pm 2}\left(\delta \Phi\right)=0\, ,
\end{equation}
where $\mathcal{E}_{\delta\Psi,\pm 2}\left(\cdot\right)$ and $\mathcal{E}_{\delta\Phi,\pm 2}\left(\cdot\right)$ are linear differential operators.
In principle, these equations involve all perturbed variables. However, a simplification occurs if the background geometry corresponds to a corrected vacuum (Ricci flat) Petrov-D solution. By this we mean that the solution only departs perturbatively from the vacuum Petrov type D due to, say, higher-derivative corrections. This is in particular the case for rotating black holes in the theory \req{eq:EFTofGR}. 

As a bookkeeping parameter, we introduce $\lambda$ to keep track of the departure from vacuum Petrov type D (\textit{e.g.}, in the case of higher-derivative corrections this could be $\ell^4$ or $\ell^6$ in \req{eq:EFTofGR}). We will work at linear order in $\lambda$, since we are only interested in the leading corrections. 
Then, our background geometry will have 
\begin{equation}
\bar\Psi_{i}=\bar\Psi_{i}^{(0)}+\lambda \bar\Psi_{i}^{(1)}\, ,\quad \bar\Phi_{i}=\bar\Phi_{i}^{(0)}+\lambda \bar\Phi_{i}^{(1)}\, ,
\end{equation} 
with in particular
\begin{align}
\bar\Psi_{0}^{(0)}=\bar\Psi_{1}^{(0)}=\bar\Psi_{3}^{(0)}=\bar\Psi_{4}^{(0)}&=0\, , \label{eqn:petrovDexplicitweyl}\\ 
\bar\phi_{00}^{(0)}=\bar\phi_{01}^{(0)}=\bar\phi_{02}^{(0)}=\bar\phi_{21}^{(0)}=\bar\phi_{11}^{(0)}=\bar\phi_{22}^{(0)}&=0\, ,\\
\bar\kappa^{(0)}=\bar\kappa'^{(0)}=\bar\sigma^{(0)}=\bar\sigma'^{(0)}&=0\, , \label{eqn:petrovDexplicitspin}
\end{align}
on the assumption that the uncorrected background is of Petrov type D and that the frame is chosen to be adapted to the principal null directions. The corrections to these quantities will nevertheless be non-vanishing in general.  

When we evaluate the linearized equation \req{linearE1} on this type of background, one can see that the operators $\mathcal{E}_{\delta\Phi, \pm 2}$ become of order $\lambda$, which we denote by 
\begin{equation}
\mathcal{E}_{\delta\Phi, \pm 2}\left(\delta \Phi\right) =\lambda \mathcal{E}^{(1)}_{\delta\Phi, \pm 2}\left(\delta \Phi\right)\,  .
\end{equation}
On the other hand, 
\begin{align}
\mathcal{E}_{\delta \Psi,+2}\left(\delta \Psi_{i}\right)&=\mathcal{D}^{(0)}_{2}\left(\delta \Psi_{0}\right)+\lambda\mathcal{E}^{(1)}_{\delta \Psi, +2}\left(\delta \Psi_{i}\right)\, ,\\
\mathcal{E}_{\delta \Psi,-2}\left(\delta \Psi_{i}\right)&=\mathcal{D}^{(0)}_{-2}\left(\delta \Psi_{4}\right)+\lambda\mathcal{E}^{(1)}_{\delta \Psi,-2}\left(\delta \Psi_{i}\right)\, ,
\end{align}
where $\mathcal{D}^{(0)}_{\pm 2}$ are the Teukolsky operators for $\delta\Psi_{0,4}$ on the Kerr background,

\begin{equation}
\begin{aligned}
\mathcal{D}^{(0)}_{+2} &=2 \left[(\th - 4 \rho - \rho^*)(\th'-\rho') - (\edth-4\tau-\tau'{}^*)(\edth'-\tau') -3\Psi_2\right]\big|_{\rm Kerr}\, ,\\
\mathcal{D}^{(0)}_{-2} &=2 \left[(\th' - 4 \rho' - \rho'{}^*)(\th-\rho) - (\edth'-4\tau'-\tau{}^*)(\edth-\tau) -3\Psi_2\right]\big|_{\rm Kerr}\, .
\end{aligned}
\end{equation}
Then, to first order in $\lambda$, the equations become
\begin{align}
\mathcal{D}^{(0)}_{+2}\left(\delta \Psi_{0}\right)+\lambda\left[\mathcal{E}^{(1)}_{\delta \Psi,+2}\left(\delta \Psi_{i}\right)+\mathcal{E}^{(1)}_{\delta\Phi, + 2}\left(\delta \Phi\right)\right]&=0\, ,\\
\mathcal{D}^{(0)}_{-2}\left(\delta \Psi_{4}\right)+\lambda\left[\mathcal{E}^{(1)}_{\delta \Psi,-2}\left(\delta \Psi_{i}\right)+\mathcal{E}^{(1)}_{\delta\Phi, - 2}\left(\delta \Phi\right)\right]&=0 \, .
\end{align}
Now, for every perturbed quantity, its value will be the one on Kerr plus a linear correction, 

\begin{equation}
\delta \Phi=\delta\Phi^{(0)}+\lambda \delta\Phi^{(1)}\, .
\end{equation}
Thus, at first order in $\lambda$, the previous equations are equivalent to 

\begin{align}
	\mathcal{D}^{(0)}_{+2}\left(\delta \Psi_{0}\right)+\lambda\left[\mathcal{E}^{(1)}_{\delta \Psi,+2}\left(\delta \Psi_{i}^{(0)}\right)+\mathcal{E}^{(1)}_{\delta\Phi, + 2}\left(\delta \Phi^{(0)}\right)\right]&=0\, ,\\
	\mathcal{D}^{(0)}_{-2}\left(\delta \Psi_{4}\right)+\lambda\left[\mathcal{E}^{(1)}_{\delta \Psi,-2}\left(\delta \Psi_{i}^{(0)}\right)+\mathcal{E}^{(1)}_{\delta\Phi, - 2}\left(\delta \Phi^{(0)}\right)\right]&=0 \, .
\end{align}

Note that we intentionally do not split  $\delta \Psi_{0,4}=\delta \Psi_{0,4}^{(0)}+\lambda\delta \Psi_{0,4}^{(1)}$, as we will write an equation for the complete variables $\delta \Psi_{0,4}$. 

In fact, every zeroth-order perturbed quantity can be determined in terms of the Teukolsky variables $\delta\Psi_{0}^{(0)}$ and $\delta\Psi_{4}^{(0)}$, this is
\begin{equation}\label{PhitoPsi}
\delta\Phi^{(0)}=\delta\Phi^{(0)}(\delta\Psi_{0}^{(0)},\delta\Psi_{4}^{(0)}) =\delta\Phi^{(0)}(\delta\Psi_{0},\delta\Psi_{4})+\mathcal{O}(\lambda)\, .
\end{equation}
Therefore, at the end of the day we should be able to write the equations explicitly as 
\begin{equation}\label{goalequations}
\begin{aligned}
\mathcal{D}_{+2}^{(0)}\left(\delta \Psi_{0}\right)+\lambda\left[\mathcal{E}^{(1)}_{\delta \Psi_0,+2}\left(\delta \Psi_{0}\right)+\mathcal{E}^{(1)}_{\delta \Psi_4,+2}\left(\delta \Psi_{4}\right)\right]&=0\, ,\\ 
\mathcal{D}_{-2}^{(0)}\left(\delta \Psi_{4}\right)+\lambda\left[\mathcal{E}^{(1)}_{\delta \Psi_0,-2}\left(\delta \Psi_{0}\right)+\mathcal{E}^{(1)}_{\delta \Psi_4,-2}\left(\delta \Psi_{4}\right)\right]&=0\, .
\end{aligned}
\end{equation}
These are two coupled equations for the two variables $\delta\Psi_{0,4}$. In order to obtain these equations, we first need to obtain explicitly the relations \req{PhitoPsi}. We will achieve this by finding the leading-order (general relativistic) metric perturbations associated to solutions to the Teukolsky equations in a procedure known as metric reconstruction.

\subsection{Implementation}
In practice, to obtain the equations \req{goalequations} we proceed as follows. For every quantity $\Phi=\{\Psi_i,e_a,\gamma_{abc}\}$, we must find its explicit expression as 

\begin{equation}\label{phidec}
\Phi=\bar\Phi+\delta \Phi(\delta\Psi_{0},\delta\Psi_{4})\, ,
\end{equation}
where $\bar\Phi$ is the background value of the corresponding quantity, consisting of its Einstein gravity value plus a correction, 
\begin{equation}
\bar\Phi=\bar\Phi^{(0)}+\lambda \bar\Phi^{(1)}\, ,
\end{equation}
and $\delta \Phi(\delta\Psi_{0},\delta\Psi_{4})$ is the solution of the perturbed quantity in the Kerr background expressed in terms of $\delta\Psi_{0}$ and $\delta\Psi_{4}$. We remark that the term $\delta\Phi(\delta\Psi_{0},\delta\Psi_{4})$ involves no higher-derivative corrections; it is the perturbed quantity as obtained in Einstein gravity. 

We then insert \req{phidec} into \req{eqn:universalteukolsky0} and \req{eqn:universalteukolsky4} keeping the terms linear in the fluctuations $\delta\Phi$ to first order in $\lambda$. In this process we can also reduce the number of derivatives in the terms proportional to $\lambda$ by making use of the zeroth-order Teukolsky equations $\mathcal{D}_{\pm 2}^{(0)}\left(\delta \Psi_{0,4}\right)=0$. The result is then the two equations \req{goalequations}. In order to carry out this computation we need two ingredients: 
\begin{enumerate}
	\item[(1)] the Newman-Penrose description of the background geometry (with higher-derivative corrections), 
	
	\item[(2)] the solution for every perturbed quantity in the Einstein gravity case (without corrections) expressed in terms of $\delta\Psi_0$ and $\delta\Psi_4$.
\end{enumerate}
We address these points in the next section.  

There is a last observation that must be made. In general, it will be convenient to work with complex metric perturbations. However, this comes at the price that complex conjugates in the NP formalism are not actual complex conjugates. Instead, the conjugate variables become independent and they satisfy their own equations. We need to consider in particular the equations for $\delta\Psi_{0}^{*}$ and $\delta\Psi_{4}^{*}$. Thus, together with  \req{eqn:universalteukolsky0} and \req{eqn:universalteukolsky4}, we also need to consider their ``conjugated'' versions, which yield two equations for $\delta\Psi_{0}^{*}$ and $\delta\Psi_{4}^{*}$ analogous to \req{goalequations}. Therefore, at the end we will obtain four coupled linear equations for $\delta\Psi_{0}$, $\delta\Psi_{4}$, $\delta\Psi_{0}^{*}$ and $\delta\Psi_{4}^{*}$.

\section{Ingredients for the computation}\label{sec:computation} 
In this section, we start moving from the abstract conceptual outline as spelled out in the previous section and developed along similar lines as \cite{Li:2022pcy,Hussain:2022ins} to the specific and explicit computation. 
\subsection{Newman-Penrose description of the background geometry}\label{subsec:background}

The solutions of \req{eq:EFTofGR}, perturbatively in the higher-derivative couplings, take the form
\begin{equation}\bar
g_{\mu\nu}=\bar g^{(0)}_{\mu\nu}+\lambda \bar g^{(1)}_{\mu\nu}\, ,
\end{equation}
where $\bar  g^{(0)}_{\mu\nu}$ is a solution of vacuum Einstein equations (\textit{i.e.}, a Ricci flat metric) and $\bar  g^{(1)}_{\mu\nu}$ is the correction at linear order in the higher-derivative couplings which, as in the previous section, we keep track of through the bookkeeping parameter $\lambda$. In the case of the rotating black hole solutions, we use the following ansatz to capture the deviations with respect to Kerr \cite{Cano:2019ore},

\begin{equation}\label{rotatingmetric}
\begin{aligned}
d\bar s^2=&-\left(1-\frac{2 M r}{\Sigma}-\lambda H_1\right)dt^2-\left(1+\lambda H_2\right)\frac{4 M a r (1-x^2)}{\Sigma}dtd\phi\\
&+\left(1+\lambda H_3\right)\Sigma\left(\frac{dr^2}{\Delta}+\frac{dx^2}{1-x^2}\right)\\
&+\left(1+\lambda H_4\right)\left(r^2+a^2+\frac{2 M  r a^2(1-x^2)}{\Sigma}\right)(1-x^2)d\phi^2\, ,
\end{aligned}
\end{equation}
where 
\begin{equation}
\Sigma=r^2+a^2x^2\, ,\quad \Delta=r^2-2Mr+a^2\, ,
\end{equation} 
and $x=\cos\theta$.\footnote{We remark that it is  computationally faster to use the algebraic $x$ variable as opposed to the trigonometric  $\theta$ variable.} The metric $\bar  g^{(0)}_{\mu\nu}$ is in this case the Kerr metric in Boyer-Lindquist coordinates, while $\bar  g^{(1)}_{\mu\nu}$ is parametrized by the four functions $H_{i}(x,r)$, 

\begin{equation}\label{gmunu1}
\begin{aligned}
\bar g^{(1)}_{\mu\nu}dx^{\mu}dx^{\nu}=&H_1dt^2-H_2\frac{4 M a r (1-x^2)}{\Sigma}dtd\phi+H_3\Sigma\left(\frac{dr^2}{\Delta}+\frac{dx^2}{1-x^2}\right)\\
&+H_4\left(r^2+a^2+\frac{2 M r a^2(1-x^2)}{\Sigma}\right)(1-x^2)d\phi^2\, .
\end{aligned}
\end{equation}
The solution depends on the two constants $M$ and $a$, which represent the total mass and specific angular momentum as long as the $H_{i}$ functions satisfy appropriate boundary conditions \cite{Cano:2019ore}, which we assume. We also introduce the dimensionless spin parameter $\chi=a/M$. Here we consider solutions with $|\chi|<1$, as the extremal limit $|\chi|\sim 1$ poses additional difficulties and should be studied separately. 
For $|\chi|<1$, the solutions for the $H_i$ functions can be expressed as a convergent power series in $\chi$, and they take the form\footnote{In order to determine the convergence of the spin expansion one can perform standard convergence tests, like the root test, with a high-order expansion of the solution ---  in our case, we used the solution to order $\chi^{30}$. These tests indicate that the power series is indeed convergent in the exterior of the black hole, with a radius of convergence $|\chi_{\rm max}|\sim 1$. Furthermore, in the case of certain observables the convergence of the spin expansion can be checked explicitly as the full series can be summed up analytically \cite{Reall:2019sah,Cano:2022wwo}.}

\begin{equation}\label{chiexp}
H_{i}=\sum_{n=0}^{\infty}\chi^n\sum_{p=0}^n\sum_{q=0}^{q_{\rm max}(n,p)}H_{i}^{(n,p,q)}x^pr^{-q}\, ,
\end{equation}
for certain coefficients $H_{i}^{(n,p,q)}$ that one can compute. We have computed these solutions to order $\mathcal{O}(\chi^{30})$, which we believe would suffice for most purposes up to spins $\chi\sim 0.9$. For smaller spins, much less terms are required to obtain high accuracy. The availability of this high-order solution will not yet be fully exploited here, since we will only make use of a $\mathcal{O}(\chi^6)$ expansion in Section~\ref{sec:QNM6}, but it is important for future applications.  
The solution could otherwise be obtained numerically by solving the system of linear partial differential equations satisfied by the $H_{i}$ functions, but we find this less practical. 

Let us also take note that the event horizon of the metric \req{rotatingmetric} is located at 
\begin{equation}\label{eq:rhorizon}
r_{+}=M+\sqrt{M^2-a^2}\, ,
\end{equation}
and this is not modified by the higher-derivative corrections. This is because we are making a coordinate choice in \req{rotatingmetric} that guarantees that the horizon is always located at \req{eq:rhorizon}.

Let us now provide the Newman-Penrose description of this spacetime. We again consider a null tetrad $\tensor{e}{_{a}^{\mu}}$ that we denote by 
\begin{equation}
\tensor{e}{_{1}^{\mu}}= l^{\mu}\, ,\quad \tensor{e}{_{2}^{\mu}}=n^{\mu}\, ,\quad \tensor{e}{_{3}^{\mu}}=m^{\mu}\, ,\quad \tensor{  e}{_{4}^{\mu}}=\bar{m}^{\mu}\,  ,
\end{equation}
and that satisfies
\begin{equation}
\bar g_{\mu\nu}=-2l_{(\mu}n_{\nu)}+2m_{(\mu}\bar{m}_{\nu)}\, .
\end{equation}
Just like for the metric, the tetrad vectors are given by their Kerr value plus a correction, 
\begin{equation}
\tensor{e}{_{a}^{\mu}}=\tensor{e}{_{(0)a}^{\mu}}+\lambda \tensor{e}{_{(1)a}^{\mu}}\, .
\end{equation}
For the tetrad in the uncorrected Kerr geometry, we use the Kinnersley tetrad \cite{Pound:2021qin}
\begin{align}\label{frameKerr}
l_{(0)}^{\mu}&=\left(\frac{r^2+a^2}{\Delta}\, ,1\, ,0\, ,\frac{a}{\Delta}\right)\, ,\\
n_{(0)}^{\mu}&=\left(r^2+a^2\, ,-\Delta \, ,0 \, ,a\right)\frac{1}{\Sigma}\, ,\\
m_{(0)}^{\mu}&=\left(ia\sqrt{1-x^2}\, ,0\, ,-\sqrt{1-x^2}\, ,\frac{i}{\sqrt{1-x^2}}\right)\frac{1}{\sqrt{2}\zeta^{*}}\, ,\\
\bar{m}_{(0)}^{\mu}&=\left(-ia\sqrt{1-x^2}\, ,0\, ,-\sqrt{1-x^2}\, ,-\frac{i}{\sqrt{1-x^2}}\right)\frac{1}{\sqrt{2}\zeta}\, ,
\end{align}
where 
\begin{equation}
\zeta=r-iax\, .
\end{equation} 
As is well-known, this frame has the property that it is aligned with the principal null directions of the Kerr metric. Therefore, in this frame, one in particular has \eqref{eqn:petrovDexplicitweyl} and \eqref{eqn:petrovDexplicitspin}
\begin{align}
	\bar\Psi_{0}^{(0)}=\bar\Psi_{1}^{(0)}=\bar\Psi_{3}^{(0)}=\bar\Psi_{4}^{(0)}&=0\, , \\ 
	\bar\kappa^{(0)}=\bar\kappa'^{(0)}=\bar\sigma^{(0)}=\bar\sigma'^{(0)}&=0\, , 
\end{align}
as used in the previous section. Then, the correction to the tetrad can be found from the correction to the metric $\bar g^{(1)}_{\mu\nu}$ in \req{gmunu1} as

\begin{equation}\label{tetrad1}
\tensor{e}{_{(1)a}^{\mu}}=-\frac{1}{2}\bar g^{(1)}_{\alpha\beta}\bar g^{(0)\mu\alpha}\tensor{e}{_{(0)a}^{\beta}}\, .
\end{equation}
There is additional gauge freedom in the choice of $\tensor{e}{_{(1)a}^{\mu}}$, corresponding to infinitesimal Lorentz transformations of the uncorrected frame. One may use this freedom to set to zero some of the components of the spin connection or some of the Weyl scalars (in particular, $\Psi_{1}$ and $\Psi_{3}$). However, this makes the form of the tetrad much more complicated, so in practice it seems more efficient to work with the simple choice given by \req{tetrad1}. This will be the background tetrad we use throughout.  


Once the tetrad is determined, obtaining the coefficients of the spin connection as well as the Weyl scalars, that we need for our computations,\footnote{Ricci curvature, although non-vanishing, is not needed.} is just an straightforward task that we carry out with the help of software.
\subsection{Metric reconstruction of Kerr perturbations}
\label{subsec:reconstruction}
In order to find the master equations \req{goalequations} we also need to express every perturbed quantity on the Kerr background in terms of the Teukolsky variables. The most efficient way to accomplish this is by reconstructing the metric perturbation. From the metric perturbation, the remaining perturbed quantities can be straightforwardly computed. 

Following \cite{Dolan:2021ijg}, one can reconstruct the metric perturbation on the Kerr background as follows

\begin{equation}
M \partial_{t} h_{\mu\nu}=-\frac{1}{3}\nabla_{\beta}\left[\zeta^4\nabla_{\alpha}\tensor{\mathcal{C}}{_{(\mu}^{\alpha}_{\nu)}^{\beta}}\right]-\frac{1}{3}\nabla_{\beta}\left[(\zeta^{*})^4\nabla_{\alpha}\tensor{\bar{\mathcal{C}}}{_{(\mu}^{\alpha}_{\nu)}^{\beta}}\right]\, ,
\end{equation}
where 

\begin{align}\label{Ctensor}
\mathcal{C}_{\mu\alpha\nu\beta}&=4\left( \psi_0 n_{[\mu}\bar{m}_{\alpha]}n_{[\nu}\bar{m}_{\beta]}+ \psi_4 l_{[\mu}m_{\alpha]}l_{[\nu}m_{\beta]}\right)\, ,\\
\bar{\mathcal{C}}_{\mu\alpha\nu\beta}&=4\left(\psi_0^{*} n_{[\mu}m_{\alpha]}n_{[\nu}m_{\beta]}+\psi_4^{*} l_{[\mu}\bar{m}_{\alpha]}l_{[\nu}\bar{m}_{\beta]}\right)\, .
\end{align}
The variables $\psi_{0,4}$ and $\psi_{0,4}^{*}$ satisfy the Teukolsky equations, but they are in general different from the Weyl scalars $\delta\Psi_{0,4}$,  $\delta\Psi_{0,4}^{*}$. They are, however, proportional to them, as we show below. Let us also note that we will be working with a complex metric perturbation. This implies that conjugate quantities in the NP formalism are not complex conjugates anymore, but actually independent quantities. Therefore, $\psi_{0,4}$ and $\psi_{0,4}^{*}$ have to be treated as independent variables.
 
Since these variables satisfy Teukolsky equations, they can be separated as
\begin{align}
\psi_0&=e^{-i\omega t+i m \phi} R_{2}(r)S_{2}(x)\, ,\\
\psi_0^{*}&=e^{-i\omega t+i m \phi} R_{2}^{*}(r)S_{-2}(x)\, ,\\
\psi_4&=e^{-i\omega t+i m \phi} \zeta^{-4} R_{-2}(r)S_{-2}(x)\, ,\\
\psi_4^{*}&=e^{-i\omega t+i m \phi} (\zeta^{*})^{-4} R_{-2}^{*}(r)S_{2}(x)\, ,
\end{align}
where the $S_{s}(x)$, $R_{s}(r)$ and $R^{*}_{s}(r)$ functions satisfy the angular and radial Teukolsky equations
\begin{align}\label{angularequation}
\frac{d}{dx}\left[(1-x^2)\frac{dS_{s}}{dx}\right]+\left[(a\omega)^2 x^2-2sa\omega x+ B_{lm}-\frac{(m+s x)^2}{1-x^2}\right]S_{s}&=0\, ,\\
\mathfrak{D}_{s}^2R_{s}=\Delta^{-s+1}\frac{d}{dr}\left[\Delta^{s+1}\frac{dR_{s}}{dr}\right]+V R_{s}&=0\, ,\\
\mathfrak{D}_{s}^2R_{s}^{*}=\Delta^{-s+1}\frac{d}{dr}\left[\Delta^{s+1}\frac{dR_{s}^{*}}{dr}\right]+V R_{s}^{*}&=0\, ,
\label{Teukoperator}
\end{align}
with
\begin{equation}
\begin{aligned}
V&=(am)^2+\omega^2 \left(a^2+r^2\right)^2-4 a m M r \omega+i s \left(2 a m (r-M)-2 M \omega \left(r^2-a^2\right)\right)\\
&+\Delta\left(-a^2 \omega^2+s-B_{lm}+2 i r s \omega\right)\, ,
\end{aligned}
\end{equation}
and angular separation constants $B_{lm}$.\footnote{These are related to the more conventional definition of angular separation constants in the literature, denoted $_{s}A_{lm}$, by $B_{lm}=  {_s}A_{lm}+s$. However, the advantage of using $B_{lm}$ is that they are the same for $s=+2$ and $s=-2$ and hence we do not need to distinguish between the two cases.} 
Note that the conjugate variables $R_{s}^{*}$ satisfy the same equations as $R_{s}$. 
Now, since the time dependence is separated as $e^{-i\omega t}$, we can express the metric perturbation explicitly as

\begin{equation}\label{hmunureconstructed}
 h_{\mu\nu}=-\frac{i}{3M\omega}\nabla_{\beta}\left[\zeta^4\nabla_{\alpha}\tensor{\mathcal{C}}{_{(\mu}^{\alpha}_{\nu)}^{\beta}}\right]-\frac{i}{3M\omega}\nabla_{\beta}\left[(\zeta^{*})^4\nabla_{\alpha}\tensor{\bar{\mathcal{C}}}{_{(\mu}^{\alpha}_{\nu)}^{\beta}}\right]\, ,
\end{equation}
We have checked by direct computation that this metric perturbation indeed satisfies the linearized Einstein's equations $R_{\mu\nu}=0$. 

From $h_{\mu\nu}$ we can obtain the perturbation of the NP frame, 
\begin{equation}\label{tetrad2}
\delta\tensor{e}{_{a}^{\mu}}=-\frac{1}{2}h_{\alpha\beta}\bar g^{(0)\mu\alpha}\tensor{\bar e}{_{(0)a}^{\beta}}\, ,
\end{equation}
where $g^{(0)}_{\mu\alpha}$ is the Kerr metric and $\tensor{\bar e}{_{(0)a}^{\beta}}$ is the frame \req{frameKerr}. It is then straightforward (but computationally heavy) to obtain the perturbation of the spin connection $\delta\gamma_{abc}$ and of the Weyl scalars $\delta\Psi_{i}$, $\delta\Psi_{i}^{*}$ which are the quantities we need to evaluate the universal Teukolsky equations \req{eqn:universalteukolsky0}, \req{eqn:universalteukolsky4} and their conjugates.  

The missing crucial step is to establish the link between the metric variables $\psi_{i}$ and the Teukolsky variables.
To this end, it is important to observe the following relations between the four radial variables $R_{s}$ and $R_{s}^{*}$. First of all, since $R_{s}$ and the ``conjugates'' $R_{s}^{*}$ satisfy the same equation (and, in the case of QNMs, the same boundary conditions), it follows that they must be proportional, and therefore,

\begin{equation}\label{conjugaterelation}
\begin{aligned}
R_{+2}^{*}(r)&=q_{+2} R_{+2}(r)\, ,\\
R_{-2}^{*}(r)&=q_{-2} R_{-2}(r)\, ,
\end{aligned}
\end{equation}
for certain constants $q_{\pm 2}$. We will refer to these constants as polarization parameters, since they determine the polarization of the perturbations. This will become clear in section~\ref{sec:QNM6}. On the other hand, $R_{+2}$ and $R_{-2}$ can be related by means of Starobinsky-Teukolsky (ST) identities, which we review in the appendix \ref{app:STidentities}. These read
\begin{equation}\label{STidentities}
\begin{aligned}
R_{-2}&=C_{+2}\Delta^2\left(\mathcal{D}_{0}\right)^4\left(\Delta^2 R_{+2}\right)\, ,\\
R_{+2}&=C_{-2}\left(\mathcal{D}^{\dagger}_{0}\right)^{4}R_{-2}\, ,
\end{aligned}
\end{equation}
where $\mathcal{D}_{0}$ and $\mathcal{D}^{\dagger}_{0}$ are the operators 

\begin{equation}
\begin{aligned}
\mathcal{D}_{0}&=\partial_{r}+\frac{i\left(\omega(r^2+a^2)-m a\right)}{\Delta}\, ,\\
\mathcal{D}^{\dagger}_{0}&=\partial_{r}-\frac{i\left(\omega(r^2+a^2)-m a\right)}{\Delta}\, .
\end{aligned}
\end{equation}
The two proportionality constants $C_{\pm 2}$ are not independent but related by
\begin{equation}\label{Cproduct}
C_{+2}C_{-2}=\frac{1}{\mathcal{K}^2}\, ,
\end{equation}
where 
\begin{equation}
\mathcal{K}^2=D_{2}^2+144 M^2 \omega^2\, ,
\end{equation}
and $D_{2}$ is the ST constant for the angular functions, given by \req{D2value}.
Thus, in sum, the relations \req{conjugaterelation} and \req{STidentities} imply that the full metric perturbation is determined once we know one of the $R_{s}$ or $R_{s}^{*}$ and the constants $q_{s}$ and $C_{s}$.

When we use all of these relations, together with the ST identities for the angular functions --- see appendix \ref{app:STidentities} --- we obtain that the the Weyl scalars equal the metric variables up to proportionality constants, 
\begin{equation}\label{metricweylrelation}
\delta\Psi_{2-s}=P_{s}\psi_{2-s}\, ,\quad \delta\Psi_{2-s}^{*}=P_{s}^{*}\psi_{2-s}^{*}\, , \quad s=\pm 2\, .
\end{equation}
The constants $P_{s}$, $P_{s}^{*}$ depend on the polarization parameters $q_{s}$ and ST constants $C_{s}$ and are given by
\begin{equation}\label{eq:Pconstants}
\begin{aligned}
P_{+2}&=\frac{1}{2}+\frac{i D_2 q_{+2}}{24 M \omega }-\frac{i C_{+2} q_{-2}\mathcal{K}^2}{6 M\omega }\, ,\\
P_{-2}&=\frac{1}{2}-\frac{i D_2 q_{-2}}{24 M \omega }+\frac{i C_{-2} q_{+2}\mathcal{K}^2}{96 M\omega }\, ,\\
P_{+2}^{*}&=\frac{1}{2}+\frac{i D_2}{24 M \omega  q_{+2}}-\frac{i C_{+2} \mathcal{K}^2}{6M \omega  q_{+2}}\, ,\\
P_{-2}^{*}&=\frac{1}{2}-\frac{i D_2}{24 M \omega  q_{-2}}+\frac{i C_{-2} \mathcal{K}^2}{96M \omega  q_{-2}}\, .
\end{aligned}
\end{equation}

Note that there are two choices for $q_{s}$ and $C_{s}$ for which $P_{s}=P_{s}^{*}=1$, corresponding to $q_{+2}=q_{-2}=\pm 1$ with 
\begin{equation}
C_{+2}=\frac{1}{4(D_{2}-q_{+2}12 i M \omega )}\, ,\quad C_{-2}=\frac{4}{D_{2}-q_{+2}12 i M \omega}\, .
\end{equation}
However, we will not restrict ourselves to this choice of constants. On the one hand, the choices $q_{+2}=q_{-2}=\pm 1$ correspond to modes of definite parity, and this is not suitable for theories that violate parity, in whose case modes of odd and even parity become mixed.
On the other hand, by allowing the ST constants to be general we will be able to perform a strong consistency check on our results in section \ref{sec:QNM6}.

\section{Master radial equations}\label{sec:masterradial}
In the previous section, we have provided the details needed to obtain explicitly the pair of partial differential equations \req{goalequations} for the Teukolsky variables and their conjugate versions. In this section we study these equations explicitly and we show how they can be effectively separated into a set of four decoupled radial equations.
\subsection{Evaluation and separation of Teukolsky equations}\label{subsec:evalandsep}
Let us assume that we are performing perturbation theory on the background of a rotating black hole in the theory \req{eq:EFTofGR} --- although most of the discussion in this section can be applied in generality to any other theory.  Our starting point is the two perturbed Weyl scalars $\delta\Psi_{0,4}$ and their conjugates $\delta\Psi_{0,4}^{*}$. We can always separate the dependence in the $t$ and $\phi$ coordinates since these are isometric coordinates. Without loss of generality, we can additionally decompose the Weyl scalars in spin-weighted spheroidal harmonics, since these are a basis of functions, and write
\begin{equation}\label{deltapsiexpansion}
\begin{aligned}
\delta \Psi_0&=e^{-i\omega t+i m \phi} \sum_{l} P_{2}^{lm} R^{lm}_{2}(r)S^{lm}_{2}(x)\, ,\\
\delta \Psi_4&=e^{-i\omega t+i m \phi}  \sum_{l} \zeta^{-4} P^{lm}_{-2}R^{lm}_{-2}(r)S^{lm}_{-2}(x)\, ,\\
\delta \Psi_0^{*}&=e^{-i\omega t+i m \phi} \sum_{l} P^{*lm}_{2}R^{*lm}_{2}(r)S^{lm}_{-2}(x)\, ,\\
\delta \Psi_4^{*}&=e^{-i\omega t+i m \phi}  \sum_{l} (\zeta^{*})^{-4}P^{*lm}_{-2}R^{*lm}_{-2}(r)S^{lm}_{+2}(x)\, .
\end{aligned}
\end{equation}
This decomposition depends on the four sets of radial functions $R^{lm}_{s}(r)$, $R^{*lm}_{s}(r)$. Also, here we are including the constants $P^{lm}_{s}$ and $P^{*lm}_{s}$ that we found in \req{metricweylrelation}. There is no loss of generality in doing this, as we can always reabsorb these constants into the radial variables. However, the computation is clearer in this way. 

Now, the logic to find the radial Teukolsky equations goes as follows. We start with a Weyl tensor that we have decomposed as \req{deltapsiexpansion}. In order to obtain the equations \req{goalequations} from \req{eqn:universalteukolsky0} and \req{eqn:universalteukolsky4} we need the metric that produced this Weyl tensor. Of course, we do not know the answer for higher-derivative gravities. However, we know the answer for Einstein gravity: every mode in \req{deltapsiexpansion} comes from a metric perturbation given by \req{hmunureconstructed}. Therefore, in the case of perturbative higher-derivative corrections, the metric perturbation associated to the Weyl tensor \req{deltapsiexpansion} will correspond to \req{hmunureconstructed} plus $\mathcal{O}(\lambda)$ terms. But these are irrelevant in order to obtain \req{goalequations}, since all the terms coming from the metric reconstruction are already of order $\lambda$. Hence, we simply need to use \req{hmunureconstructed}. 

Using the expansion \req{deltapsiexpansion}, the metric reconstruction \req{hmunureconstructed} and the background solution \req{rotatingmetric} in \req{eqn:universalteukolsky0} and \req{eqn:universalteukolsky4}, and expanding to linear order in the fluctuations and in the higher-derivative corrections, yields equations of the form
\begin{equation}\label{nonradial}
\begin{aligned}
 \frac{\zeta^{s-2}}{\Delta\Sigma}\sum_{l}\Bigg[&- S^{lm}_{s}P^{lm}_{s} \mathfrak{D}_{s}^2R^{lm}_{s}\\
 &+\lambda\left(f^{lm}_{s,0,0}R^{lm}_{s}S^{lm}_{s}+f^{lm}_{s,1,0}\frac{R^{lm}_{s}}{dr}S^{lm}_{s}+f^{lm}_{s,0,1}R^{lm}_{s}\frac{S^{lm}_{s}}{dx}+f^{lm}_{s,1,1}\frac{R^{lm}_{s}}{dr}\frac{S^{lm}_{s}}{dx}\right)\\
  &+\lambda\left(h^{lm}_{s,0,0}R^{*lm}_{s}S^{lm}_{-s}+h^{lm}_{s,1,0}\frac{R^{*lm}_{s}}{dr}S^{lm}_{-s}+h^{lm}_{s,0,1}R^{*lm}_{s}\frac{S^{lm}_{-s}}{dx}+h^{lm}_{s,1,1}\frac{R^{*lm}_{s}}{dr}\frac{S^{lm}_{-s}}{dx}\right)\\
 &+\lambda\left(g^{lm}_{s,0,0}R^{lm}_{-s}S^{lm}_{-s}+g^{lm}_{s,1,0}\frac{R^{lm}_{-s}}{dr}S^{lm}_{-s}+g^{lm}_{s,0,1}R^{lm}_{-s}\frac{S^{lm}_{-s}}{dx}+g^{lm}_{s,1,1}\frac{R^{lm}_{-s}}{dr}\frac{S^{lm}_{-s}}{dx}\right)\\
 &+\lambda\left(j^{lm}_{s,0,0}R^{*lm}_{-s}S^{lm}_{s}+j^{lm}_{s,1,0}\frac{R^{*lm}_{-s}}{dr}S^{lm}_{s}+j^{lm}_{s,0,1}R^{*lm}_{-s}\frac{S^{lm}_{s}}{dx}+j^{lm}_{s,1,1}\frac{R^{*lm}_{-s}}{dr}\frac{S^{lm}_{s}}{dx}\right)\Bigg]=0\, ,
\end{aligned}
\end{equation}
for $s=\pm2$, plus another two conjugate equations.
Here $\mathfrak{D}_{s}^2$ is the radial Teukolsky operator defined in \req{Teukoperator} and $f_{s,i,j}^{lm}$, $h_{s,i,j}^{lm}$, $g_{s,i,j}^{lm}$ and $j_{s,i,j}^{lm}$ are functions of $r$ and $x$ as well as the black hole spin $a$, the angular separation constants $B_{lm}$ and the frequency $\omega$. These functions additionally depend on the $H_{i}$ functions of the corrected Kerr metric \req{rotatingmetric}. They are given by very lengthy and non-illuminating expressions that we have obtained explicitly for the six-derivative theories \req{cubiclagrangian} . Note that in the terms proportional to $\lambda$ only first derivatives appear, as we have used the zeroth-order radial and angular Teukolsky equations to reduce the derivatives of higher order. For this reason, the functions $f_{s,i,j}^{lm}$, $h_{s,i,j}^{lm}$, $g_{s,i,j}^{lm}$ and $j_{s,i,j}^{lm}$ depend on the angular separation constants $B_{lm}$.

Let us now make the following two observations. First, the spin-weighted spheroidal harmonics satisfy the orthogonality relations\footnote{We remark that, addition to orthogonality, this relation defines our convention for their normalization.}
\begin{equation}
2\pi\int_{-1}^{1}dx S_{s}^{lm}(x;a\omega)S^{l'm}_{s}(x; a\omega)=\delta_{ll'}\, ,
\end{equation}
which hold even for complex $\omega$. We can thus project the equations \req{nonradial} into $S^{l'm}_{s}$ and obtain an infinite system of radial equations labelled by the number $l'$, 
\begin{equation}\label{radial0}
\begin{aligned}
 -P^{l'm}_{s}\mathfrak{D}_{s}^2R^{l'm}_{s}+\sum_{l}\lambda&\left(f^{l'lm}_{s,0}R^{lm}_{s}+f^{l'lm}_{s,1}\frac{dR^{lm}_{s}}{dr}+h^{l'lm}_{s,0}R^{*lm}_{s}+h^{l'lm}_{s,1}\frac{dR^{*lm}_{s}}{dr}\right.\\
 &\left.+g^{l'lm}_{s,0}R^{lm}_{-s}+g^{l'lm}_{s,1}\frac{dR^{lm}_{-s}}{dr}+j^{l'lm}_{s,0}R^{*lm}_{-s}+j^{l'lm}_{s,1}\frac{dR^{*lm}_{-s}}{dr}\right)=0\, ,
\end{aligned}
\end{equation}
where

\begin{equation}\label{fgintegrals}
\begin{aligned}
f_{s,i}^{l'lm}&=2\pi \int_{-1}^{1}dx S^{l'm}_{s}\left(f^{lm}_{s,i,0}S^{lm}_{s}+f^{lm}_{s,i,1}\frac{S^{lm}_{s}}{dx}\right)\, ,\\
h_{s,i}^{l'lm}&=2\pi \int_{-1}^{1}dx S^{l'm}_{s}\left(h^{lm}_{s,i,0}S^{lm}_{-s}+h^{lm}_{s,i,1}\frac{S^{lm}_{-s}}{dx}\right)\, ,\\
g_{s,i}^{l'lm}&=2\pi \int_{-1}^{1}dx S^{l'm}_{s}\left(g^{lm}_{s,i,0}S^{lm}_{-s}+g^{lm}_{s,i,1}\frac{S^{lm}_{-s}}{dx}\right)\, ,\\
j_{s,i}^{l'lm}&=2\pi \int_{-1}^{1}dx S^{l'm}_{s}\left(j^{lm}_{s,i,0}S^{lm}_{s}+j^{lm}_{s,i,1}\frac{S^{lm}_{s}}{dx}\right)\, .
\end{aligned}
\end{equation}
These are functions of $r$ only. 

Second, as the quasinormal modes of the Kerr black hole consist of a single term with a fixed $l$ and $m$, in the corrected QNMs the sum in \req{deltapsiexpansion} will contain a leading term while the rest of the terms will be of order $\lambda$. This is, for the QNM that we would label with the numbers $l_0$ and $m$, we will have $R_{s}^{l_0m}=\mathcal{O}(1)$ and $R_{s}^{lm}=\mathcal{O}(\lambda)$ for $l\neq l_0$.  Therefore, at first order in $\lambda$, the equation \req{radial0} becomes 

\begin{equation}
\begin{aligned}
-P^{l'm}_{s}\mathfrak{D}_{s}^2R^{l'm}_{s}+\lambda&\left(f^{l'l_0m}_{s,0}R^{l_0m}_{s}+f^{l'l_0m}_{s,1}\frac{dR^{l_0m}_{s}}{dr}+h^{l'l_0m}_{s,0}R^{*l_0m}_{s}+h^{l'l_0m}_{s,1}\frac{dR^{*l_0m}_{s}}{dr}\right.\\
&\left.+g^{l'l_0m}_{s,0}R^{l_0m}_{-s}+g^{l'l_0m}_{s,1}\frac{dR^{l_0m}_{-s}}{dr}+j^{l'l_0m}_{s,0}R^{*l_0m}_{-s}+j^{l'l_0m}_{s,1}\frac{dR^{*l_0m}_{-s}}{dr}\right)=0\, ,
\end{aligned}
\end{equation}
and similarly for the conjugate equations. 
For $l'=l_0$ these are in total four equations that only involve the variables $R_{s}^{l_0m}$, $R_{s}^{*l_0m}$. Dropping the labels for clarity, we find the master radial equations

\begin{equation}\label{radial1}
\begin{aligned}
-P_{+2}\mathfrak{D}_{+2}^2R_{+2}+\lambda&\left(f_{+2,0}R_{+2}+f_{+2,1}\frac{dR_{+2}}{dr}+h_{+2,0}R^{*}_{+2}+h_{+2,1}\frac{dR^{*}_{+2}}{dr}\right.\\
&\left.+g_{+2,0}R_{-2}+g_{+2,1}\frac{dR_{-2}}{dr}+j_{+2,0}R^{*}_{-2}+j_{+2,1}\frac{dR^{*}_{-2}}{dr}\right)=0\, ,\\
-P_{-2}\mathfrak{D}_{-2}^2R_{-2}+\lambda&\left(f_{-2,0}R_{-2}+f_{-2,1}\frac{dR_{-2}}{dr}+h_{-2,0}R^{*}_{-2}+h_{-2,1}\frac{dR^{*}_{-2}}{dr}\right.\\
&\left.+g_{-2,0}R_{+2}+g_{-2,1}\frac{dR_{+2}}{dr}+j_{-2,0}R^{*}_{+2}+j_{-2,1}\frac{dR^{*}_{+2}}{dr}\right)=0\, ,
\end{aligned}
\end{equation}
together with their conjugate counterparts.

We still need to compute the integrals in \req{fgintegrals} in order to determine the functions $f^{llm}_{s,i}$, $h^{llm}_{s,i}$, $g^{llm}_{s,i}$,  and $j^{llm}_{s,i}$. This could be achieved numerically if a numerical solution for the corrected background is known. However, a possibility to obtain an analytic result consists in performing an expansion in the black hole spin $\chi=a/M$. 
On the one hand, we know analytically the $\chi^n$ expansion of the functions $H_i$ determining the background black hole metric. On the other hand, the spin-weighted spheroidal harmonics $S_{s}^{lm}(x; a\omega)$ and the angular separation constants $B_{lm}$ have known analytic expansions for small $\gamma=a\omega$ \cite{Berti:2005gp}.\footnote{Strictly speaking, one should perform a double series expansion both in $\chi$ and in $\gamma$, but for simplicity here we perform an overall expansion in $\chi$ using that $\gamma=M\omega \chi$. This could be problematic only if $\omega$ is large, so that $|\gamma|>>1$, because in that case one may need to include more terms in the $\gamma$ expansion. However, the result should converge anyways if enough terms are included in the overall $\chi$ expansion. Besides, for the most relevant QNMs, corresponding to the fundamental modes and a few overtones of the first few harmonics $l=2,3,...$, we actually have $|\gamma|<\chi$. }

When we expand the integrand of \req{fgintegrals} in powers of $\chi$, we find that every term can indeed be integrated analytically. We thus are able to obtain explicit expressions for the functions

\begin{equation}
f^{llm}_{s,i}(r; a,\omega)\, ,\quad h^{llm}_{s,i}(r; a,\omega)\, ,\quad g^{llm}_{s,i}(r; a,\omega)\, ,\quad j^{llm}_{s,i}(r; a,\omega)
\end{equation}
as series expansions in $a$. By including enough terms in these expansions, one should be able to obtain an accurate result, even for moderate or large spins. 

The computation of the functions $f^{llm}_{s,i}$, $h^{llm}_{s,i}$,$g^{llm}_{s,i}$,  and $j^{llm}_{s,i}$ governing the equations \req{radial1} for the theory \req{cubiclagrangian} is one of the most important results of this paper. 

\subsection{Decoupling of the radial equations}\label{subsec:decouple}
So far we have been able to reduce the problem of studying black hole perturbations to a system of four coupled radial equations \req{radial1} (and their conjugate versions) for the variables $R_{s}$ and $R_{s}^{*}$. This problem can already be tackled with different methods in order to find the quasinormal modes \cite{McManus:2019ulj}, but it is still much more involved than the case of a decoupled equation.  
The idea to decouple these equations is to use the Starobinsky-Teukolsky identities \req{STidentities} as well as the relations with the conjugate variables \req{conjugaterelation}.  In fact, those relations are already implicitly assumed in \req{radial1} through the constants $P_{s}$. 

We note that, upon using the zeroth-order Teukolsky equations, the ST relationships \req{STidentities} can be expressed in the form of a first order operator, 
\begin{equation}\label{STidentities3}
\begin{aligned}
R_{-2}&=C_{+2}\left(A_{+2}R_{+2}+B_{+2}\frac{dR_{+2}}{dr}\right)\, ,\\
R_{+2}&=C_{-2}\left(A_{-2}R_{-2}+B_{-2}\frac{dR_{-2}}{dr}\right)\, ,
\end{aligned}
\end{equation}
for given functions $A_{\pm 2}$, $B_{\pm 2}$. Taking a derivative in these expressions and using again the Teukolsky equations we obtain similar relations for the derivatives $dR_{\pm 2}/dr$.  Combining these relations with  \req{conjugaterelation}, we can always express any of the $R_{s}$ or $R_{s}^{*}$ in terms of any of the other variables.

All of these relations hold at zeroth order in $\lambda$ because the radial variables satisfy the zeroth-order Teukolsky equations. Therefore, we are allowed to assume those relations in the terms proportional to $\lambda$ in \req{radial1}. The result is four decoupled equations of the form

\begin{equation}\label{radial2}
\begin{aligned}
 -P_{s}\mathfrak{D}_{s}^2R_{s}+\lambda\left[f_{s,0}R_{s}+f_{s,1}\frac{dR_{s}}{dr}\right]&=0\, ,\\
 -P_{s}^{*}\mathfrak{D}_{s}^2R_{s}^{*}+\lambda\left[f_{s,0}^{*}R_{s}^{*}+f_{s,1}^{*}\frac{dR_{s}^{*}}{dr}\right]&=0\, ,
\end{aligned}
\end{equation}
for $s=\pm2$. The price to pay for the decoupling is that the functions $f_{s,i}$, $f^{*}_{s,i}$ now depend on the polarization parameters $q_{s}$ and the ST constants $C_{s}$. Let us take note that the dependence on these parameters is of the form 
\begin{align}
f_{s,i}&=f_{s,i,1}+q_{s}f_{s,i,2}+C_{s} f_{s,i,3}+C_{s}q_{-s} f_{s,i,3}\, ,\\
f_{s,i}^{*}&=f_{s,i,1}^{*}+q_{s}^{-1}f_{s,i,2}^{*}+C_{s} q_{s}^{-1} f_{s,i,3}^{*}+C_{s}q_{-s}q_{s}^{-1} f_{s,i,3}^{*}\, ,
\end{align}
where each of the $f_{s,i,j}$ and $f_{s,i,j}^{*}$ are functions or $r$. We recall that the coefficients $P_{s}$ and $P_{s}^{*}$ \req{eq:Pconstants} also depend on these constants.

Now, besides the two polarization parameters $q_{s}$ and one of the $C_{s}$, there is a fourth undetermined constant in the problem: the frequency $\omega$. In order to obtain the QNMs, we must proceed as follows. We can solve each equation in \req{radial2} independently to find the shift in Kerr's QNM frequency, 
\begin{equation}
\omega_{s}=\omega_{\rm Kerr}+\lambda \delta\omega_{s}\, \quad \omega_{s}^{*}=\omega_{\rm Kerr}+\lambda \delta\omega_{s}^{*}\, ,
\end{equation} 
where $\omega_{s}$ and $\omega_{s}^{*}$ denote the frequency obtained from the corresponding equation. The shifts $\delta\omega_{s}$ and $\delta\omega_{s}^{*}$ are functions of the various parameters and we can see that their dependence on them is of the form
\begin{equation}\label{deltaomegaCq}
\begin{aligned}
\delta\omega_{s}&=\frac{1}{P_{s}}\left[\delta\omega_{s,1}+q_{s}\delta\omega_{s,2}+C_{s}\delta\omega_{s,3}+C_{s}q_{-s}\delta\omega_{s,4}\right]\, ,\\
\delta\omega_{s}^{*}&=\frac{1}{q_{s}P_{s}^{*}}\left[q_{s}\delta\omega_{s,1}^{*}+\delta\omega_{s,2}^{*}+C_{s}\delta\omega_{s,3}^{*}+C_{s}q_{-s}\delta\omega_{s,4}^{*}\right]\, ,
\end{aligned}
\end{equation}
where $\delta\omega_{s,i}$ and $\delta\omega_{s,i}^{*}$ are coefficients that we have to determine numerically. 
Now, for a QNM, the same frequency must be a solution of all the equations. Therefore, the following must hold in a QNM solution
\begin{equation}\label{allomegasarethesame}
\delta\omega_{+2}(q_{s},C_{s})=\delta\omega_{-2}(q_{s},C_{s})=\delta\omega_{+2}^{*}(q_{s},C_{s})=\delta\omega_{-2}^{*}(q_{s},C_{s})\, .
\end{equation}
These equations fix the value of the polarization parameters $q_{s}$ allowing us to obtain the QNM frequencies.  
On the other hand, as we show explicitly in the next section, the value of the ST constants $C_s$ remains arbitrary because the frequencies are actually independent of these constants. This is a highly non-trivial fact that cannot be easily concluded from the form of the equations. Nevertheless, our numeric results confirm this, as will be discussed in the next section.  This is a strong self-consistency test of our method: it tells us that the choice of ST constants represents a choice of gauge to reconstruct the metric perturbation from the Teukolsky variables. The same behavior is found in general relativity, where one can reconstruct the metric perturbation in the ingoing or outgoing radiation gauges, or more generally, as a linear combination of both \cite{Pound:2021qin, Dolan:2021ijg}. 

\section{Quasinormal mode frequencies in six-derivative gravity}\label{sec:QNM6}
As an application of our method, we compute the $(l,m)=(2,2)$ and $(3,3)$  fundamental QNM frequencies of the six-derivative theory \req{cubiclagrangian} to linear order in the higher-derivative couplings. By following the procedure outlined in the previous sections we have obtained the radial equations \req{radial2} for those modes at order $a^6$ and $a^4$, respectively.\footnote{In the ancillary files we provide these equations truncated at order $a^2$.} One last technical obstacle arises in the computation: we observe that the corrections to the Teukolsky equations in \req{radial2} have a singular character at the horizon because $f_{s,i}$ and $f_{s,i}^{*}$ behave as $f_{s,0}\sim (r-r_{+})^{-1}$ and $f_{s,1}\sim (r-r_{+})^{0}$.
This behavior, which arises due to the fact that we have reduced the order of the equations of motion by imposing the zeroth-order equations, is problematic for the following reason. Observe that we want our variables to behave as $R_{s}\sim (r-r_{+})^{\gamma}$ near the horizon, for a certain $\gamma$. In that case we have $\mathfrak{D}_{s}^2R_s\sim  (r-r_{+})^{\gamma}$, but due to the behavior of the $f_{s,i}$ functions, the terms proportional to $\lambda$ in \req{radial2} behave in turn as $(r-r_{+})^{\gamma-1}$. This indicates that the variables  $R_{s}$, $R_{s}^{*}$ are not appropriate in the presence of corrections.\footnote{In the case of $s=+2$ equation, the terms proportional to $\lambda$ actually behave as $(r-r_{+})^{\gamma}$ once we use the value of $\gamma$ from the uncorrected Teukolsky equation, so the equation is secretly regular. This is not the case for the $s=-2$ equation, though.} In fact, we can redefine the variables as\footnote{We suspect the origin of these singularities lies in the choice of corrected NP frame \req{tetrad1}, which is probably singular at the horizon. A redefinition of the radial variables is therefore equivalent to choosing a different NP frame.}
\begin{equation}
\begin{aligned}
R_{s}&=\hat{R}_{s}+\lambda \left(\alpha_{s}\hat{R}_{s}+\beta_{s}\frac{d\hat{R}_{s}}{dr}\right)\, ,\\
R_{s}^{*}&=\hat{R}_{s}^{*}+\lambda \left(\alpha_{s}^{*}\hat{R}_{s}^{*}+\beta_{s}^{*}\frac{d\hat{R}_{s}^{*}}{dr}\right)\, ,
\end{aligned}
\end{equation}
for radial-dependent coefficients $\alpha_{s}$, $\beta_{s}$, $\alpha_{s}^{*}$, $\beta_{s}^{*}$.
The equation for the new variables $\hat{R}_{s}$, $\hat{R}_{s}^{*}$ can again be rewritten --- after making use of the zeroth-order Teukolsky equation in the terms proportional to $\lambda$ --- in the form of \req{radial2}, with new functions $\hat{f}_{s,i}$, $\hat{f}^{*}_{s,i}$ that depend on the $\alpha_{s}$, $\beta_{s}$, $\alpha_{s}^{*}$, $\beta_{s}^{*}$ functions. It suffices to choose these functions as
\begin{equation}
\alpha_{s}(r)=\frac{a_s}{r-r_{+}}\, ,\quad \beta_{s}(r)=b_{s}\, ,
\end{equation}
for constants $a_{s}$, $b_{s}$, and analogously for $\alpha_{s}^{*}$ and $\beta_{s}^{*}$.  By choosing the coefficients $a_s$, $b_s$ appropriately, we can remove the divergences and obtain well-behaved equations for the variables $\hat{R}_{s}$, with the functions $\hat{f}_{s,i}$ satisfying $\hat{f}_{s,0}\sim (r-r_{+})^0$, $\hat{f}_{s,1}\sim (r-r_{+})$ for $r\rightarrow r_{+}$. In fact, with this choice we find that  these functions take a polynomial form 
\begin{equation}
\hat{f}_{s,0}(r)=M^2\sum_{n=0}^{N}\left(\frac{r}{M}\right)^{4-n}\hat{f}_{s,0,n}\, , \quad \hat{f}_{s,1}(r)=M(r-r_{+})\sum_{n=0}^{N}\left(\frac{r}{M}\right)^{3-n}\hat{f}_{s,1,n}\
\end{equation}
where $\hat{f}_{s,i,n}$ are coefficients and $N$ increases with the order of the $a$-expansion.   Thus, the resulting equations 
\begin{equation}\label{radial3}
\begin{aligned}
 -P_{s}\mathfrak{D}_{s}^2\hat{R}_{s}+\lambda\left[\hat f_{s,0}\hat R_{s}+f_{s,1}\frac{d\hat R_{s}}{dr}\right]&=0\, ,\\
 -P_{s}^{*}\mathfrak{D}_{s}^2\hat{R}_{s}^{*}+\lambda\left[\hat f_{s,0}^{*}\hat{R}_{s}^{*}+\hat f_{s,1}^{*}\frac{d\hat{R}_{s}^{*}}{dr}\right]&=0\, ,
\end{aligned}
\end{equation}
now have the same behavior as the zeroth-order Teukolsky equation for $r\rightarrow r_{+}$ and for $r\rightarrow\infty$ and we can impose the usual boundary conditions for QNMs. 
At $r=r_{+}$ the solution must behave as
\begin{equation}\label{Rsnh}
R_{s}\sim \left(\frac{r}{r_{+}}-1\right)^{\gamma_s}\, ,\quad \gamma_s=-s-\frac{i}{r_{+}^2-a^2}\left[2\omega M r_{+}-m a\right]+\lambda \delta\gamma_{s}\, ,
\end{equation}
where the exponent corresponds to modes falling toward the black hole horizon and it takes the Kerr value plus a theory-dependent correction $\delta\gamma_{s}$. 
At infinity $r\rightarrow\infty$ we look for solutions corresponding to outgoing radiation and hence $R_{s}$ behaves as 
\begin{equation}\label{Rsinf}
R_{s}\sim r^{-1-2s} e^{-i\omega(1+\lambda\delta) r} \, ,
\end{equation}
where $\delta$ is again a theory-dependent coefficient. 

In order to find the QNM frequencies we have applied two different numerical approaches. The first one consists in performing a direct numerical integration of \req{radial3} with the boundary conditions \req{Rsnh}, \req{Rsinf}, implementing a shooting-method strategy in order to find $\omega$. The second one is based on the approach of \cite{Zimmerman:2014aha,Mark:2014aja,Hussain:2022ins}, in the spirit of computing spectral shifts of a perturbed Hermitian operator. 
The results we report below were obtained from the first method, but we checked that the relative difference between the two methods for the shift in the QNM frequencies is smaller than $2\%$ for $|\chi| \leq 0.12$ and remains this small on average for the complete range of values of angular momentum we study. Let us discuss the cases of the even- and odd-parity theories independently.

\subsection{Even-parity corrections}
We consider Einstein gravity supplemented with the even-parity six-derivative operator in \req{cubiclagrangian}. This will introduce corrections to the Kerr QNM frequencies that we can express as
\begin{equation}\label{eqn:shifteven}
\omega=\omega^{\rm Kerr}+\frac{\ell^4\lambda_{\rm ev}}{M^5}\delta\omega\, ,
\end{equation}
so that $\delta\omega$ is a dimensionless coefficient that depends on the dimensionless spin $\chi$. Our goal is to determine this coefficient.

In the case of parity-preserving corrections, perturbations can be naturally decoupled into modes of odd and even parity. For the background metric \req{rotatingmetric}, a natural parity transformation corresponds to 
\begin{equation}\label{eq:paritytransf}
(x,\phi)\rightarrow (-x,\phi+\pi)\, .
\end{equation}
One can see that this has the effect on the NP frame $(m_{\mu},\bar{m}_{\mu})\rightarrow -(\bar{m}_{\mu}, m_{\mu})$.  In addition, the spheroidal harmonics verify $S_{s}(-x)=\pm S_{-s}(x)$. Therefore, up to a global sign, the parity transformation \req{eq:paritytransf} is equivalent to $R_{s}\leftrightarrow R_{s}^{*}$ at the level of the metric perturbation \req{hmunureconstructed}. Hence, we conclude that the modes of definite parity are those with 
\begin{equation}
R_{s}=\pm R_{s}^{*}\, ,
\end{equation}
or in other words, this fixes the polarization parameters $q_{s}$ to be $q_{+2}=q_{-2}=\pm1$. The $+$ sign corresponds to polar perturbations while the $-$ sign to axial ones. When $q_{s}=\pm 1$, we check that, indeed, the conjugate equations in \req{radial3} are identical to the non-conjugate ones. Thus, we only need to consider the equations for $\hat R_{+2}$ and $\hat{R}_{-2}$, and only the ST constants $C_{s}$ remain unspecified. By solving these equations and taking into account \req{deltaomegaCq} and \req{eq:Pconstants}, we obtain that the shifts in the QNM frequencies for each polarization $\pm$ take the form

\begin{equation}
\delta\omega_{s}^{\pm}=\sigma_{s}^{\pm}\frac{1+\gamma_{s}^{\pm} C_{s}}{1+\rho_{s}^{\pm}C_{s}}\, ,
\end{equation}
for coefficients $\sigma_{s}^{\pm}$, $\gamma_{s}^{\pm}$, $\rho_{s}^{\pm}$ that depend on the dimensionless angular momentum $\chi$ and that we determine numerically. Then, one may think that the ST constants are determined by the condition that the two shifts are the same (because all the equations should give the same QNM frequency), so that we would have the equation $\delta\omega_{+2}=\delta\omega_{-2}$ for the ST constants. However, the numerical results indicate strongly that 
\begin{equation}
\gamma_{s}^{\pm}=\rho_{s}^{\pm}\, ,
\end{equation}
meaning that the QNM frequencies are, in fact, independent of these constants, which can be chosen arbitrarily. Furthermore, and even more importantly, the numerical results also indicate that $\delta\omega_{+2}^{\pm}=\delta\omega_{-2}^{\pm}$.  

\begin{figure}[t!]
	\centering
	\includegraphics[width=0.48\textwidth]{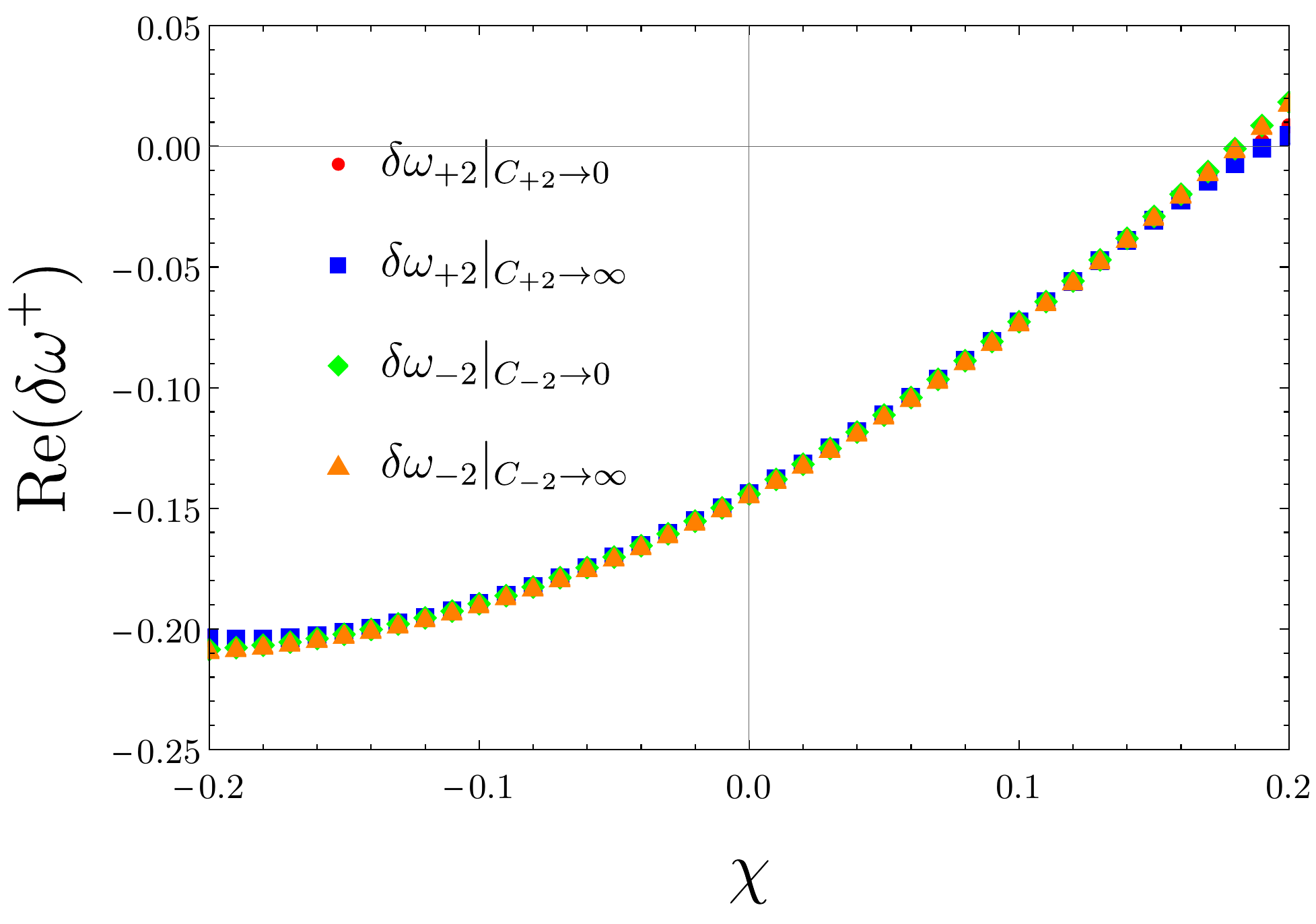}
	\includegraphics[width=0.48\textwidth]{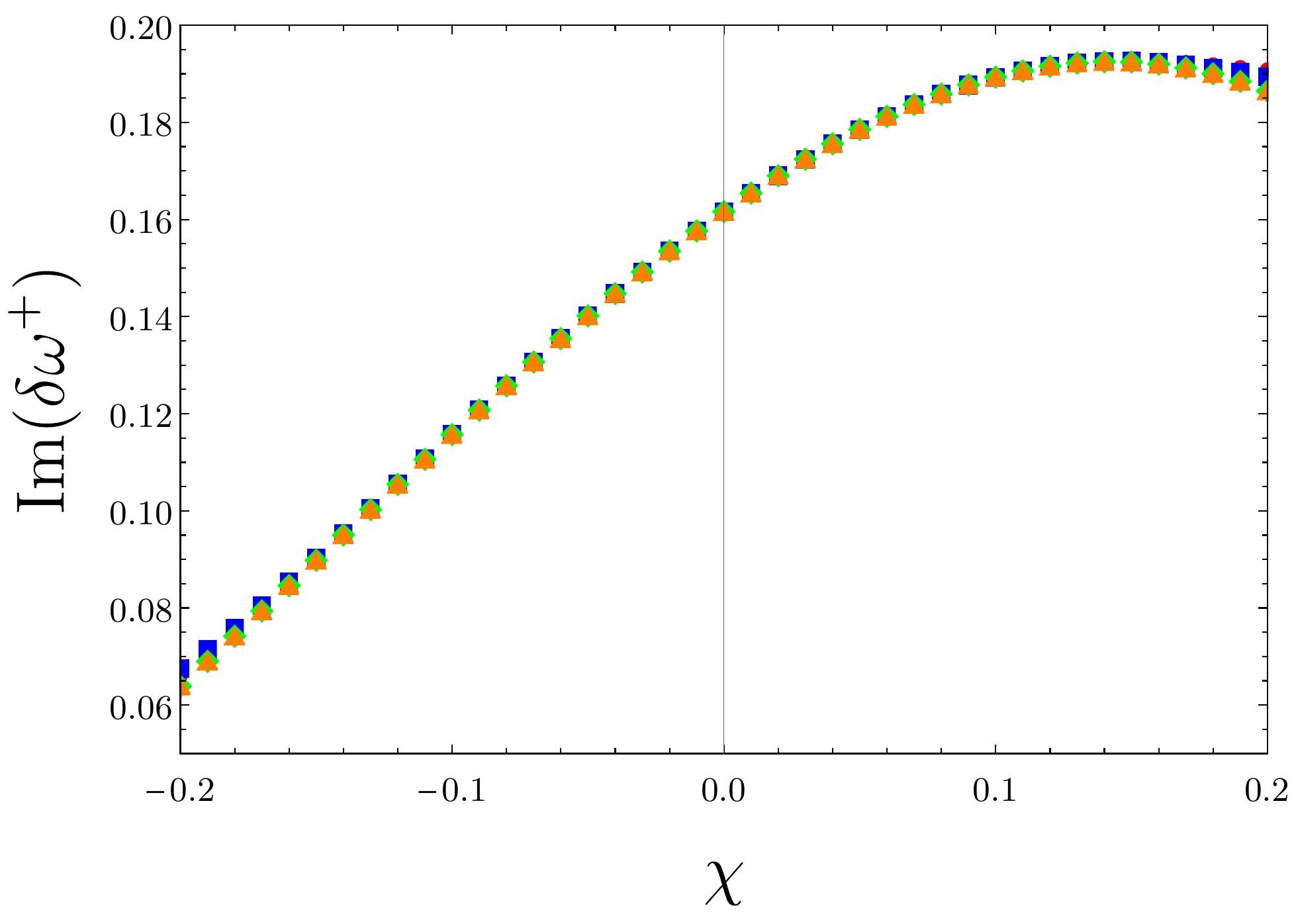}
	\includegraphics[width=0.48\textwidth]{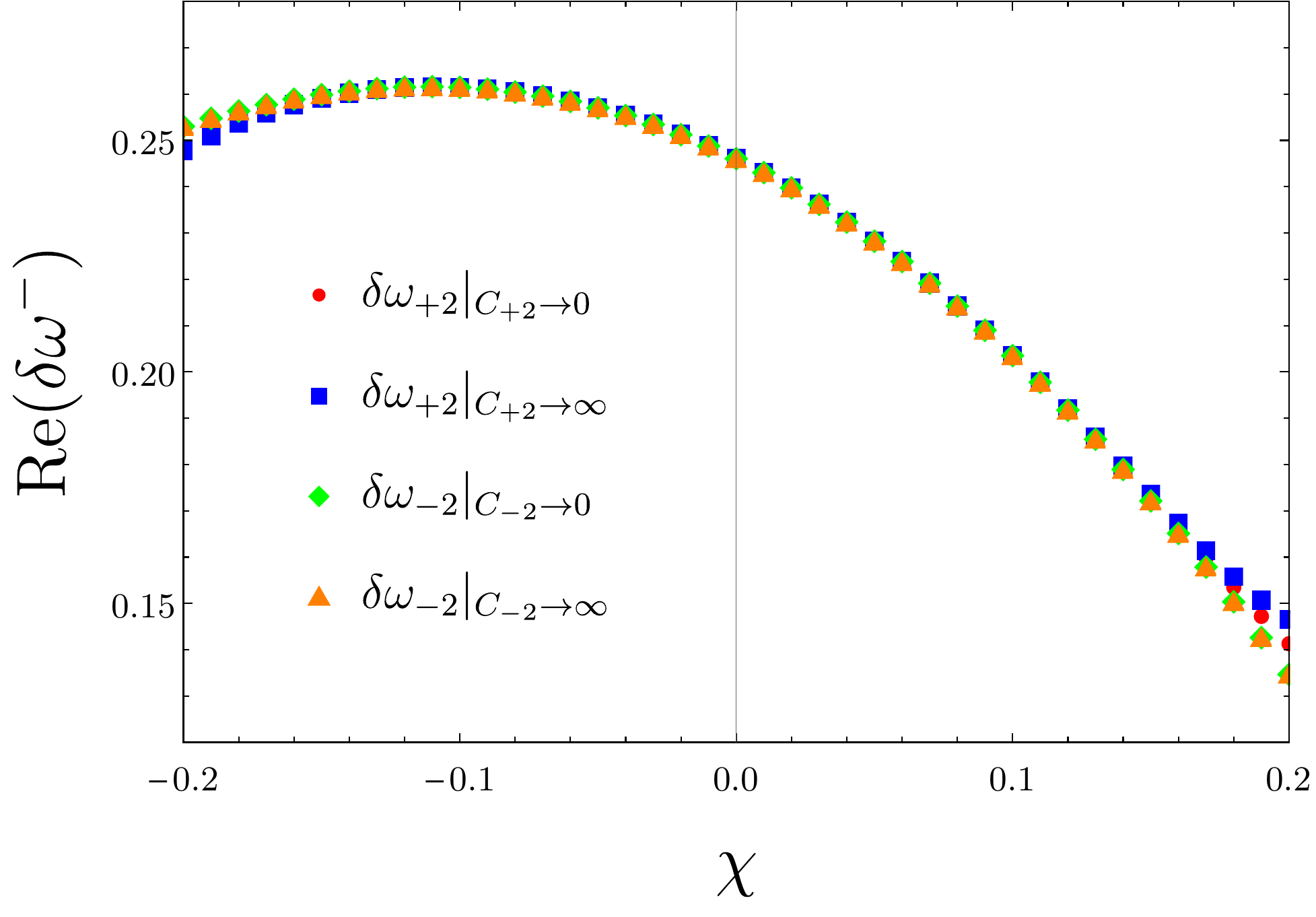}
	\includegraphics[width=0.48\textwidth]{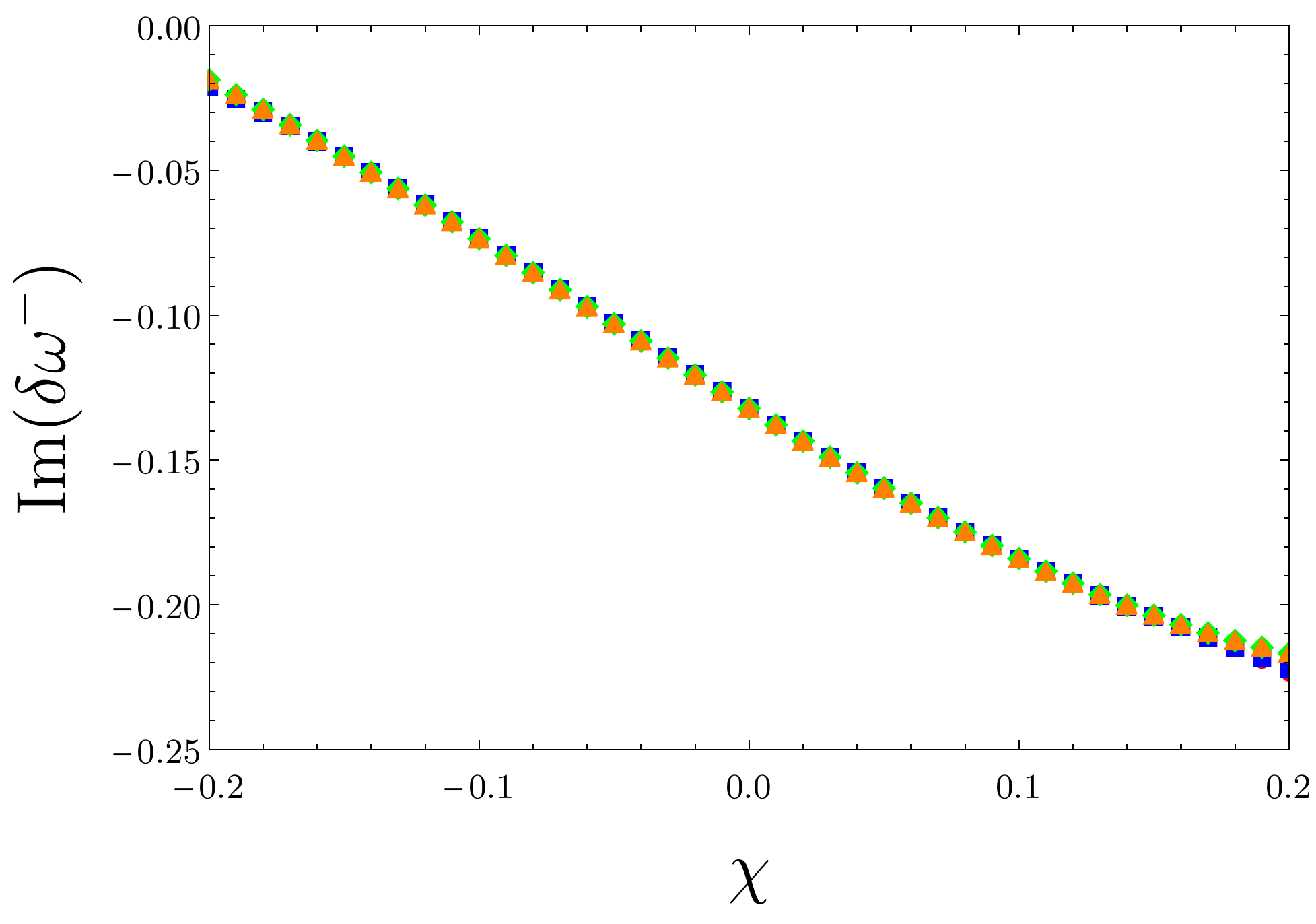}
	\caption{Shifts in the polar (top row) and axial (bottom row) QNMs frequencies with $(l,m)=(2,2)$, as defined in \eqref{eqn:shifteven}, due to the even-parity six-derivative correction in \eqref{cubiclagrangian}. We compute these shifts in four different ways using an $\mathcal{O}(\chi^6)$ expansion of the equations \req{radial3} and observe that all the frequencies are approximately the same. This indicates that the result is independent of the ST constants and that the computation is consistent. }
	\label{fig:evenl2}
\end{figure}

To illustrate this, we compute the shifts $\delta\omega_{s}$ for $C_{s}=0$ and $C_{s}\rightarrow\infty$. Let us study first the case of the $(l,m)=(2,2)$ modes. In Fig.~\ref{fig:evenl2} we show the four different results,
\begin{equation}
\delta\omega_{+2}\big|_{C_{+2}=0}\,, \quad \delta\omega_{+2}\big|_{C_{+2}=\infty}\,, \quad \delta\omega_{-2}\big|_{C_{-2}=0}\,, \quad \delta\omega_{-2}\big|_{C_{-2}=\infty}\,,
\end{equation} 
for each polarization obtained from a numerical integration of \req{radial3} expanded at order $\chi^6$. We observe that all the frequencies are in fact almost identical for all the values of $\chi$ we are plotting. Especially, the two estimations for $\delta\omega_{-2}$ are indistinguishable in these plots, indicating that they are independent of $C_{-2}$. These two values remain very close to each other even for larger values of the spin, up to $|\chi|\sim 0.4$. 
On the other hand, we observe that the two estimations for $\delta\omega_{+2}$ depart slightly from themselves and from $\delta\omega_{-2}$ already at $|\chi|\sim 0.2$. These small discrepancies can be explained by the truncation of the spin expansion and by the different character of the $s=+2$ and $s=-2$ equations. In fact, we expect the equations to yield the same QNM frequencies only when the complete series in $\chi$ is included, but when the series in truncated at a given order we can only expect the results to agree for small enough $\chi$. 
In this regard, the equation for $s=+2$ seems to be quite sensitive to the order of the expansion as it converges more slowly than the $s=-2$ equation, which turns out to be more stable. However, the results from both equations show convergence as more terms are added in the spin expansion.  Thus, we expect that by including additional orders in $\chi$ all the estimations will converge even for larger spins --- in principle, we expect convergence for all spins $|\chi|<1$, although many terms may be needed to obtain a good accuracy for black holes close to extremality. 

In order to perform a quantitative comparison, we can fit the numerical results to a polynomial in $\chi$.  Performing a quadratic fit in the interval $-0.1\le \chi\le 0.1$ we obtain the following result for the polar modes
\begin{equation}\label{eq:evenl2plus}
\begin{aligned}
\delta\omega_{+2}^{+}\big|_{C_{+2}=0}&=(-0.144+0.162 i)+(0.586\, +0.375 i) \chi +(1.290\, -0.909 i) \chi ^2\, ,\\
\delta\omega_{+2}^{+}\big|_{C_{+2}=\infty}&=(-0.144+0.162 i)+(0.586\, +0.375 i) \chi +(1.290\,
   -0.909 i) \chi ^2\, ,\\
\delta\omega_{-2}^{+}\big|_{C_{-2}=0}&=(-0.144+0.162 i)+(0.586\, +0.375 i) \chi +(1.294\,
   -0.911 i) \chi ^2\, ,\\
\delta\omega_{-2}^{+}\big|_{C_{-2}=\infty}&=(-0.144+0.162 i)+(0.586\, +0.375 i) \chi +(1.294\,
   -0.911 i) \chi ^2\, ,
\end{aligned}
\end{equation}
while for the axial ones we get

\begin{equation}\label{eq:evenl2minus}
\begin{aligned}
\delta\omega_{+2}^{-}\big|_{C_{+2}=0}&=(0.246\, -0.132 i)-(0.289\, +0.560 i) \chi -(1.360\,
   -0.337 i) \chi ^2\, ,\\
\delta\omega_{+2}^{-}\big|_{C_{+2}=\infty}&=(0.246\, -0.132 i)-(0.289\, +0.560 i) \chi -(1.360\,
   -0.337 i) \chi ^2\, ,\\
\delta\omega_{-2}^{-}\big|_{C_{-2}=0}&=(0.246\, -0.132 i)-(0.289\, +0.559 i) \chi -(1.358\,
   -0.338 i) \chi ^2\, ,\\
\delta\omega_{-2}^{-}\big|_{C_{-2}=\infty}&=(0.246\, -0.132 i)-(0.289\, +0.559 i) \chi -(1.359\,
   -0.338 i) \chi ^2 \, .
\end{aligned}
\end{equation}
These results pass several consistency tests. First, the agreement of the $\mathcal{O}(\chi^0)$ and $\mathcal{O}(\chi^1)$ coefficients is essentially perfect. In addition, those coefficients match with good accuracy the results obtained in \cite{Cano:2021myl} following a direct metric perturbation approach. In fact, in that case we got
\begin{equation}
\begin{aligned}
\delta\omega^{+}&=-0.137+0.161 i+(0.612+0.376 i)\chi+\mathcal{O}(\chi^2)\, ,\\ \delta\omega^{-}&=0.244-0.130 i-(0.288+572 i)\chi+\mathcal{O}(\chi^2)\, .
\end{aligned}
\end{equation}
which coincide with the results above with an accuracy better than $5\%$ for $\delta\omega^{+}$\footnote{\label{foot}We suspect that most of this error comes from the results in \cite{Cano:2021myl}, since the result in \req{eq:evenl2plus}, obtained in four different ways, seems to be very robust. Additionally, we can quote the results of \cite{deRham:2020ejn} for static black holes, which in our conventions yield $\delta\omega^{+}\approx-0.144+0.163 i$ and $\delta\omega^{-}\approx 0.246-0.133 i$, matching our results here with an error less than $1\%$.} and around $1\%$ for $\delta\omega^{-}$. Finally, the agreement among the $\mathcal{O}(\chi^2)$ coefficients is also excellent. The small discrepancies can be easily attributed to the numerical precision and to the different behavior of the equations for a finite truncation of the spin series.  

\begin{figure}[t!]
	\centering
	\includegraphics[width=0.48\textwidth]{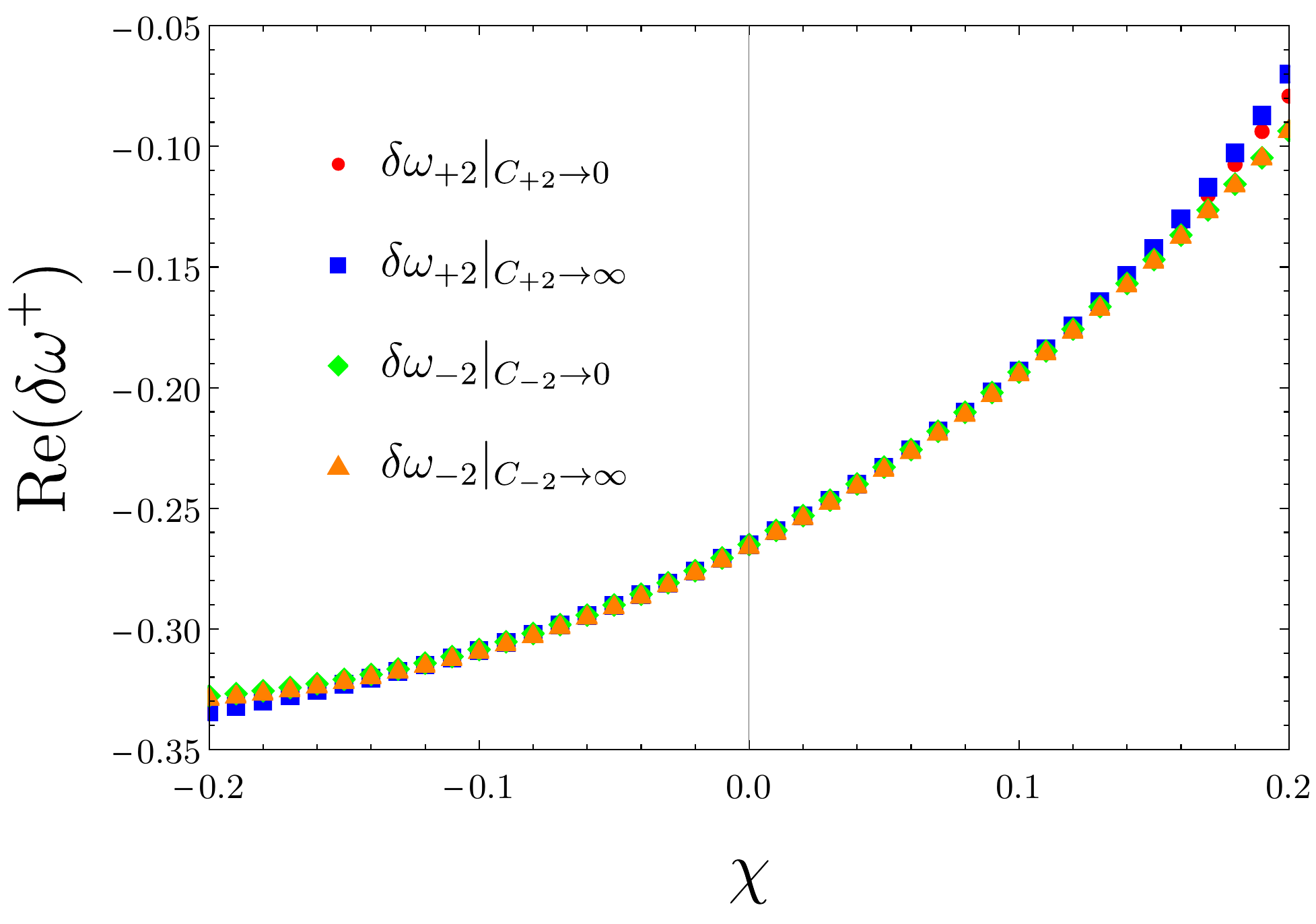}
	\includegraphics[width=0.48\textwidth]{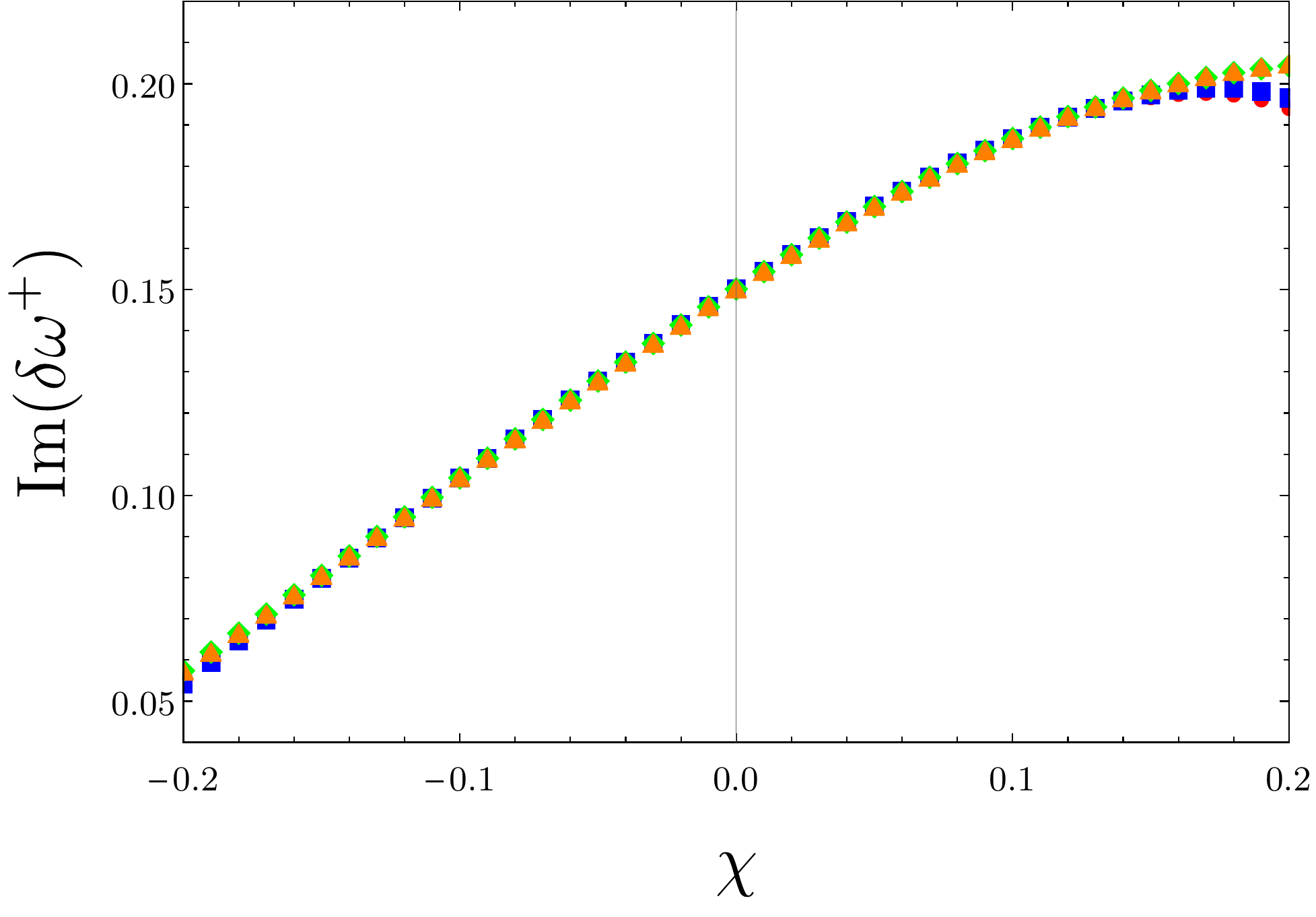}
	\includegraphics[width=0.48\textwidth]{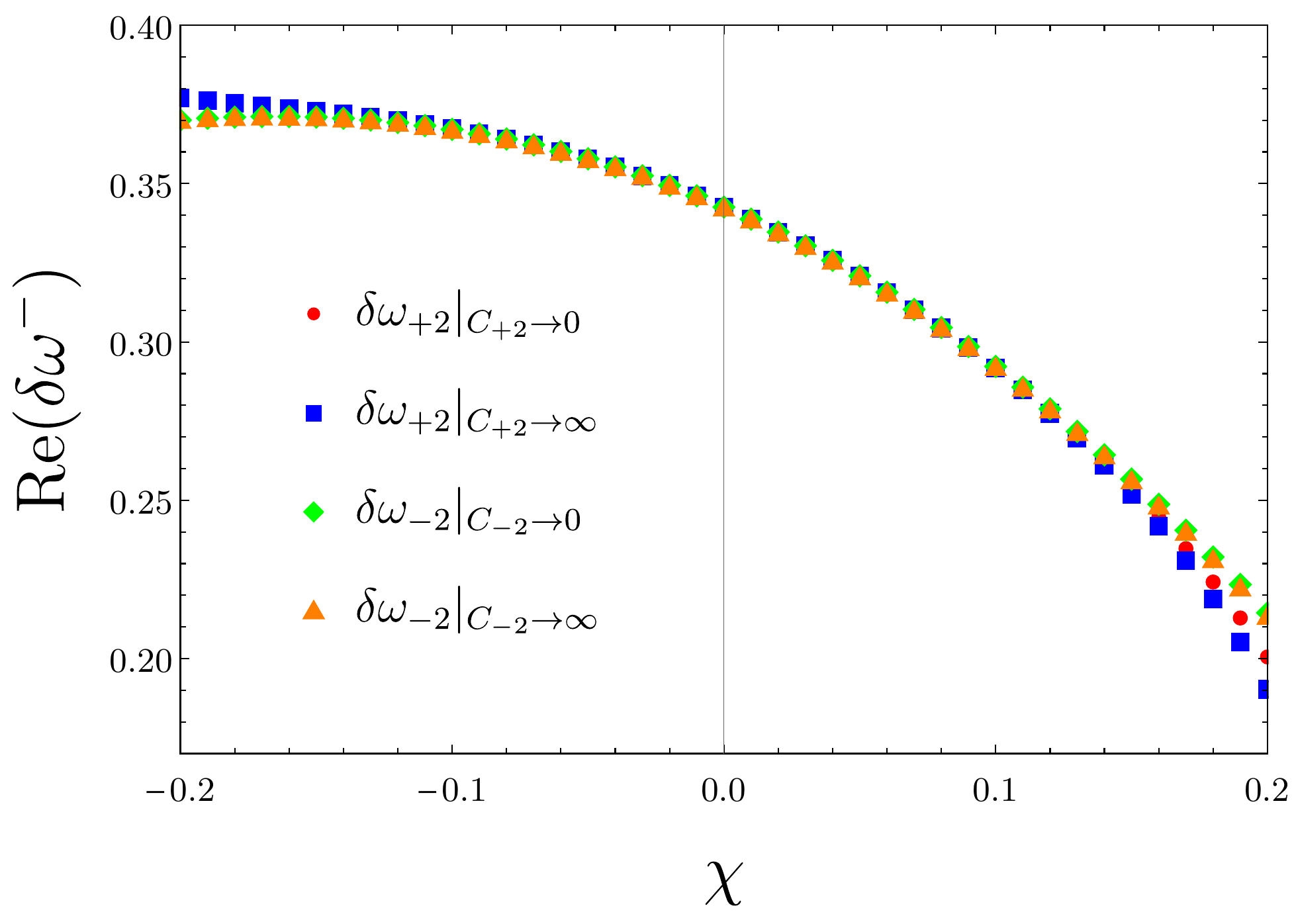}
	\includegraphics[width=0.48\textwidth]{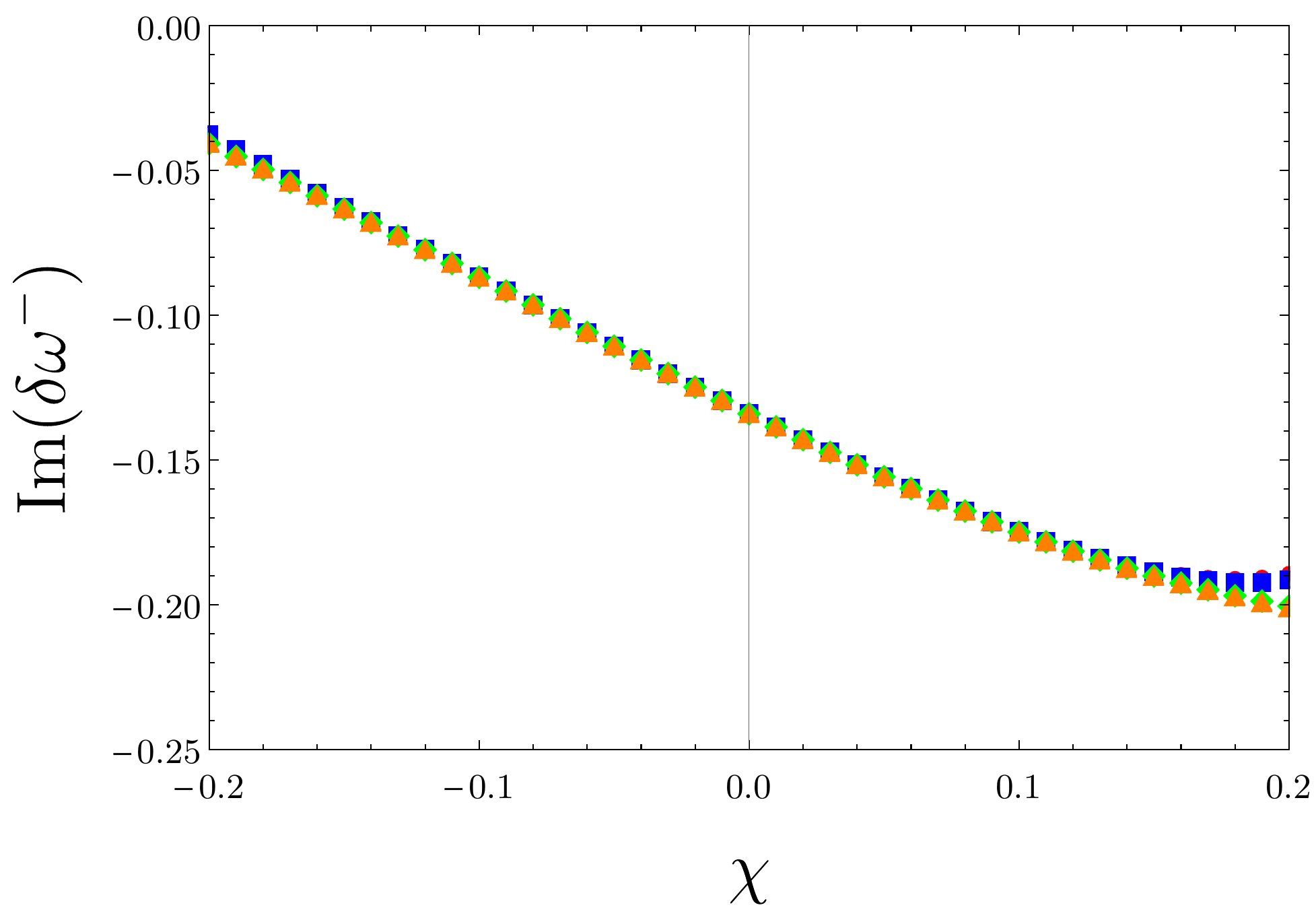}
	\caption{Shifts in the polar (top row) and axial (bottom row) QNMs frequencies with $(l,m)=(3,3)$, as defined in \eqref{eqn:shifteven}, due to the even-parity six-derivative correction in \eqref{cubiclagrangian}.  We compute the shifts in four different ways using an $\mathcal{O}(\chi^4)$ expansion of the equations \req{radial3}.}
	\label{fig:evenl3}
\end{figure}

As a further test, we have also computed the $(l,m)=(3,3)$ modes. The convergence in the spin expansion seems to be faster than for $l=2$ and in fact, we obtain a similar accuracy using an expansion of order $\chi^4$ instead of $\chi^6$. The four different estimations for the frequencies are shown in Fig.~\ref{fig:evenl3} where we again observe that all of them remain very close to each other.

Fitting these results to a quadratic polynomial in the range $-0.1\le \chi\le 0.1$ yields
\begin{equation}\label{eq:evenl3plus}
\begin{aligned}
\delta\omega_{+2}^{+}\big|_{C_{+2}=0}&=(-0.265+0.150 i)+(0.575\, +0.418 i) \chi +(1.405\,
   -0.473 i) \chi ^2\, ,\\
\delta\omega_{+2}^{+}\big|_{C_{+2}=\infty}&=(-0.265+0.150 i)+(0.576\, +0.418 i) \chi +(1.411\,
   -0.473 i) \chi ^2\, ,\\
\delta\omega_{-2}^{+}\big|_{C_{-2}=0}&=(-0.265+0.150 i)+(0.573\, +0.418 i) \chi +(1.398\,
   -0.463 i) \chi ^2\, ,\\
\delta\omega_{-2}^{+}\big|_{C_{-2}=\infty}&=(-0.265+0.150 i)+(0.573\, +0.418 i) \chi +(1.401\,
   -0.463 i) \chi ^2\, ,
\end{aligned}
\end{equation}
and 
\begin{equation}\label{eq:evenl3minus}
\begin{aligned}
\delta\omega_{+2}^{-}\big|_{C_{+2}=0}&=(0.343\, -0.134 i)-(0.373\, +0.444 i) \chi -(1.295\,
   -0.319 i) \chi ^2\, ,\\
\delta\omega_{+2}^{-}\big|_{C_{+2}=\infty}&=(0.343\, -0.134 i)-(0.374\, +0.445 i) \chi -(1.301\,
   -0.318 i) \chi ^2\, ,\\
\delta\omega_{-2}^{-}\big|_{C_{-2}=0}&=(0.343\, -0.134 i)-(0.372\, +0.445 i) \chi -(1.291\,
   -0.316 i) \chi ^2\, ,\\
\delta\omega_{-2}^{-}\big|_{C_{-2}=\infty}&=(0.343\, -0.134 i)-(0.372\, +0.445 i) \chi -(1.294\,
   -0.315 i) \chi ^2 \, .
\end{aligned}
\end{equation}
We observe excellent agreement among all the coefficients, including those of $\chi^2$, and again the results to linear order reproduce the ones previously obtained in \cite{Cano:2021myl}, which we rewrite here for convenience

\begin{equation}
\begin{aligned}
\delta\omega^{+}&=-0.258+0.151 i+(0.597+0.426 i)\chi+\mathcal{O}(\chi^2)\, ,\\ \delta\omega^{-}&=0.340-0.132 i-(0.369+0.447 i)\chi+\mathcal{O}(\chi^2)\, .
\end{aligned}
\end{equation}

These results make us confident that the frequencies are indeed independent of the choice of ST constants and that both equations $s=2$ and $s=-2$ yield the same frequencies. These highly non-trivial properties, together with the fact that we reproduce the results at linear order in the spin of \cite{Cano:2021myl},  provide a strong test of the validity of our approach and computations. 

\subsection{Odd-parity corrections}
Let us now consider the parity-breaking cubic interaction in \req{cubiclagrangian}, and write the corresponding shift in the QNM frequencies as
\begin{equation}\label{eqn:shiftodd}
\omega=\omega^{\rm Kerr}+\frac{\ell^4\lambda_{\rm odd}}{M^5}\delta\omega\, .
\end{equation}
Since this theory does not preserve parity, we can no longer decouple polar and axial modes, and these are inevitably mixed. Thus, the polarization parameters $q_{s}$ have to be determined at the same time as the frequencies by solving \req{allomegasarethesame}. Now, based on our experience with the parity-preserving corrections, we expect the frequencies to be independent of the choice of ST constants, since these represent, essentially, a redundancy in our description of perturbations. 
To test this, we solve the equations \req{allomegasarethesame} again in the limits $C_{s}\rightarrow 0$ and $C_{s}\rightarrow \infty$, in which case the $s=2$ and $s=-2$ equations decouple. For instance, if we set  $C_{+2}=0$, then the condition $\delta\omega_{+2}=\delta\omega_{+2}^{*}$ gives us an equation for $q_{+2}$ which has two solutions. Naturally these correspond to the two possible polarizations for QNMs, and in this case $q_{+2}\neq \pm 1$, indicating that these modes do not have a definite parity.
If instead we set $C_{+2}\rightarrow \infty$ then the condition $\delta\omega_{+2}=\delta\omega_{+2}^{*}$ gives us an equation for $q_{-2}$ which again has two solutions. A similar discussion applies for the case of $\delta\omega_{-2}=\delta\omega_{-2}^{*}$ for $C_{-2}=0$ and $C_{-2}\rightarrow\infty$. 

\begin{figure}[t!]
	\centering
	\includegraphics[width=0.48\textwidth]{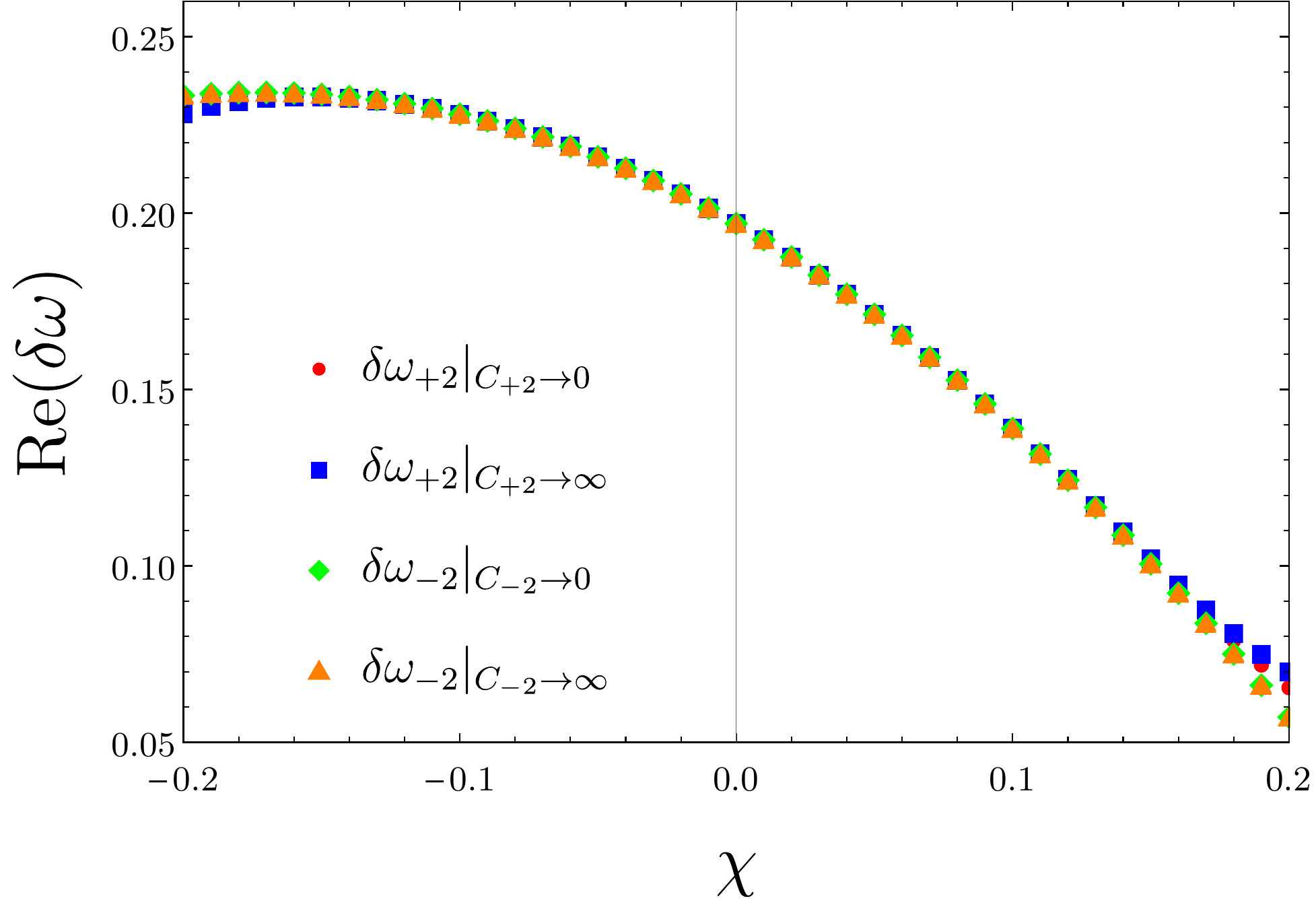}
	\includegraphics[width=0.48\textwidth]{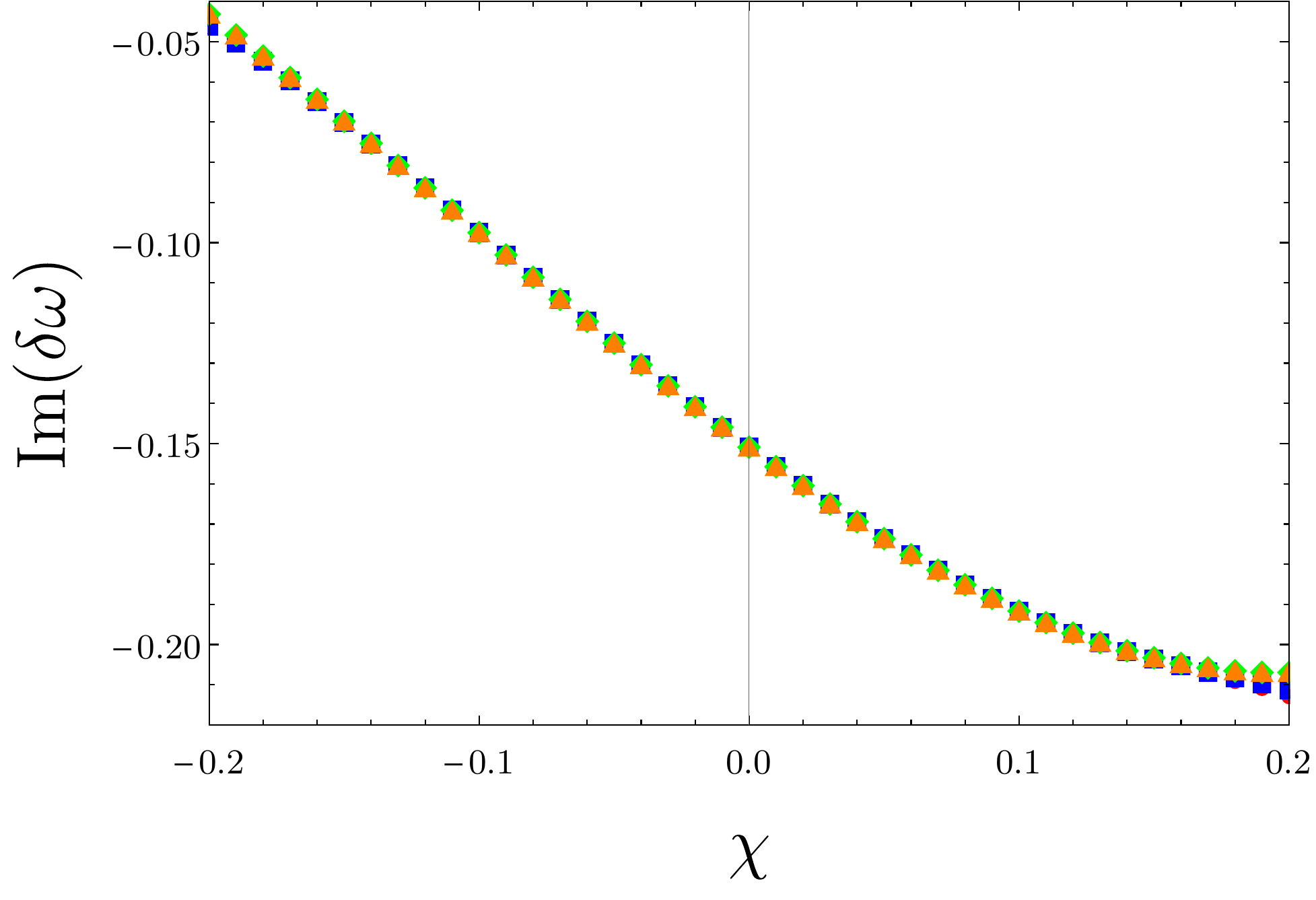}
	\caption{Shift in the $(l,m)=(2,2)$ QNM frequency, as defined in \eqref{eqn:shiftodd}, due to the odd-parity six-derivative correction in \req{cubiclagrangian}. We compute the shift in four different ways using an $\mathcal{O}(\chi^6)$ expansion of the equations \req{radial3}. The four estimations coincide to a very good approximation, providing a self-consistency test of our method.}
	\label{fig:oddl2}
\end{figure}

\begin{figure}[t!]
	\centering
	\includegraphics[width=0.48\textwidth]{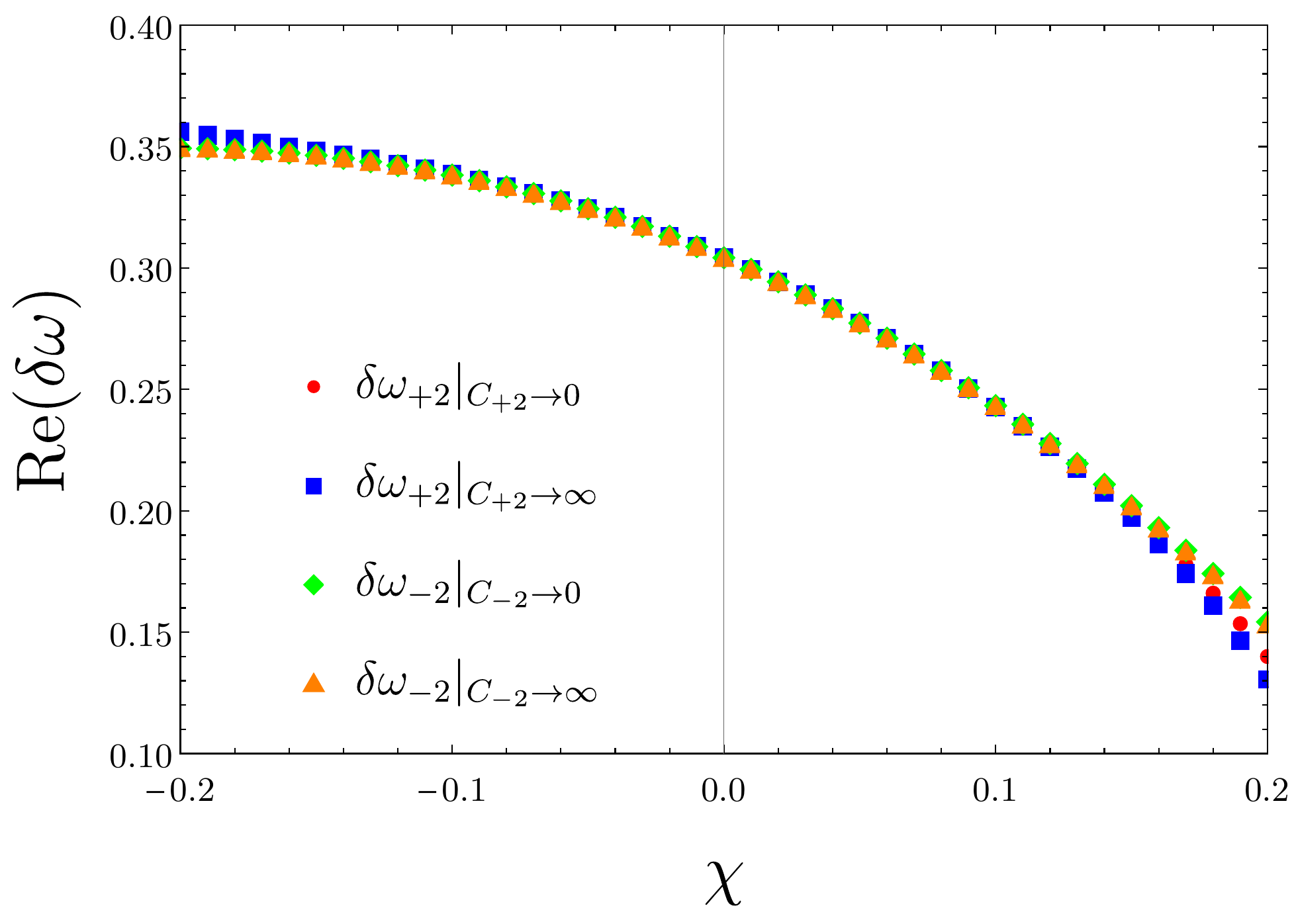}
	\includegraphics[width=0.48\textwidth]{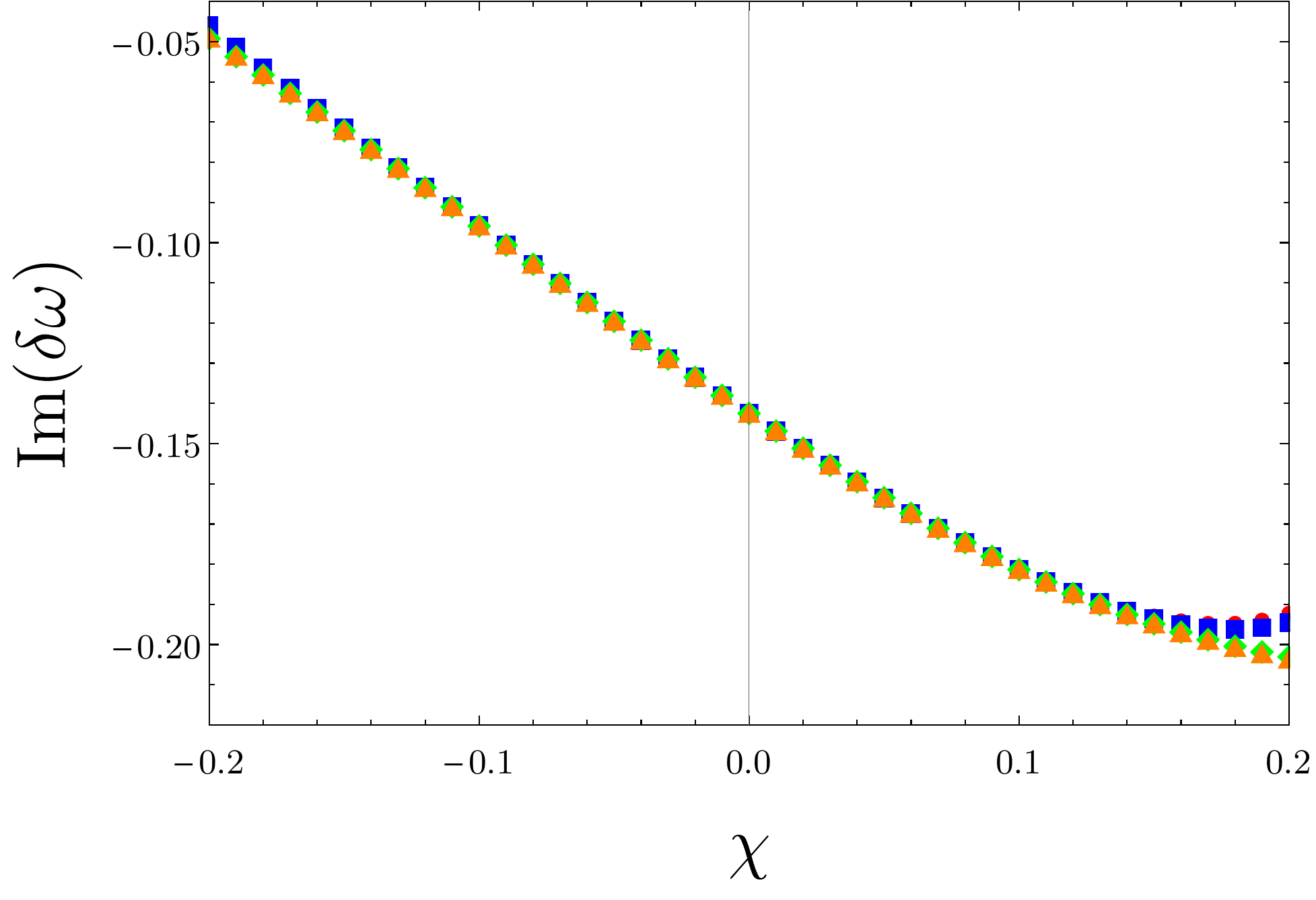}
	\caption{Shift in the $(l,m)=(3,3)$ QNM frequency, as defined in \eqref{eqn:shiftodd}, due to the odd-parity six-derivative correction in \req{cubiclagrangian}. We compute the shift in four different ways using an $\mathcal{O}(\chi^4)$ expansion of the equations \req{radial3}. The four estimations coincide to a very good approximation, providing a self-consistency test of our method. }
	\label{fig:oddl3}
\end{figure}

Proceeding in this way, we obtain four independent estimations for the shifts in the QNM frequencies. If everything is consistent, the four results should agree. As a first check, we observe that the two different polarizations $\delta\omega^{\pm}$ obtained in each case satisfy
\begin{equation}
\delta\omega^{+}=-\delta\omega^{-}\, .
\end{equation}
This is expected for parity-breaking corrections since they couple odd- and even-parity modes \cite{McManus:2019ulj,Cano:2021myl}. Thus, we show only the shifts with $\text{Re}(\delta\omega)>0$, as the other polarization simply has a shift $-\delta\omega$.
In Fig.~\ref{fig:oddl2} we show these shifts for the $(l,m)=(2,2)$ modes while in Fig.~\ref{fig:oddl3} we show the $(l,m)=(3,3)$ ones.

We observe that the four different estimations of the QNM frequencies match with very high accuracy. As before, the matching between the frequencies obtained from the $s=-2$ equation is almost perfect while the equation $s=+2$ yields slightly worse results, for the same reasons we explained above in the case of the parity-preserving corrections.

Performing a quadratic fit in the interval $-0.1<\chi<0.1$ for the $(l,m)=(2,2)$ modes yields
\begin{equation}\label{eq:oddl2}
\begin{aligned}
\delta\omega_{+2}\big|_{C_{+2}=0}&=(0.197\, -0.151 i)-(0.447\, +0.506 i) \chi -(1.371\,
   -0.291 i) \chi ^2\, ,\\
\delta\omega_{+2}\big|_{C_{+2}=\infty}&=(0.197\, -0.151 i)-(0.447\, +0.506 i) \chi -(1.373\,
   -0.289 i) \chi ^2\, ,\\
\delta\omega_{-2}\big|_{C_{-2}=0}&=(0.197\, -0.151 i)-(0.445\, +0.506 i) \chi -(1.356\,
   -0.293 i) \chi ^2\, ,\\
\delta\omega_{-2}\big|_{C_{-2}=\infty}&=(0.197\, -0.151 i)-(0.445\, +0.505 i) \chi -(1.354\,
   -0.296 i) \chi ^2\, ,
\end{aligned}
\end{equation}
and in the case of the $(l,m)=(3,3)$ modes we obtain
\begin{equation}\label{eq:oddl3}
\begin{aligned}
\delta\omega_{+2}\big|_{C_{+2}=0}&=(0.304\, -0.143 i)-(0.475\, +0.432 i) \chi -(1.354\,
   -0.396 i) \chi ^2\, ,\\
\delta\omega_{+2}\big|_{C_{+2}=\infty}&=(0.304\, -0.143 i)-(0.475\, +0.433 i) \chi -(1.359\,
   -0.395 i) \chi ^2\, ,\\
\delta\omega_{-2}\big|_{C_{-2}=0}&=(0.304\, -0.143 i)-(0.473\, +0.433 i) \chi -(1.348\,
   -0.389 i) \chi ^2\, ,\\
\delta\omega_{-2}\big|_{C_{-2}=\infty}&=(0.304\, -0.143 i)-(0.473\, +0.433 i) \chi -(1.347\,
   -0.388 i) \chi ^2\, ,
\end{aligned}
\end{equation}
finding excellent agreement in all the coefficients.  Finally, these results compare well with those in \cite{Cano:2021myl}, which we reproduce here for convenience,\footnote{As already observed in Footnote~\ref{foot}, we suspect that our results here are more accurate than those in \cite{Cano:2021myl} and hence the small discrepancies can probably be attributed to that reference.}

\begin{equation}
\begin{aligned}
\delta\omega_{2,2}&=0.192-0.151 i-(0.466+0.414 i)\chi+\mathcal{O}(\chi^2)\, ,\\ \delta\omega_{3,3}&=0.304-0.144 i-(0.493+0.438 i)\chi+\mathcal{O}(\chi^2)\, .
\end{aligned}
\end{equation}

\section{Conclusions}\label{sec:conclusions}
We have put forward a general framework to analyze linearized excitations of rotating black holes that depart perturbatively from a Kerr spacetime. These rotating black holes arise, for instance, in theories of gravity beyond general relativity. Our approach, schematically summarized in Figure \ref{fig:overview}, has enabled us to reduce the study of such gravitational waves propagating on these black hole backgrounds to a system of four radial equations that stem from the effective separation of the universal Teukolsky equations, introduced in section \ref{sec:Teukolsky}. 

The underlying perturbation theory philosophy, to exploit the vicinity of the algebraically special Kerr black hole, is in many ways evident and has been developed previously to various degrees of generality \cite{Cano:2020cao,Li:2022pcy,Hussain:2022ins,Ghosh:2023etd}. Yet, to put this into practice has proven a huge technical challenge, which lay at the heart of this work. We have provided a detailed end-to-end strategy to implement the computation and we have presented the first results that emerge from this. 

This entire computation is the synthesis of much previous work on the subject. On the one hand, the corrected Kerr background must be known, for which we used the results of \cite{Cano:2019ore} that expresses the corrections to the Kerr metric analytically as a series in the black hole spin. On the other hand, it is necessary to reconstruct the metric perturbation (on the uncorrected Kerr background) from the Weyl scalars. By using the results of \cite{Dolan:2021ijg} we provided an explicit map between metric variables and the Teukolsky variables $\delta\Psi_{s}$ --- see section \ref{subsec:reconstruction}. Finally, we showed how one can effectively separate and decouple the Teukolsky equation by projecting it onto the spin-weighted spheroidal harmonics, as successfully applied by \cite{Cano:2020cao} for scalar perturbations. 
The series expansion in the black hole angular momentum is not strictly necessary, yet it turns out to be very important in practice. The reason is that, contrary to the Kerr metric in general relativity, rotating black hole backgrounds are typically not known exactly. Nevertheless, a high-order expansion in spin allows us to obtain fully analytic radial equations, postponing the use of numerical methods until the very last step --- the resolution of these equations.  

We applied the approach just outlined explicitly to the case of the six-derivative EFT in \req{cubiclagrangian}, in order to obtain the corrections to the QNMs. As we discussed, the QNMs are determined by the frequency $\omega$, by the polarization parameters $q_{\pm 2}$, and by the Starobinsky-Teukolsky (ST) constants $C_{\pm 2}$. 
In the case of even-parity higher-derivative corrections, one can decouple perturbations of odd and even parity, which fixes the polarization parameters to be $q_{+2}=q_{-2}=\pm 1$. Given this choice, one should be able to obtain the QNM frequency by solving any of the radial Teukolsky equations. Therefore, in order for the result to be consistent, the two equations (for $s=\pm2$) should indeed yield the same frequencies.
On the other hand, parity-violating higher-derivative corrections mix modes of odd and even parity and one has to determine the QNM frequencies simultaneously with the polarization parameters $q_{s}$ by imposing that all radial equations for a given polarization have the same frequencies. In both cases, the ST constants should be irrelevant. They represent a choice of gauge in which one reconstructs the metric perturbation. Thus, the QNM frequencies should actually be independent of the ST constants. This is highly non-trivial and provides a self-consistent check of our results. 

In fact, our results in Section \ref{sec:QNM6} pass three consistency checks with very high accuracy:
\begin{enumerate}
\item The $s=+2$ and $s=-2$ equations give the same QNM frequencies.
\item The shifts in the QNM frequencies are independent of the choice of Starobinsky-Teukolsky constants. 
\item We reproduce the results of \cite{Cano:2021myl} for static and slowly-rotating black holes at linear order in the spin obtained from a metric perturbation approach.
\end{enumerate}
Any one of these tests is, taken on its own, an impressive consistency check, when one considers the extensive calculations, expansions, transformations and the various numerical methods involved in obtaining the results. Given that they are all satisfied to a good approximation, we are confident that we have achieved a successful method to compute spectral shifts of QNMs of rotating black holes beyond general relativity.

We note that two of the above tests are based on redundancies in choices associated \textit{e.g.} to gauge freedom. In addition to consistency checks, such redundancies allow some internal estimate of the precision we achieve. Moreover, the choices turn out to be not equally amenable to the combination of expansions and numerics that are used. For example, we observed that the $s=2$ equation is quite sensitive to the order of the spin expansion, making it less precise. The $s=-2$ equation is much more stable. Indeed, the frequencies computed from the $s=-2$ equation pass the second test above with high precision even for high values of the angular momentum. The exact source of these discrepancies should be understood better such that the entire approach can be tuned to the optimal choices. 

It is rather straightforward to apply the computational strategy given here more generally in order to compute perturbative corrections to the QNM spectrum of Kerr black holes. Natural extensions include going further in the EFT of gravity, \textit{e.g.}, by including eight-derivative terms in the effective action, as well as going to higher values of the black hole spin. For the latter, one simply needs to include enough terms in the angular momentum expansion so as to get an accurate result \cite{Cano:2023qqm}, a process that is now entirely algorithmic. We will report on a thorough analysis of the shifts in the QNM frequencies in these higher-derivative extensions elsewhere \cite{QNMbeyondKerr}.

Next, with a little more effort, our method can be generalized to theories with additional fields, such as Einstein-scalar-Gauss-Bonnet gravity \cite{Kanti:1995vq} or dynamical Chern-Simons gravity \cite{Campbell:1990fu,Alexander:2009tp}. In these theories, QNMs of static \cite{Cardoso:2009pk,Blazquez-Salcedo:2016enn,Blazquez-Salcedo:2017txk} and slowly-rotating black holes up to, respectively,  second \cite{Pierini:2021jxd,Pierini:2022eim} and first order \cite{Wagle:2021tam,Srivastava:2021imr} in the angular momentum  have been computed using a metric perturbation approach. In those cases, one needs to take into account that the stress-energy tensor in the universal Teukolsky equations \req{eqn:universalteukolsky0}, \req{eqn:universalteukolsky4} depends on a scalar field. Hence, one should be able to obtain a system of radial equations involving the Teukolsky variables and the scalar field along the lines of \cite{Li:2022pcy,Hussain:2022ins}. It would be of particular interest to see if matching results could be obtained with the methods presented in those works and with the previous results in \cite{Pierini:2021jxd,Wagle:2021tam,Srivastava:2021imr,Pierini:2022eim}. Indeed, despite our consistency checks, it would be highly desirable to achieve converging results from different groups using different approaches. This could include the analysis of specific limits, such as the geometrical optics limit \cite{Yang:2012he,Silva:2019scu,Glampedakis:2019dqh,Kuntz:2019zef,Bryant:2021xdh,Volkel:2022khh,Chen:2022ynz,Yagi:2022vys,Fransen:2023eqj},  as well as entirely different methods, such as numerical relativity beyond general relativity \cite{Okounkova:2017yby,Okounkova:2018pql,Witek:2018dmd,Okounkova:2020rqw,Okounkova:2019zjf,Okounkova:2019dfo,Ripley:2019irj,Kovacs:2020ywu,Figueras:2021abd,East:2020hgw,Ripley:2022cdh,Okounkova:2022grv,Cayuso:2023aht}. Finally, advances in the latter, together with our work and post-Newtonian/post-Minkowskian results \cite{Endlich:2017tqa,Julie:2019sab,Bern:2020uwk,AccettulliHuber:2020dal,Julie:2022qux,Goldberger:2022ebt} could ultimately be combined into full theory-based, inspiral-merger-ringdown waveform templates. The availability of such waveforms for a broad class of gravitational theories would help carve out a better motivated theoretical prior on the space of (potentially) physical waveforms and would represent a milestone in our ability to search for physics beyond general relativity using gravitational wave observations. 

\section*{Acknowledgments}
KF thanks Yanbei Chen and especially Dongjun Li for useful discussions. The work of PAC is supported by a postdoctoral fellowship from the Research Foundation - Flanders (FWO grant 12ZH121N). KF gratefully acknowledges the support of the Heising-Simons Foundation grant \#2021-2819. TH acknowledges support from the PRODEX grant LISA - BEL (PEA 4000131558), the FWO Research Project G0H9318N and the inter-universitary project iBOF/21/084.

\appendix

\section{(Newman/Geroch-Held)-Penrose}
\label{app:NPGHP}

Here we gather our conventions and useful identities of the Newman-Penrose (NP) and Geroch-Held-Penrose (GHP) formalism. Where possible we follow \cite{Pound:2021qin}. Recall we have introduced an NP-frame
\begin{equation}
	g_{\mu\nu}=-2l_{(\mu}n_{\nu)}+2m_{(\mu}\bar{m}_{\nu)}\, .
\end{equation}
In terms of the frame, the spin connection is defined as
\begin{equation}
	\gamma_{abc} = e_a{}^{\mu}e_c{}^{\nu}\nabla_{\nu}e_{b}{}_{\mu} \, .
\end{equation}
We have additionally defined the conventionally named spin coefficients in \eqref{eqn:spincoefficients} as well as the projections of the curvature components on the NP frame in \eqref{eqn:appintro:curv} and \eqref{eqn:appintro:curv2}. The main difference between the NP and GHP approaches is that the latter remains covariant with respect to the local (``type III'') rotations \eqref{eqn:intro:typeIII}. This is theoretically appealing and leads to more compact expressions but for our final computational purposes we will translate back to the more explicit NP formulas. The conventional GHP derivatives which preserve covariant behavior with respect to the local frame transformation \eqref{eqn:intro:typeIII} were defined in \eqref{eqn:GHPderivatives}  \\
The key starting point to derive the universal Teukolsky equations in Section \ref{sec:Teukolsky} are the Bianchi identities. However, before presenting these, note the relations between spin coefficients and curvatures in GHP form
\begin{subequations}\label{eqn:GHPriemann}
	\bea
	& & \phi_{00} = \th\rho - \edth'\kappa - \rho^2+\tau'\kappa+\tau\kappa^*-|\sigma|^2\, , \label{eqn:intro:NPa} \\
	& & \Psi_0 = \th\sigma-\edth\kappa - (\rho+\rho^*)\sigma+(\tau\!+\!\tau'^*)\kappa\;, \label{eqn:intro:NPb}\\
	& & \phi_{01}+\Psi_1 = -\th'\kappa + \th\tau - (\tau-\tau'{}^*)\rho-(\tau^*-\tau')\sigma \, , \label{eqn:intro:NPc}\\
	& & \phi_{01}-\Psi_1 = -\edth'\sigma + \eth\rho - (\rho-\rho^*)\tau + (\rho'-\rho'{}^*)\kappa \, ,  \label{eqn:intro:NPd}\\
	& & \phi_{02} = -\th'\sigma+\edth\tau - \tau^2-\kappa\kappa'{}^* + \rho'\sigma + \rho\sigma'{}^*\;, \label{eqn:intro:NPe}\\
	& & \Psi_2 + \frac{1}{12}R = -\th'\rho+\edth'\tau+\rho\rho'{}^*+\sigma\sigma'-|\tau|^2-\kappa\kappa'\, .   \label{eqn:intro:NPf}
	\eea
\end{subequations}
These are simply a translation from 
\be\label{eqn:appintro:curvstart}
R_{abcd}  = \nabla_{c} \gamma_{abd} - \nabla_{d} \gamma_{abc} - \gamma_{abf}\eta^{fe}\left(\gamma_{c e d} - \gamma_{dec}\right) + \gamma_{afc} \eta^{fe} \gamma_{e bd}- \gamma_{afd} \eta^{fe} \gamma_{e bc}\, ,
\ee
into the GHP language. Note that \eqref{eqn:GHPriemann} is (Eq.~4.12.32) of \cite{penrose1984spinors} despite the different choices for the metric signature and Riemann curvature.\footnote{It differs from \cite{BenTov:2017kyf} by oppositely defined spin coefficients. Their appendix also contains a useful discussion on sign conventions.} On the other hand it is (2.37) in \cite{Aksteiner:2014zyp} for a vacuum spacetime as well as (43) of \cite{Pound:2021qin} for a vacuum Petrov D spacetime (with an appropriately aligned tetrad). Similarly translating 
\be
\nabla_{[e}R_{a b] c d} + 2 \eta^{i j} \gamma_{a i [e}R_{j b] c d}+ 2 \eta^{i j} \gamma_{c i [e}R_{a b] j d} = 0 \, ,
\ee
one finds
\begin{subequations}\label{eqn:GHPbianchi}
	\be\label{eqn:app:bianchi36}
	\begin{aligned}
		\th\Psi_1-\edth'\Psi_0-\th\phi_{01}+\edth\phi_{00} = &- \tau'\Psi_0 + 4\rho \Psi_1-3\kappa \Psi_2+\tau'{}^* \phi_{00} \\ &-2\rho^*\phi_{01}-2\sigma \phi_{01}^*+2\kappa \phi_{11}+\kappa^* \phi_{02} \, ,
	\end{aligned}
	\ee
	\be\label{eqn:app:bianchi37}
	\begin{aligned}
		\th \Psi_2 - \edth'\Psi_1-\edth'\phi_{01}+\th'\phi_{00} &+\frac{1}{12}\th R =  \sigma'\Psi_0 -2 \tau'\Psi_1+3 \rho \Psi_2- 2\kappa \Psi_3 \\ &+\rho'^*\phi_{00}-2\tau^*\phi_{01}-2\tau \phi_{01}^*+2\rho \phi_{11}+\sigma^* \phi_{02} \, ,
	\end{aligned}
	\ee
	\be\label{eqn:app:bianchi39p}
	\begin{aligned}
		\th'\Psi_0-\edth\Psi_1-\edth\phi_{01}+\th\phi_{02} = & \rho'\Psi_0 - 4\tau \Psi_1+3\sigma \Psi_2+\sigma'{}^* \phi_{00} \\ &-2\tau'{}^*\phi_{01}-2\kappa \phi_{21}^*+2\sigma \phi_{11}+\rho^* \phi_{02} \, ,
	\end{aligned}
	\ee
	\be\label{eqn:app:bianchi38p}
	\begin{aligned}
		\th'\Psi_1-\edth \Psi_2-\th'\phi_{01}+\edth'\phi_{02} &-\frac{1}{12}\edth R =  -\kappa'\Psi_0 +2 \rho'\Psi_1-3 \tau \Psi_2- 2\sigma \Psi_3 \\ &+\kappa'^*\phi_{00}-2\rho'{}^*\phi_{01}-2\rho \phi_{21}^*+2\tau \phi_{11}+\tau^* \phi_{02} \, ,
	\end{aligned}
	\ee
\end{subequations}
\begin{subequations}\label{eqn:GHPbianchicontracted}
	\be\label{eqn:app:bianchi40}
	\begin{aligned}
		\th\phi_{11}+\th'\phi_{00}-\edth\phi_{01}^*-\edth'\phi_{01} &+\frac{1}{8}\th R = (\rho'+\rho'{}^*)\phi_{00}+2(\rho+\rho^*)\phi_{11}-(\tau'+2\tau^*)\phi_{01} \\ 
		&-(2\tau+\tau'{}^*)\phi_{01}^*-\kappa^* \phi_{21}^*-\kappa \phi_{21} + \sigma \phi_{02}^* +\sigma^* \phi_{02} \, ,
	\end{aligned}
	\ee
	\be\label{eqn:app:bianchi41}
	\begin{aligned}
		\th\phi_{21}^*+\th'\phi_{01}-\edth\phi_{11}-\edth'\phi_{02} &+\frac{1}{8}\edth R = (\rho'+2\rho'{}^*)\phi_{01}+(2\rho+\rho^*)\phi_{21}^*-(\tau'+\tau^*)\phi_{02} \\ 
		&-2(\tau+\tau'{}^*)\phi_{11}-\kappa'{}^* \phi_{00}-\kappa \phi_{22} + \sigma \phi_{21} +\sigma'{}^* \phi_{01}^* \, .
	\end{aligned}
	\ee
\end{subequations}
Importantly, these include (2.2) and (2.3) in \cite{teukolsky1973}, but they were also presented in full as (4.12.36)-(4.12.41) of \cite{penrose1984spinors} as well as in special cases in (2.39) of \cite{Aksteiner:2014zyp}, and (44) of \cite{Pound:2021qin}. They are, in short, well known in the literature but several conventions are in circulation.
The GHP formalism has a natural symmetry exchanging $l \leftrightarrow n$ and $m \leftrightarrow \bar{m}$. The associated equations are implicit in \ref{eqn:GHPbianchi} as what is known in this language as the ``primed'' versions of these equations. We will not present all of them explicitly but as a relevant example note that this is how \eqref{eqn:app:bianchi36} or \eqref{eqn:intro:bianchi36}
	\begin{equation} 
	\begin{aligned}
		\th\Psi_1-\edth'\Psi_0-\th\phi_{01}+\edth\phi_{00} = &- \tau'\Psi_0 + 4\rho \Psi_1-3\kappa \Psi_2+\tau'{}^* \phi_{00} -2\rho^*\phi_{01}-2\sigma \phi_{01}^*\\
		&+2\kappa \phi_{11}+\kappa^* \phi_{02} \, ,
	\end{aligned}
\end{equation}
is related to \eqref{eqn:intro:bianchi36p}
	\begin{equation}
		\begin{aligned}
			\th'\Psi_3-\edth\Psi_4-\th'\phi_{21}+\edth'\phi_{22} = &- \tau\Psi_4 + 4\rho' \Psi_3-3\kappa' \Psi_2+\tau{}^* \phi_{22} -2\rho'{}^*\phi_{21}-2\sigma' \phi_{21}^*\\
			&+2\kappa' \phi_{11}+\kappa'{}^* \phi_{02}{}^* \,  . \\
		\end{aligned}
	\end{equation}
A last GHP-ingredient that is useful for our derivations are the GHP-commutation relations is
\begin{equation}\label{eqn:GHPcommute}
	\begin{aligned}
	\left[\th, \th'\right] &= (\tau^*-\tau')\edth + (\tau-\tau'{}^*)\edth'+ (p+q)F_{B}{}_{1 2} +(p-q) F_{\Omega}{}_{1 2}  \, ,   \\ 
	\left[\th, \edth\right] &=  \rho^* \edth + \sigma \edth' - \tau'{}^* \th -\kappa \th'+(p+q)F_{B}{}_{1 3} + (p-q)F_{\Omega}{}_{1 3} \, , \hspace{1.5cm}\\
	\left[\edth, \edth'\right] &= (\rho'{}^*-\rho')\th + (\rho-\rho^*)\th'+(p+q) F_{B}{}_{3 4} +(p-q) F_{\Omega}{}_{3 4}  \, .  
	\end{aligned}
\end{equation}
when acting on an object of weight $\{p,q\}$. Here, we have used a notation associated to the more standard form of a covariant derivative 
\begin{equation}
\Dbar_{\mu} = -n_{\mu} \th  - l_{\mu} \th' + \bar{m}_{\mu} \edth + m_{\mu} \edth' \, ,
\end{equation}
written in terms of the connections $B_{\mu}$ and $\Omega_{\mu}$
\begin{equation}
\Dbar_{\mu} = \nabla_{\mu}+(p+q)B_{\mu}+(p-q) \Omega_{\mu} \, ,
\end{equation}
and defining the associated curvatures as
\begin{equation}
[\Dbar_{\mu},\Dbar_{\nu}] V = \left(\left(p+q\right) F_{B}{}_{\mu \nu}  + \left(p-q\right) F_{\Omega}{}_{\mu \nu}\right) V \, ,
\end{equation}
where $V$ has weight $\{p, q\}$. In terms of the gauge fields this becomes
\begin{equation}
F_{B}{}_{\mu \nu} = \nabla_{\mu} B_{\nu} - \nabla_{\nu}B_{\mu} \, , \quad F_{\Omega}{}_{\mu \nu} = \nabla_{\mu} \Omega_{\nu} - \nabla_{\nu}\Omega_{\mu} \, .
\end{equation}
Spelled out explicitly, the commutation relations are
\begin{equation}\label{eqn:GHPcommuteexpl}
\begin{aligned}
\left[\th, \th'\right] &= (\tau^*-\tau')\edth + (\tau-\tau'{}^*)\edth'-p(\kappa \kappa'-\tau \tau'+\Psi_2+\phi_{11}-\frac{1}{24}R)  \\ &  -q(\kappa^* \kappa'{}^*-\tau^* \tau'{}^*+\Psi_2^*+\phi_{11}-\frac{1}{24}R) \, , \\ 
\left[\th, \edth\right] &=  \rho^* \edth + \sigma \edth' - \tau'{}^* \th -\kappa \th'-p(\rho' \kappa -\tau'\sigma+\Psi_1)  \\  & 
-q(\sigma'{}^* \kappa^* -\rho^* \tau'{}^*+\phi_{01}) \, , \\
\left[\edth, \edth'\right] &= (\rho'{}^*-\rho')\th + (\rho-\rho^*)\th'+p(\rho \rho'-\sigma \sigma'+\Psi_2-\phi_{11}-\frac{1}{24}R)  \\ &  -q(\rho^* \rho'{}^*-\sigma^* \sigma'{}^*+\Psi_2^*-\phi_{11}-\frac{1}{24}R)   \, ,
\end{aligned}
\end{equation}
as can again be compared against (4.12.33)-(4.12.35) in \cite{penrose1984spinors}. 

To arrive at the generally valid equations \eqref{eqn:universalteukolsky0} and \eqref{eqn:universalteukolsky4}, one first needs to identify the relevant combination of these identities that are used to derive the Teukolsky equation. Let us therefore briefly reduce to this case. That is, we assume a vacuum Petrov type D spacetime and a frame aligned along the principal null directions of this spacetime such that all Weyl curvature components but $\Psi_2$ vanish. Similarly, $\kappa=\sigma=\kappa'=\sigma'=0$ and all $\phi_{ij}$ vanish. The only nontrivial surviving equations from \eqref{eqn:GHPriemann}, \eqref{eqn:GHPbianchi}, and \eqref{eqn:GHPbianchicontracted} are (equation (43) of \cite{Pound:2021qin})
\begin{equation}\label{eqn:DspinpetrovD}
	\begin{aligned}
	\th \rho = \rho^2 \, , \quad 	\edth \tau &= \tau^2 \, , \quad  \th \tau = (\tau-\tau'{}^*)\rho\, , \quad \edth \rho = (\rho-\rho{}^*)\tau \, , \\
	\Psi_2  &= -\th'\rho+\edth'\tau+\rho\rho'{}^*-|\tau|^2 \, ,
	\end{aligned}
\end{equation}
and (equation (44) of \cite{Pound:2021qin})
\begin{equation}\label{eqn:GHPDPsi2petrovD}
	\th \Psi_2 = 3 \rho \Psi_2 \, , \quad  \edth \Psi_2 = 3 \tau \Psi_2 \, ,
\end{equation}
as well as their primed and conjugated versions. It is the entirely trivialized (but non-contracted) Bianchi identities \eqref{eqn:app:bianchi36} and \eqref{eqn:app:bianchi39p} that will be the key players in the perturbation analysis. Finally, the commutation relations \eqref{eqn:GHPcommuteexpl} become (equation (45) of \cite{Pound:2021qin}) 
\begin{equation}\label{eqn:GHPcommuteexplpetrovD}
	\begin{aligned}
		\left[\th, \th'\right] &= (\tau^*-\tau')\edth + (\tau-\tau'{}^*)\edth'-p(\Psi_2-\tau \tau')  \\ &  -q(\Psi_2^*-\tau^* \tau'{}^*) \, , \\ 
		\left[\th, \edth\right] &=  \rho^* \edth  - \tau'{}^* \th  
		+q \rho^* \tau'{}^* \, , \\
		\left[\edth, \edth'\right] &= (\rho'{}^*-\rho')\th + (\rho-\rho^*)\th'+p(\rho \rho'+\Psi_2)    -q(\rho^* \rho'{}^*+\Psi_2^*)   \, ,
	\end{aligned}
\end{equation}

 Consider perturbations of order $\epsilon$ around such a background\footnote{This perturbation parameter $\epsilon$ should not be confused with the NP $\epsilon$. The latter will only appear implicitly here in the GHP-derivatives.} and let
\begin{equation}
	\Psi_{i}=\bar\Psi_{i}+\epsilon \delta \Psi_{i}\, , \quad \gamma_{abc}=\bar\gamma_{abc}+\epsilon  \delta \gamma_{abc}\, .
\end{equation} 
For this derivation, which is mainly motivational, we will not keep track of the ``sources'' $\phi_{ab}$, so assume these still vanish. On account of the vacuum Petrov type D assumption for the unperturbed case 
\begin{align}
	\bar\Psi_{0}=\bar\Psi_{1}=\bar\Psi_{3}=\bar\Psi_{4}&=0\, ,\\ 
	\phi_{00}=\phi_{01}=\phi_{02}=\phi_{21}=\phi_{11}=\phi_{22}=R&=0\, ,\\
	\bar\kappa=\bar\kappa'=\bar\sigma=\bar\sigma'&=0\, .
\end{align} 
We remark that the GHP derivatives will also be perturbed but only their leading order will play a role so we do not burden our notation even further by indicating this explicitly. Consider $\th$ acting on \eqref{eqn:app:bianchi39p}. The first contribution comes in at order $\epsilon$ and is given by
\begin{equation}\label{eqn:39ptemp}
		\th \th' \delta \Psi_0-\th \edth \delta \Psi_1 =  \th\left(\bar \rho'\delta \Psi_0\right) - 4\th\left(\bar \tau \delta \Psi_1\right)+3 \th\left(\delta\sigma \bar \Psi_2\right) \, .
\end{equation}
Similarly acting with $\edth$ on \eqref{eqn:app:bianchi36} yields
\begin{equation}\label{eqn:36temp}
		\edth\th \delta\Psi_1-\edth\edth'\delta\Psi_0 = - \edth\left(\bar \tau'\delta\Psi_0\right) + 4 \edth\left(\bar \rho \delta\Psi_1\right)-3\edth\left(\delta \kappa \bar \Psi_2\right)\, .
\end{equation}
Now commute $\th$ and $\edth$ in \eqref{eqn:39ptemp} in the term $\th \edth \delta \Psi_1$. To leading order, one can use \eqref{eqn:GHPcommuteexplpetrovD} and finds
\begin{equation}\label{eqn:39ptemp2}
	\th \th' \delta \Psi_0-\edth \th \delta \Psi_1 - \left(\bar \rho^* \edth  - \bar \tau'{}^* \th   \right) \delta \Psi_1 =  \th\left(\bar \rho'\delta \Psi_0\right) - 4\th\left(\bar \tau \delta \Psi_1\right)+3 \th\left(\delta\sigma \bar \Psi_2\right) \, .
\end{equation}
Here, we have additionally used the GHP-weight
\begin{equation}
		w_{\rm GHP}(\Psi_1)= 	w_{\rm GHP}\left( C_{\alpha \beta \mu \nu} l^{\alpha}n^{\beta}l^{\mu}m^{\nu}\right) =  \{2,0\} \, .
\end{equation}
In \eqref{eqn:36temp}, which came from \eqref{eqn:app:bianchi36} by acting with $\edth$, one can replace $\edth \Psi_1$ by using \eqref{eqn:app:bianchi39p}, while conversely in \eqref{eqn:39ptemp2}, which came from \eqref{eqn:app:bianchi39p} by acting with $\th$, one can replace $\th \Psi_1$ using \eqref{eqn:app:bianchi36} (as well as replacing the additional $\edth \Psi_1$ term from the commutation relation). One finds respectively 
\begin{equation}\label{eqn:36temp2}
	\edth\th \delta\Psi_1-\edth\edth'\delta\Psi_0 = - \edth\left(\bar \tau'\delta\Psi_0\right) + 4 \delta \Psi_1 \edth \bar \rho  + 4 \bar \rho \left(	\th'\delta\Psi_0- \bar \rho'\delta \Psi_0 + 4 \bar \tau \delta \Psi_1-3\delta \sigma \Psi_2 \right) -3\edth\left(\delta \kappa \bar \Psi_2\right)\, ,
\end{equation}
and
\begin{equation}\label{eqn:39ptemp3}
	\begin{aligned}
	\th \th' \delta \Psi_0&-\edth \th \delta \Psi_1 - \bar \rho^* \left(\th'\delta\Psi_0- \bar \rho'\delta \Psi_0 + 4 \bar \tau \delta \Psi_1-3\delta \sigma \bar \Psi_2 \right) = \\ & \th\left(\bar \rho'\delta \Psi_0\right) - 4\delta \Psi_1\th \bar \tau -(4 \bar \tau+ \bar \tau'{}^*)\left(\edth'\delta \Psi_0- \bar \tau'\delta\Psi_0 + 4 \bar \rho \delta\Psi_1-3\delta\kappa \bar \Psi_2\right)+3 \th\left(\delta\sigma \bar \Psi_2\right) \, .
	\end{aligned}
\end{equation}
Now simplify further, by using \eqref{eqn:GHPDPsi2petrovD} to replace $\edth \bar\Psi_2 = 3\bar\tau \bar\Psi_2$, and  $\th \bar\Psi_2 = 3\bar\rho \bar \Psi_2$, \eqref{eqn:DspinpetrovD} to replace $\th \bar \tau = (\bar \tau - \bar \tau'{}^*)\bar \rho$ and $\edth \bar \rho = (\bar \rho - \bar \rho{}^*)\bar \tau$, as well as \eqref{eqn:intro:NPb} to relate $\edth \delta \kappa$ with $\th \delta \sigma$ using $\delta \Psi_0$
\begin{equation}
\edth\delta\kappa = -\delta \Psi_0 +\th \delta\sigma - (\bar \rho+\bar \rho^*)\delta\sigma+(\bar \tau\!+\! \bar \tau'^*)\delta \kappa\; .
\end{equation}
The results after some extra rewriting are 
\begin{equation}\label{eqn:36temp3}
	\begin{aligned}
	\left(\edth\th + 4  \bar \rho{}^* \bar \tau  -20 \bar \rho \bar \tau \right)\delta\Psi_1+3\left( \th  - \bar \rho^*  + 3\bar \rho\right) \bar\Psi_2  \delta\sigma + 3(4 \bar \tau\!+\! \bar \tau'^*)\bar \Psi_2 \delta \kappa   = \\ \left(\edth\edth' - \edth \bar \tau'-  \bar \tau' \edth  + 4 \bar \rho \left(	\th'- \bar \rho'\right) +3 \bar\Psi_2 \right)\delta\Psi_0  \, ,
		\end{aligned}
\end{equation}
and
\begin{equation}\label{eqn:39ptemp4}
	\begin{aligned}
		-\left(\edth \th + 4 \bar \rho^* \bar \tau - 20 \bar \rho \bar \tau\right) \delta \Psi_1-3  \left(\th - \bar \rho^* +3\bar\rho \right) \bar \Psi_2 \delta\sigma  -3 \left(4\bar \tau+\bar \tau'{}^* \right)\bar \Psi_2  \delta\kappa  = \\ \left( - \th \th' + \bar \rho^* \left(\th'- \bar \rho'\right)+ \th \bar \rho' + \bar \rho' \th  -(4\bar \tau+ \bar \tau'{}^*)\left(\edth'- \bar \tau' \right)\right)\delta\Psi_0   \, .
	\end{aligned}
\end{equation}
Written in this way, it is obvious that the sum of both equations will, of all the perturbed quantities, only depend on $\delta \Psi_0$. Thus we have found the (sourceless) Teukolsky equation for $\delta \Psi_0$
\begin{equation}\label{eqn:teukolsky}
-\left[\left(\th -4 \bar\rho-\bar\rho^*\right)\left(\th'-\bar\rho'\right)-\left(\edth -4 \bar\tau-\bar\tau'{}^*\right)\left(\edth'-\bar\tau'\right) -3 \bar \Psi_2\right]\delta\Psi_0  = 0 \, ,
\end{equation}
as can be confirmed against (58) of \cite{Pound:2021qin}. The logic in the main text to derive the universal Teukolsky equations \eqref{eqn:universalteukolsky0} and \eqref{eqn:universalteukolsky4} is simply to consider the same combinations and perform the same manipulations but without dropping any terms. \\

Finally, given the dependence of the GHP derivatives on the GHP weight of the fields they are acting upon, to insert actual coordinate expressions, it is useful to write these in terms of the more explicit NP forms
\begin{equation}\label{eqn:universalteukolskyNP0}
\cO^{(0)}_{2}\left(\Psi_0\right) +\cO^{(1)}_{2}\left(\Psi_1\right)+ \cO^{(2)}_{2}\left(\Psi_0\right)   = 8 \pi \left(\cT^{(0)}_{2} +\cT^{(1)}_{2} +\cT^{(2)}_{2}\right) \, ,
\end{equation}
with
\begin{subequations}
	\bea
	\cO^{(0)}_{2} &=& 2 \left(-\beta '^*+\tau '^*+3 \beta +4 \tau
	\right) \bar{m}^{\mu } \nabla {}_\mu  -2 m^{\nu } \nabla {}_\nu \bar{m}^{\mu } \nabla {}_\mu \\   &+& 2  \left(-l^{\mu } \nabla {}_\mu \left(\rho '\right)+4 l^{\mu
	} \nabla {}_\mu \left(\epsilon '\right)-4 m^{\mu } \nabla {}_\mu \left(\beta
	'\right)+m^{\mu } \nabla {}_\mu \left(\tau '\right) \right) \nn  &+& 2\left( 4 \beta ' \tau '^*+\tau ' \beta '^*-4
	\beta ' \beta '^*+\rho ' \rho ^*-\tau ' \tau '^*-4 \epsilon ' \rho ^*+4 \epsilon '
	\epsilon ^*-\rho ' \epsilon ^*+16 \tau  \beta ' \right. \nn &+& \left. 12 \beta  \beta '-3 \beta  \tau '+4 \rho
	\rho '-4 \tau  \tau '-3 \Psi _2-16 \rho  \epsilon '-12 \epsilon  \epsilon '+3 \epsilon 
	\rho '\right) \nn &+& 2 \left(-\rho ^*+\epsilon ^*-4
	\rho -3 \epsilon \right) n^{\mu } \nabla {}_\mu +2 l^{\nu } \nabla {}_\nu n^{\mu } \nabla {}_\mu
	\nn &+& 2  \left(4 \epsilon '-\rho '\right) l^{\mu } \nabla {}_\mu
	+2  \left(\tau '-4 \beta '\right) m^{\mu } \nabla {}_\mu 
	\, , \nn
	\cO^{(1)}_{2} &=& -8  \bar{m}^{\mu } \nabla {}_\mu (\sigma )-8 \sigma  \bar{m}^{\mu } \nabla
	{}_\mu  \\&+& 8 n^{\mu } \nabla {}_\mu (\kappa )+8 \kappa 
	n^{\mu } \nabla {}_\mu  \nn &-& 8 \sigma   \beta ^*-8 \kappa 
	\rho '^*+8 \sigma  \tau ^*+8 \kappa  
	\epsilon '^*-40 \sigma  \beta '+20 \Psi _1 +40 \kappa 
	\epsilon '   , \nn 
	\cO^{(2)}_{2} &=& 6  \left(\kappa  \kappa '-\sigma  \sigma '\right) \, , \\   \nn
	\cT^{(0)}_{2}  &=& T_{lm} \left\lbrace-2 l^{\mu } \nabla {}_\mu \left(\tau '^*\right)-2 m^{\mu } \nabla {}_\mu
	\left(\rho ^*\right)-2 l^{\mu } \nabla {}_\mu (\beta )-2 m^{\mu } \nabla {}_\mu (\epsilon
	) -  2 \rho ^* \beta '^*-2 \epsilon  \beta '^* \right. \\ &+&  \left. 8 \beta  \rho ^*  - 2 \beta  \epsilon ^*+8 \rho
	\tau '^*+4 \rho ^* \tau '^* + 8 \tau  \rho ^*+8 \epsilon  \tau '^* - 2 \epsilon ^* \tau
	'^*+8 \beta  \rho +12 \beta  \epsilon +8 \tau  \epsilon \right\rbrace \nn &+& T_{mm} \left\lbrace l^{\mu } \nabla {}_\mu \left(\rho ^*\right)-2 l^{\mu } \nabla
	{}_\mu \left(\epsilon ^*\right)+2 l^{\mu } \nabla {}_\mu (\epsilon )-\left(\rho
	^*\right)^2-4 \rho  \rho ^*+8 \rho  \epsilon ^*-5 \epsilon  \rho ^*\right. \nn &+& \left.3 \rho ^* \epsilon
	^*-2 \left(\epsilon ^*\right)^2 +  8 \epsilon  \epsilon ^*-6 \epsilon ^2-8 \rho  \epsilon
	\right\rbrace + l^{\mu } \nabla {}_\mu
	\left(T_{lm}\right) \left(\beta '^*-3 \tau '^*-5 \beta -4 \tau
	\right) \nn &+& l^{\mu } \nabla {}_\mu
	\left(T_{mm}\right) \left(2 \rho ^*-3 \epsilon ^*+4 \rho +5 \epsilon \right)  + m^{\mu } \nabla {}_\mu \left(T_{ll}\right) \left(-3
	\beta '^*+2 \tau '^*+5 \beta +4 \tau \right) \nn &+& T_{ll} \left\lbrace-2 m^{\mu } \nabla
	{}_\mu \left(\beta '^*\right)+m^{\mu } \nabla {}_\mu \left(\tau '^*\right)+2 m^{\mu }
	\nabla {}_\mu (\beta )+3 \beta '^* \tau '^*+8 \tau  \beta '^*-2 \left(\beta
	'^*\right)^2+8 \beta  \beta '^* \right. \nn &-& \left. 5 \beta  \tau '^*-\left(\tau '^*\right)^2-4 \tau  \tau
	'^*-6 \beta ^2-8 \beta  \tau \right\rbrace + m^{\mu }\nabla {}_\mu \left(T_{lm}\right) \left(-3 \rho ^*+\epsilon ^*-4 \rho -5
	\epsilon \right)  \nn &+& l^{\mu } \nabla {}_\mu
	\left(m^{\nu } \nabla {}_\nu \left(T_{lm}\right)\right)+m^{\mu } \nabla {}_\mu
	\left(l^{\nu } \nabla {}_\nu \left(T_{lm}\right)\right)-l^{\mu } \nabla {}_\mu
	\left(l^{\nu } \nabla {}_\nu \left(T_{mm}\right)\right)-m^{\mu } \nabla {}_\mu
	\left(m^{\nu } \nabla {}_\nu \left(T_{ll}\right)\right)
	\, , \nn
	\cT^{(1)}_{2}  &=& T_{m \bar{m}} \left\lbrace 2 l^{\mu } \nabla {}_\mu (\sigma )-2 \sigma  \rho ^*+2 \sigma  \epsilon
	^*-2 \rho  \sigma -\Psi _0-6 \sigma  \epsilon \right\rbrace \\ &+& T_{l \bar{m}} \left\lbrace 2 \sigma 
	\left(-\beta '^*+\tau '^*+3 \beta +\tau \right)-2 m^{\mu } \nabla {}_\mu (\sigma
	)\right\rbrace \nn &+& \sigma  \left(l^{\mu } \nabla {}_\mu \left(T_{m \bar{m}}\right)+l^{\mu } \nabla
	{}_\mu \left(T_{l n }\right)\right) - 2 \sigma  m^{\mu } \nabla {}_\mu \left(T_{l
		\bar{m}}\right) \nn &+& \kappa 
	\left(m^{\mu } \nabla {}_\mu \left(T_{m \bar{m}}\right)+m^{\mu } \nabla {}_\mu
	\left(T_{l n}\right)\right) + 3 \sigma  \bar{m}^{\mu } \nabla {}_\mu \left(T_{lm}\right) \nn &-& 3 \kappa  \bar{m}^{\mu } \nabla {}_\mu
	\left(T_{mm}\right)+l^{\mu } \sigma '^* \nabla {}_\mu
	\left(T_{ll}\right) - 3 \sigma  n^{\mu } \nabla {}_\mu \left(T_{ll}\right)+3 \kappa  n^{\mu }
	\nabla {}_\mu \left(T_{lm}\right) \nn &+& T_{ll} \left\lbrace l^{\mu } \nabla {}_\mu \left(\sigma
	'^*\right)+3 \sigma  \rho '^*-4 \rho  \sigma '^*-\rho ^* \sigma '^*-6 \sigma  \epsilon
	'^*-3 \epsilon  \sigma '^*+\epsilon ^* \sigma '^*-6 \sigma  \epsilon '\right\rbrace \nn&+& T_{l n}
	\left\lbrace 2 l^{\mu } \nabla {}_\mu (\sigma )-2 \sigma  \rho ^*+2 \sigma  \epsilon ^*-2 \rho 
	\sigma -\Psi _0-6 \sigma  \epsilon \right\rbrace \nn &+& T_{mm} \left\lbrace m^{\mu } \nabla {}_\mu
	\left(\kappa ^*\right)+\kappa ^* \beta '^*-6 \kappa  \beta ^*-3 \beta  \kappa ^*-\kappa
	^* \tau '^*-4 \tau  \kappa ^*+3 \kappa  \tau ^*-6 \kappa  \beta '\right\rbrace \nn &+&  \kappa
	^* m^{\mu } \nabla {}_\mu \left(T_{mm}\right)+\frac{1}{2} \left(\sigma  l^{\mu } \nabla
	{}_\mu (T)-\kappa  m^{\mu } \nabla {}_\mu (T)\right)-2 \kappa 
	l^{\mu } \nabla {}_\mu \left(T_{nm}\right)-2  T_{nm} l^{\mu } \nabla {}_\mu
	(\kappa ) \nn &+&6 T_{lm} \left\lbrace\sigma  \left(\beta '-\tau
	^*\right)-\kappa  \rho '^*+\kappa  \epsilon '\right\rbrace+2 \kappa  T_{nm} \left\lbrace\rho
	^*-\epsilon ^*+\rho +3 \epsilon \right\rbrace\, , \nn
	\cT^{(2)}_{2}  &=& \left(3 \kappa  \kappa '^*\right)T_{ll}+\left(3 \sigma  \sigma ^*\right)T_{mm} \, , 
	\eea
\end{subequations}
and
\begin{equation}\label{eqn:universalteukolskyNP4}
\cO^{(0)}_{-2}\left(\Psi_4\right) +\cO^{(1)}_{-2}\left(\Psi_3\right)+ \cO^{(2)}_{-2}\left(\Psi_4\right)   = 8 \pi \left(\cT^{(0)}_{-2} +\cT^{(1)}_{-2} +\cT^{(2)}_{-2}\right) \, ,
\end{equation} 
with
\begin{subequations}
	\bea
	\cO^{(0)}_{-2} &=&  2 \left(-\beta ^*+\tau ^*+3 \beta' +4 \tau'
	\right) m^{\mu } \nabla {}_\mu  -2 \bar{m}^{\nu } \nabla {}_\nu m^{\mu } \nabla {}_\mu \\   &+& 2  \left(-n^{\mu } \nabla {}_\mu \left(\rho \right)+4 n^{\mu
	} \nabla {}_\mu \left(\epsilon \right)-4 \bar{m}^{\mu } \nabla {}_\mu \left(\beta
	\right)+\bar{m}^{\mu } \nabla {}_\mu \left(\tau\right) \right) \nn  &+& 2\left( 4 \beta  \tau ^*+\tau  \beta ^*-4
	\beta  \beta ^*+\rho  \rho'{}^*-\tau \tau^*-4 \epsilon  \rho'{}^*+4 \epsilon 
	\epsilon'{}^*-\rho  \epsilon'{}^*+16 \tau'  \beta  \right. \nn &+& \left. 12 \beta'  \beta -3 \beta'  \tau +4 \rho'
	\rho -4 \tau'  \tau -3 \Psi _2-16 \rho'  \epsilon -12 \epsilon'  \epsilon +3 \epsilon' 
	\rho \right) \nn &+& 2 \left(-\rho'{}^*+\epsilon'{}^*-4
	\rho' -3 \epsilon' \right) l^{\mu } \nabla {}_\mu +2 n^{\nu } \nabla {}_\nu l^{\mu } \nabla {}_\mu
	\nn &+& 2  \left(4 \epsilon -\rho \right) n^{\mu } \nabla {}_\mu
	+2  \left(\tau -4 \beta \right) \bar{m}^{\mu } \nabla {}_\mu 
	\, , \nn
	\cO^{(1)}_{-2} &=&  8  l^{\mu } \nabla {}_\mu \left(\kappa '\right)+8 \kappa ' l^{\mu } \nabla
	{}_\mu \nn &-& 8 m^{\mu } \nabla {}_\mu \left(\sigma '\right)-8
	\sigma ' 	m^{\mu } \nabla {}_\mu \nn &-& 8  \sigma ' \beta '^*-8
	\kappa ' \rho ^*+8  \sigma ' \tau '^*+8 
	\kappa ' \epsilon ^*-40 \beta  \sigma '+20 \Psi _3 +40
	\epsilon  \kappa '\, , \\
	\cO^{(2)}_{-2} &=& 6 \left(\kappa \kappa'-\sigma \sigma'\right) \, , \\  \nn
	\cT^{(0)}_{-2}  &=& T_{n \bar{m}} \left\lbrace-2 n^{\mu } \nabla {}_\mu \left(\tau ^*\right)-2 \bar{m}^{\mu } \nabla {}_\mu
	\left(\rho'{} ^*\right)-2 n^{\mu } \nabla {}_\mu (\beta' )-2 \bar{m}^{\mu } \nabla {}_\mu (\epsilon'
	) -  2 \rho'{} ^* \beta ^*-2 \epsilon'  \beta ^* \right. \\ &+&  \left. 8 \beta'  \rho'{} ^*  - 2 \beta'  \epsilon'{} ^*+8 \rho'
	\tau ^*+4 \rho'{}^* \tau ^* + 8 \tau'  \rho'{} ^*+8 \epsilon'  \tau ^* - 2 \epsilon'{} ^* \tau
	^*+8 \beta'  \rho' +12 \beta'  \epsilon' +8 \tau'  \epsilon' \right\rbrace \nn &+& T_{\bar{m} \bar{m}} \left\lbrace n^{\mu } \nabla {}_\mu \left(\rho'{} ^*\right)-2 n^{\mu } \nabla
	{}_\mu \left(\epsilon'{} ^*\right)+2 n^{\mu } \nabla {}_\mu (\epsilon' )-\left(\rho'{}
	^*\right)^2-4 \rho'  \rho'{} ^*+8 \rho'  \epsilon'{} ^*-5 \epsilon'  \rho'{} ^*\right. \nn &+& \left.3 \rho'{} ^* \epsilon'{}
	^*-2 \left(\epsilon'{} ^*\right)^2 +  8 \epsilon'  \epsilon'{} ^*-6 \epsilon'{} ^2-8 \rho'  \epsilon'
	\right\rbrace + n^{\mu } \nabla {}_\mu
	\left(T_{n \bar{m}}\right) \left(\beta ^*-3 \tau ^*-5 \beta' -4 \tau'
	\right) \nn &+& n^{\mu } \nabla {}_\mu
	\left(T_{\bar{m} \bar{m}}\right) \left(2 \rho'{} ^*-3 \epsilon'{} ^*+4 \rho' +5 \epsilon' \right)  + \bar{m}^{\mu } \nabla {}_\mu \left(T_{nn}\right) \left(-3
	\beta ^*+2 \tau ^*+5 \beta' +4 \tau' \right) \nn &+& T_{nn} \left\lbrace-2 \bar{m}^{\mu } \nabla
	{}_\mu \left(\beta ^*\right)+\bar{m}^{\mu } \nabla {}_\mu \left(\tau ^*\right)+2 \bar{m}^{\mu }
	\nabla {}_\mu (\beta' )+3 \beta ^* \tau ^*+8 \tau'  \beta ^*-2 \left(\beta^*\right)^2+8 \beta'  \beta^* \right. \nn &-& \left. 5 \beta'  \tau ^*-\left(\tau ^*\right)^2-4 \tau'  \tau^*-6 \beta'{} ^2-8 \beta'  \tau' \right\rbrace + \bar{m}^{\mu }\nabla {}_\mu \left(T_{n \bar{m}}\right) \left(-3 \rho'{} ^*+\epsilon'{} ^*-4 \rho' -5
	\epsilon' \right)  \nn &+& n^{\mu } \nabla {}_\mu
	\left(\bar{m}^{\nu } \nabla {}_\nu \left(T_{n \bar{m}}\right)\right)+\bar{m}^{\mu } \nabla {}_\mu
	\left(n^{\nu } \nabla {}_\nu \left(T_{n \bar{m}}\right)\right)-n^{\mu } \nabla {}_\mu
	\left(n^{\nu } \nabla {}_\nu \left(T_{\bar{m} \bar{m}}\right)\right)-\bar{m}^{\mu } \nabla {}_\mu
	\left(\bar{m}^{\nu } \nabla {}_\nu \left(T_{nn}\right)\right)
	\, , \nn 
	\cT^{(1)}_{-2}  &=& T_{m \bar{m}} \left\lbrace 2 n^{\mu } \nabla {}_\mu (\sigma' )-2 \sigma'  \rho'{} ^*+2 \sigma'  \epsilon'{}
	^*-2 \rho'  \sigma' -\Psi _4-6 \sigma'  \epsilon' \right\rbrace \\ &+& T_{n m} \left\lbrace 2 \sigma' 
	\left(-\beta ^*+\tau ^*+3 \beta' +\tau' \right)-2 \bar{m}^{\mu } \nabla {}_\mu (\sigma'
	)\right\rbrace \nn &+& \sigma'  \left(n^{\mu } \nabla {}_\mu \left(T_{m \bar{m}}\right)+n^{\mu } \nabla
	{}_\mu \left(T_{l n }\right)\right) - 2 \sigma'  \bar{m}^{\mu } \nabla {}_\mu \left(T_{n
		m}\right) \nn &+& \kappa' 
	\left(\bar{m}^{\mu } \nabla {}_\mu \left(T_{m \bar{m}}\right)+\bar{m}^{\mu } \nabla {}_\mu
	\left(T_{l n}\right)\right) + 3 \sigma'  m^{\mu } \nabla {}_\mu \left(T_{n \bar{m}}\right) \nn &-& 3 \kappa'  m^{\mu } \nabla {}_\mu
	\left(T_{\bar{m} \bar{m}}\right)+n^{\mu } \sigma ^* \nabla {}_\mu
	\left(T_{nn}\right) - 3 \sigma'  l^{\mu } \nabla {}_\mu \left(T_{nn}\right)+3 \kappa'  l^{\mu }
	\nabla {}_\mu \left(T_{n \bar{m}}\right) \nn &+& T_{nn} \left\lbrace n^{\mu } \nabla {}_\mu \left(\sigma^*\right)+3 \sigma'  \rho ^*-4 \rho'  \sigma ^*-\rho'{} ^* \sigma ^*-6 \sigma ' \epsilon
	^*-3 \epsilon'  \sigma ^*+\epsilon'{} ^* \sigma ^*-6 \sigma'  \epsilon \right\rbrace \nn&+& T_{l n}
	\left\lbrace 2 n^{\mu } \nabla {}_\mu (\sigma' )-2 \sigma'  \rho'{} ^*+2 \sigma'  \epsilon'{} ^*-2 \rho' 
	\sigma' -\Psi _4-6 \sigma'  \epsilon' \right\rbrace \nn &+& T_{\bar{m} \bar{m}} \left\lbrace \bar{m}^{\mu } \nabla {}_\mu
	\left(\kappa'{} ^*\right)+\kappa'{} ^* \beta ^*-6 \kappa'  \beta'{} ^*-3 \beta'  \kappa'{} ^*-\kappa'{}
	^* \tau^*-4 \tau'  \kappa'{} ^*+3 \kappa'  \tau'{} ^*-6 \kappa'  \beta \right\rbrace \nn &+&  \kappa'
	{}^* m^{\mu } \nabla {}_\mu \left(T_{\bar{m} \bar{m}}\right)+\frac{1}{2} \left(\sigma'  n^{\mu } \nabla
	{}_\mu (T)-\kappa'  \bar{m}^{\mu } \nabla {}_\mu (T)\right)-2 \kappa' 
	n^{\mu } \nabla {}_\mu \left(T_{l \bar{m}}\right)-2  T_{l \bar{m}} n^{\mu } \nabla {}_\mu
	(\kappa' ) \nn &+&6 T_{n \bar{m}} \left\lbrace \sigma'  \left(\beta -\tau'
	{}^*\right)-\kappa'  \rho ^*+\kappa'  \epsilon \right\rbrace +2 \kappa'  T_{l \bar{m}} \left\lbrace \rho'
	{}^*-\epsilon'{} ^*+\rho +3 \epsilon' \right\rbrace \, , \nn
	\cT^{(2)}_{-2}  &=& \left(3 \kappa' \kappa{}^* \right) T_{nn}+\left(3 \sigma' \sigma'{}^* \right)T_{\bar{m} \bar{m}} \, .
	\eea
\end{subequations}

\section{Starobinsky-Teukolsky identities}\label{app:STidentities}
The Starobinsky-Teukolsky (ST) identities provide a relation between the Teukolsky equations of spins $\pm s$ \cite{teukolsky1972rotating,teukolsky1972rotating,Starobinsky:1973aij,Chandrasekhar:1984siy} --- see also \cite{Fiziev:2009ud} for a modern view on these identities.

Let us start by reviewing the ST identities for the angular functions\footnote{In this section we omit the $lm$ labels for the sake of clarity.} $S_{s}(x)$. Given two spin-weighted harmonic functions $S_{+2}$ and $S_{-2}$ that satisfy the equation \req{angularequation} and that are normalized according to 
\begin{equation}
2\pi\int_{-1}^{1}dx S_{s}(x)^2=1\, ,
\end{equation}
these are related by
\begin{equation}\label{angularSTidentities}
\begin{aligned}
S_{-2}&=\frac{1}{D_{2}}L_{-1}L_{0}L_{1}L_{2} S_{+2}\, ,\\
S_{+2}&=\frac{1}{D_{2}}L^{\dagger}_{-1}L^{\dagger}_{0}L^{\dagger}_{1}L^{\dagger}_{2} S_{-2}\, ,\\
\end{aligned}
\end{equation}
where $L_{n}$ and $L_{n}^{\dagger}$ are the operators
\begin{equation}
\begin{aligned}
L_{n}&= \frac{m+nx}{\sqrt{1-x^2}}-a \omega \sqrt{1-x^2}-\sqrt{1-x^2}\frac{\partial }{\partial x}\, ,\\
L_{n}^{\dagger}&= \frac{-m+nx}{\sqrt{1-x^2}}+a \omega \sqrt{1-x^2}-\sqrt{1-x^2}\frac{\partial }{\partial x}\, .
\end{aligned}
\end{equation}
The constant $D_{2}$ is in fact the same in the two identities on account of the normalization of the functions, and it is given by  
\begin{equation}\label{D2value}
\begin{aligned}
D_{2}=\Big[&\left(8+6 B_{lm}+B_{lm}^2\right)^2-8 \left(-8+B_{lm}^2 (4+B_{lm})\right) m \gamma +4 \left(8-2 B_{lm}-B_{lm}^2+B_{lm}^3\dvvtag
+2 (-2+B_{lm}) (4+3 B_{lm})
m^2\right) \gamma ^2-8 m \left(8-12 B_{lm}+3 B_{lm}^2+4 (-2+B_{lm}) m^2\right) \gamma ^3\dvtag
+2 \left(42-22 B_{lm}+3 B_{lm}^2+8 (-11+3
B_{lm}) m^2+8 m^4\right) \gamma ^4\dvtag-8 m \left(3 B_{lm}+4 \left(-4+m^2\right)\right) \gamma ^5
+4 \left(-7+B_{lm}+6
m^2\right) \gamma ^6-8 m \gamma ^7+\gamma ^8\Big]^{1/2}\, ,
\end{aligned}
\end{equation}
where $\gamma=a\omega$.

Consider now the radial ST identities. Suppose that we have two variables $R_{\pm 2}$ satisfying the corresponding radial Teukolsky equations (without corrections)
\begin{equation}
\mathfrak{D}_{+2}^2R_{+2}=0\, ,\quad \mathfrak{D}_{-2}^2R_{-2}=0\, .
\end{equation}
Then, let us define two auxiliary variables as follows 

\begin{equation}\label{STidentities1}
\begin{aligned}
\tilde{R}_{-2}&=\Delta^2\left(\mathcal{D}_{0}\right)^4\left(\Delta^2 R_{+2}\right)\, ,\\
\tilde{R}_{+2}&=\left(\mathcal{D}^{\dagger}_{0}\right)^{4}R_{-2}\, ,
\end{aligned}
\end{equation}
where $\mathcal{D}_{0}$ and $\mathcal{D}^{\dagger}_{0}$ are the operators

\begin{equation}
\begin{aligned}
\mathcal{D}_{0}&=\partial_{r}+\frac{i\left(\omega(r^2+a^2)-m a\right)}{\Delta}\, ,\\
\mathcal{D}^{\dagger}_{0}&=\partial_{r}-\frac{i\left(\omega(r^2+a^2)-m a\right)}{\Delta}\, .
\end{aligned}
\end{equation}
Then, it follows by direct computation that the variables $\tilde{R}_{\pm 2}$ also satisfy the corresponding Teukolsky equation $\mathfrak{D}_{\pm 2}^2\tilde{R}_{\pm 2}=0$. 
Furthermore, when studying black hole perturbations, we are interested in solutions that satisfy outgoing boundary conditions at infinity and at the black hole horizon. The map \req{STidentities1} preserves these conditions, meaning that if the original variables $R_{\pm 2}$ satisfy them, so do the transformed variables $\tilde{R}_{\pm 2}$. Due to the unicity of solutions of a second order ODE with fixed boundary conditions, it follows that, in fact, the transformed variables must be proportional to the original ones, $\tilde{R}_{\pm 2}\propto R_{\pm 2}$. Therefore, there must exist two constants $C_{\pm 2}$ such that 
\begin{equation}\label{STidentitiesapp}
\begin{aligned}
R_{-2}&=C_{+2}\Delta^2\left(\mathcal{D}_{0}\right)^4\left(\Delta^2 R_{+2}\right)\, ,\\
R_{+2}&=C_{-2}\left(\mathcal{D}^{\dagger}_{0}\right)^{4}R_{-2}\, ,
\end{aligned}
\end{equation}
The two constants are not independent due to the identity obtained by applying the map twice, 
\begin{equation}
R_{-2}=C_{+2}C_{-2}\Delta^2\left(\mathcal{D}_{0}\right)^4\left(\Delta^2\left(\mathcal{D}^{\dagger}_{0}\right)^{4}R_{-2}\right)\, .
\end{equation}
This allows one to conclude that
\begin{equation}\label{Cproductapp}
C_{+2}C_{-2}=\frac{1}{\mathcal{K}^2}\, ,
\end{equation}
where the constant $\mathcal{K}^2$ reads
\begin{equation}\label{eq:K2value}
\mathcal{K}^2=D_{2}^2+144 M^2 \omega^2\, .
\end{equation}

\bibliographystyle{JHEP}
\bibliography{Gravities.bib}

\end{document}